\documentclass [11pt]{scrreprt}

\usepackage{cite}

\usepackage{amsmath}
\usepackage{amssymb}
\usepackage{caption}
\usepackage{graphicx}
\usepackage{placeins}
\usepackage{scrhack}
\usepackage{subcaption}
\usepackage{afterpage}
\usepackage{nicefrac}

\usepackage{setspace}
\usepackage{gensymb}
\usepackage{textpos}
\usepackage{siunitx}

\usepackage{pythonhighlight}
\usepackage[english]{babel}

\graphicspath{{./Thesis_Images/}}
\usepackage[pdftex]{hyperref} 

\hypersetup{ 
	pdfauthor = {Tobias Peherstorfer},
	pdftitle = {Shattered pellet fragment distribution analysis},
	pdfsubject = {project thesis physics},
	pdfkeywords = {Master},
	pdfdisplaydoctitle = true,
	colorlinks, 
	citecolor=blue,
    	filecolor=blue,
    	linkcolor=blue,
    	urlcolor=blue,
	pdfborder = {0 0 0}
}

\renewcommand{\maketitle}{
\begin{titlepage}

\begin{textblock*}{5cm}(-10mm, -10mm)
     \includegraphics[width=5.5cm, height=2cm]{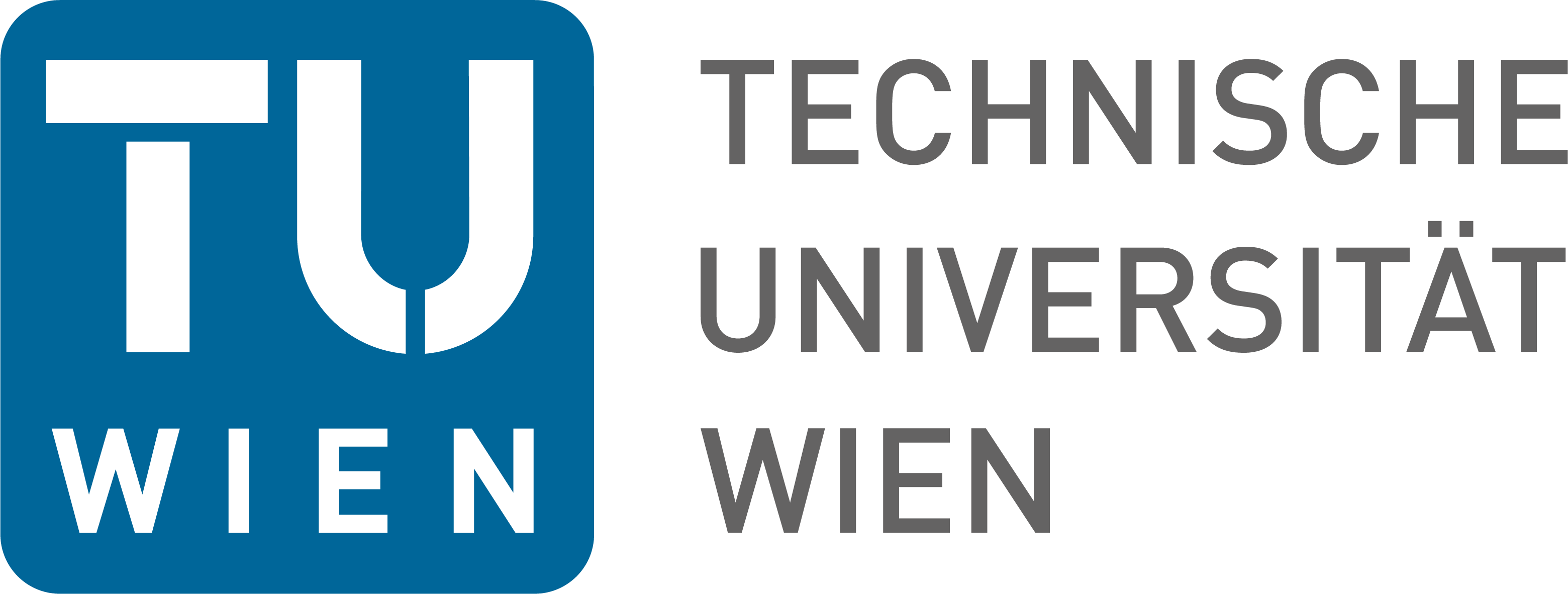}
     \end{textblock*}

\begin{textblock*}{5cm}(65mm, -10mm)
     \includegraphics[width=9cm, height=2cm]{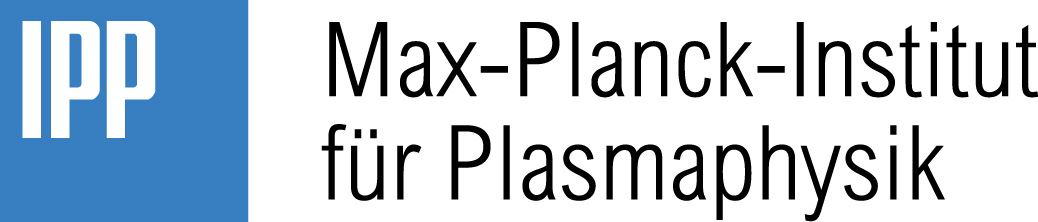}
     \end{textblock*}

\begin{singlespace*}
\begin{center}

\vspace*{2cm}
{\LARGE Diplomarbeit} \\[5ex]
{\LARGE \textbf{Fragmentation Analysis of Cryogenic Pellets for Disruption Mitigation}
\\[5ex]}
{\large
Zur Erlangung des akademischen Grades}\\[3ex] 
{\LARGE
\textbf{Diplom-Ingenieur}}
\\[3ex]
Im Rahmen des Studiums\\[3ex]
{\Large
\textbf{Technische Physik}}
\\[3ex]
eingereicht von\\[3ex]
{\Large
\textbf{Tobias Peherstorfer BSc}}
\\[1ex]
Matrikelnummer 01615523
\end{center}

\vspace*{1cm}
{\large \noindent
Ausgeführt am Institut für Angewandte Physik \\
der Fakultät für Physik der Technischen Universität Wien\\
in Zusammenarbeit mit dem Max-Planck-Institut für Plasmaphysik Garching\\[2ex]

\noindent
Betreuung:\\
Univ.-Prof. Mag.rer.nat. Dipl.-Ing. Dr.techn. Friedrich Aumayr (TU Wien)\\
Dr. Gergely Papp (IPP Garching)

\vfill\noindent
Wien, 29.08.2022 \hfill \rule[-2mm]{4cm}{0.1mm} \hfill \rule[-2mm]{4cm}{0.1mm}
}

\end{singlespace*}
\end{titlepage}}

\begin{document}
\title { Master thesis:\\
		}
\author{Tobias Peherstorfer BSc}

\maketitle
\pagenumbering{gobble}

\setcounter{page}{1} \pagenumbering{roman}

\begin{abstract}
\phantomsection
\addcontentsline{toc}{chapter}{Abstract}
\thispagestyle{plain}
\begin{center}
\Large
\textbf{Fragmentation Analysis of Cryogenic Pellets for Disruption Mitigation}
      
    \large   
    \vspace{0.4cm}
    \textbf{Tobias Peherstorfer}

    \vspace{0.9cm}
    \textbf{Abstract}
\end{center}

The lifetime and utilization of a nuclear fusion reactor like ITER depends strongly on its capabilities to mitigate damage during disruptions.
While shattered pellet injection (SPI) was chosen as the baseline mitigation method for ITER, the exact relation of pellet injection parameters and the resulting fragment distributions is not yet clear.
This knowledge is of paramount importance for optimizing the impurity deposition and disruption mitigation efficiency.
In this thesis, I present fragmentation analysis of 170 SPI pellets, with the focus on the produced fragment sizes as a function of normal impact velocity.
The experiments were carried out at the Max Planck-Institute for Plasma Physics (IPP) in Garching, Germany, using the shattered pellet injection system that is now installed on the ASDEX Upgrade tokamak.
Comparisons with a theoretical fragmentation model (Parks 2016) show that the model underestimates the amount of fragments below 0.9 mm in diameter.
Also, statistical parameters such as the mean fragment size and the standard deviation of fragment size seem to scale exponentially in the experiment and linearly in the model.
Furthermore, we found that the fragmentation induced by circular shattering geometries is less reproducible than for rectangular geometries, which might be relevant for design choices in future mitigation systems.

\end{abstract}

\newpage
\tableofcontents
\newpage

\setcounter{page}{1} \pagenumbering{arabic}

\chapter*{Acknowledgements}

I would like to thank my supervisor Gergely Papp, not only for his profound knowledge about plasma physics he shared with me, not only for his management skills that helped me focus on the right things at the right time, but also for his hospitality, and humor, which made this project much more than work for me.
I look forward to progressing our academic journey, and also to having a beer sometime soon. \par
Also, I would like to thank Friedrich Aumayr for supervising my thesis at TU Vienna and giving me the opportunity to learn and discuss in the weekly meetings of his research group. \par
Next, I thank my colleague and friend Paul Heinrich for his excellent remarks on the code, his proactiveness about trying new things and for the great time we often had via voicechat, even with loud vacuum pumps in the background.
He also greatly contributed to the work presented in this thesis, as he was responsible for conducting the integrity measurements for each investigated pellet. \par
Another person that helped me keep my sanity, even when a bug in the code would not disappear for weeks, is my friend Felix.
Thank you for being the supportive, funny fellow you are. \par
Of course none of this would have been possible without my mother and my father, who both ensured I had the means to get a good education, but never pressured me in any direction.
And, last but not least, I want to thank my girlfriend Raphaela for her love and support during the last years.
I don't think I would have made it so far without you.

\chapter{Introduction} 

Nuclear fusion has been an active field of research for over 70 years, as it represents a possibility of climate-friendly energy production that, unlike other sustainable energy sources like solar or wind power, does not rely on external parameters like weather conditions and season \cite{Euratom_H2020}.
The dream of achieving fusion energy is supported by countries all over the world, resulting in multinational projects like ITER and research support structures like EUROfusion.
The most promising approach to achieve fusion energy is magnetic confinement fusion, where hot fusion plasma is trapped inside a magnetic cage \cite{Plasmaphysik}.
ITER is a magnetic confinement fusion reactor of the tokamak type \cite{Tokamaks,Fusion_Physics_IAEA}, where charged particles are trapped in a doughnut-shaped volume of 830 cubic meters.
To reach fusion conditions inside this torus, the plasma will carry currents of several MA and be heated to temperatures of the order of 300 million \degree C, ten times hotter than the core of the Sun \cite{ITER_website}.
During normal deuterium-trituim operation, about 400 MJ of magnetic energy and 350 MJ of thermal energy will be well-contained \cite{lehnen2015disruptions}, but instabilities can lead to a so-called disruption \cite{Tokamaks}, a dramatic event where confinement is lost.
The thermal energy of the plasma can be deposited locally, damaging (e.g. melting) vessel components and can also lead to impurity transport into the plasma.
As the plasma temperature drops, its resistivity rises, causing the plasma current to decay rapidly, inducing eddy currents in the surrounding vessel structure and causing the plasma current to partially run through the reactor wall.
This imposes high electromagnetic loads on the vessel, potentially damaging components \cite{hender07mhd ,lehnen2015disruptions}.
Additionally, beams of relativistic electrons can be generated during this phase, which can cause severe melting on wall tiles and even puncture cooling structures \cite{MHD_Zohm, matthews2016melt}.\\

\noindent
Disruptions and their mitigation are a critical issue directly affecting the lifetime of nuclear fusion reactors of the size of ITER \cite{hender07mhd}. 
This is even more apparent with the advent of first plasma in ITER, currently scheduled for 2025, as the destructive effects of disruptions become even harder to control in larger machines \cite{Disruptions_Boozer}.
As these events can likely not be avoided completely, they must be mitigated to minimize damage on reactor components.
Shattered Pellet Injetion (SPI) was chosen as the baseline disruption mitigation method for ITER \cite{lehnen2015disruptions}.
Here, a mixture of hydrogen and neon is frozen into a cryogenic pellet, fired towards the plasma. To accelerate the pellet ablation, it is broken into pieces shortly before entering the plasma.
These impurities then transfer thermal energy from the plasma towards the vessel wall via homogeneous radiation to limit damage during localized energy deposition.
Furthermore, this cooling can be used to control the duration of the plasma current decay \cite{breizman2019physics}.
Here, the size, speed and spatial distribution of fragments dictates the exact assimilation of material into the plasma.
The investigation of mitigation capabilities of this method is well under way, with SPI systems installed at tokamaks like DIII-D, JET, K-STAR and, recently, ASDEX-Upgrade \cite{JET_SPI,KSTAR_SPI, DIII-D_SPI ,ASDEX_SPI_SOFT}.
However, the impact of parameters such as pellet speed, material and the angle between incoming pellet velocity and shatter tube surface (also called shatter angle) on the resulting plume characteristics are yet to be explored. \\

\noindent
In this thesis, I present fragmentation analysis of cryogenic pellets fired with the ASDEX-Upgrade SPI system.
These tests were not conducted in a tokamak, but in a dedicated laboratory test setup at the Max Planck-Institute for Plasma Physics (IPP) in Garching, Germany, with fast camera video diagnostics imaging the pellet both before and after shattering. 
I analyzed the recorded videos using computer vision software I have written in Python.
\par \noindent
In chapter~\ref{s:theoretical background} of this thesis, the concepts of nuclear fusion, magnetic confinement fusion reactors, disruptions and their mitigation are summarized shortly, along with a short rundown of existing research on SPI pellet fragmentation.
Then, in the sections~\ref{s:ASDEX-Upgrade pellet injector}~-~\ref{s:Test chamber optical design}, the experimental testing setup for the ASDEX-SPI and the optical diagnostic design are explained in detail.
In the sections~\ref{s:computer_vision}~-~\ref{s:integrity_diagnostic}, I summarize important computer vision methods and then present in detail the video diagnostic software for both the intact pellet and the fragments.
In chapter~\ref{s:Results}, the results obtained by video analysis are presented.
First, some limitations of the conducted analyses are discussed in section~\ref{s:limitations}.
Then, the scanned parameter space is presented in section~\ref{s:analysis_of_spi_pellets}.
In section~\ref{s:size_dist_res}, experimentally recorded fragment size distributions and simulated fragment size distributions for each recorded pellet are shown.
In section~\ref{s:Statistical fragment size parameters}, key statistical parameters of fragment size are presented both for the experimental and simulated case.
With this knowledge, detailed comparisons of modeled and measured fragment sizes are made in section~\ref{s:Comparison of experimental and simulated size distributions}.
Section~\ref{s:spatial} is a summary of spatial spray distributions analysis for different shatter tubes.
Motivated by unusual fragment plumes when using circular tubes, comparison of fragment sizes produced with rectangular and circular tube cross-sections is given in section~\ref{s:Comparison of rectangular and circular shatter tubes}.
Finally, a comparison of initial pellet volume and cumulative fragment volume is given in section~\ref{s:volume} and some example fragment speed distributions are discussed in seciton~\ref{s:speeds}.

\chapter{Theoretical Background}
\label{s:theoretical background}

\section{Nuclear Fusion}

Nuclear fusion describes the process, where two light nuclei (atomic mass number $ A~\leq$ 25) are joined to form a single heavier nucleus with higher average binding energy per nucleon \cite[ch. 4.1]{Fundamentals_nuclear}.
This increase in binding energy times the number of nucleons equals the energy that is released during the process. 
The goal of a nuclear fusion power plant is to harvest this energy.
One of the most promising fusion reactions for energy production is the D-T reaction between deuterium (stable hydrogen isotope with 1 proton and 1 neutron) and tritium (radioactive hydrogen isotope with 1 proton and 2 neutrons):

\begin{equation}
     ^2_1\text{D}+_1^3\negmedspace\text{T} \longrightarrow\, ^4_2\text{He} \,(3.5 \text{ MeV})+^1_0\negmedspace\text{n} \,(14.1 \text{ MeV}) \, .
     \label{e:d_t_reaction}
\end{equation}

\noindent
The reaction produces a helium nucleus (alpha particle) along with an energetic neutron.

\begin{figure}[h]
     \centering
     \includegraphics[clip, trim = 0 0 0 0, width=0.5\textwidth]{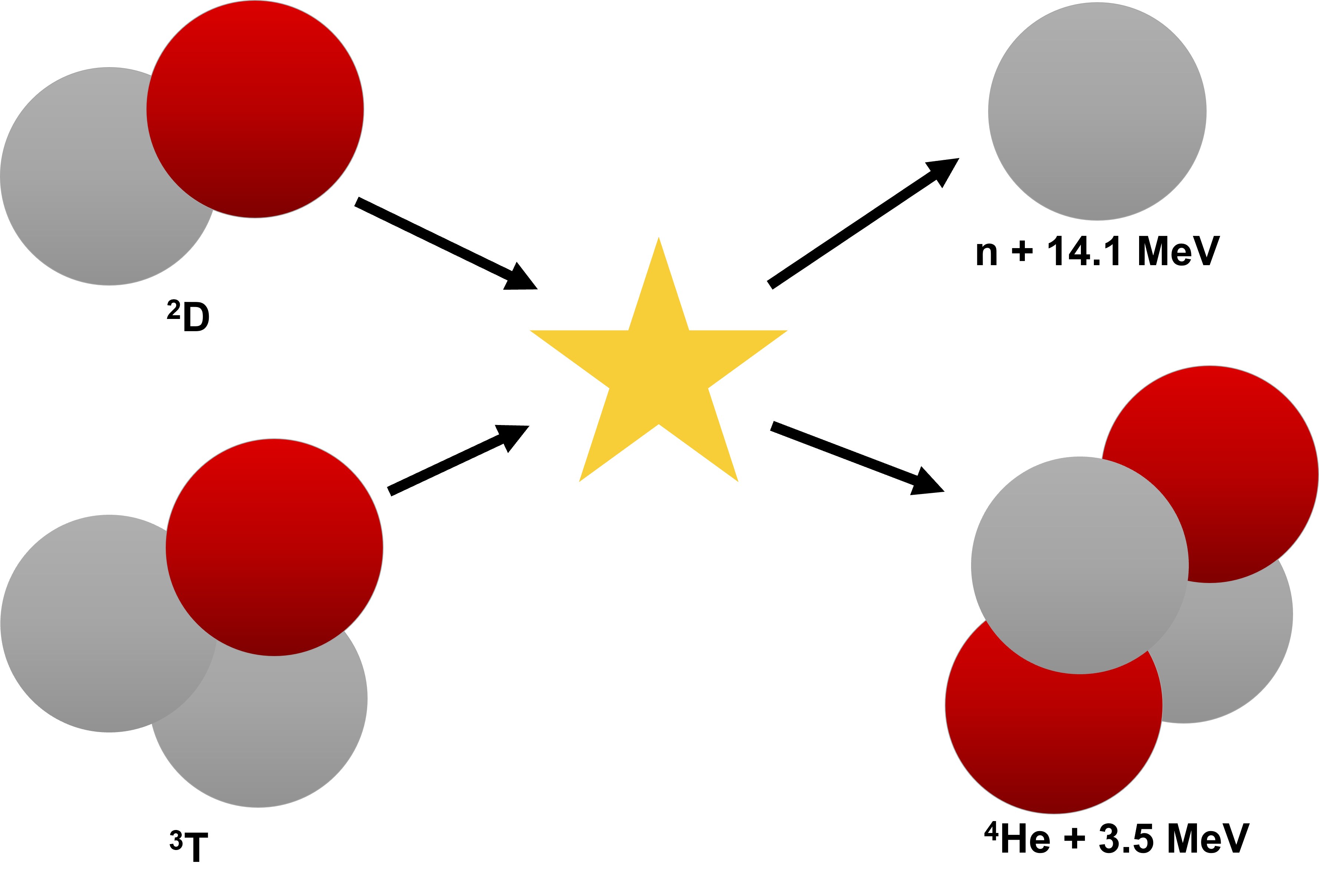}
     \captionsetup{width=\textwidth}
     \caption{Schematic of the D-T fusion reaction. Protons are colored in red, neutrons in grey.}
     \label{f:coord}
\end{figure} 

\noindent
However, nuclear fusion reactions only occur when the reactants have sufficient incident kinetic energy to overcome the repulsive Coulomb force between nuclei, often called Coulomb wall/barrier.
Therefore, a nuclear fusion power plant must provide enough energy to the fuel to achieve fusion without using more than what is gained through the reactions.
To achieve a positive energy balance, a sufficient density of fuel particles must retain high enough energy and remain in the reaction region for a sufficient amount of time \cite{Tokamaks}.
When, for example, a beam of particles is fired onto a solid target, the interacting particles lose their energy too quickly via elastic collisions and the interaction time is too short.
The most promising way of passing this Coulomb wall is increasing the temperature of the reactant materials to obtain a thermal medium \cite{Tokamaks}. 
This way, elastic collisions only lead to a redistribution of the energy within the fuel.
The required energies to obtain sufficient fuel temperatures are typically in the range of several keV (several 10 million~K)~\cite[ch. 6.7]{Fundamentals_nuclear}.
For the D-T reaction, fusion fuel will have to be heated to about 300 million K (25 keV)~\cite[ch. 1.2]{Fundamentals_IAEA}.
At these temperatures, material generally exists in the plasma state, where the electrons are stripped from their nuclei. 
Therefore, a plasma consists of the positively charged, fully or partially ionized ions and negatively charged electrons \cite{Plasmaphysik}.
The challenge in energy production via nuclear fusion is the confinement of a sufficiently hot plasma at a sufficient nuclide density for a sufficient time to allow fusion reactions to produce more energy than used to sustain and heat the plasma \cite{Tokamaks}.
One of the most prominent approaches to this problem is plasma confinement using magnetic fields in a reactor of the tokamak type. 

\section*{Magnetic Confinement in Tokamaks}

\begin{figure}[h]
     \centering
     \includegraphics[clip, trim = 100 150 100 150,width=0.8\textwidth]{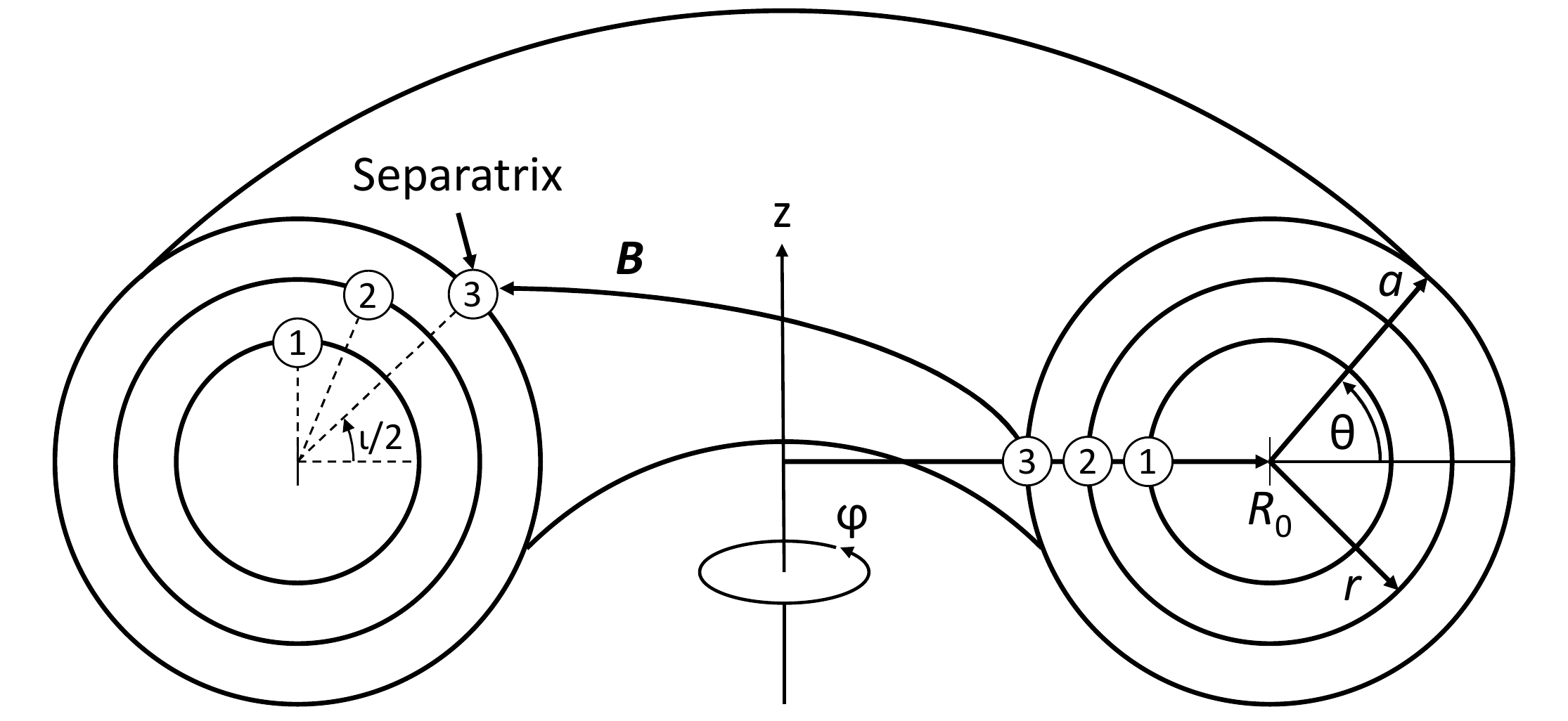}
     \captionsetup{width=0.9\textwidth}
     \caption{Toroidal coordinate system with toroidal angle $\varphi$, the poloidal angle $\theta$ and the generalized radial coordinate $r$. $R_0$ and $a$ denote the major and minor plasma radius. Adapded from Stroth \cite{Plasmaphysik}.}
     \label{f:coord}
\end{figure}

\noindent
In the presence of a magnetic field $\vec{B}$, charged particles experience the Lorentz-force $F_\text{L} = q\vec{v} \times \vec{B}$, which acts orthogonally to both the particle velocity $\vec{v}$ and the magnetic field $\vec{B}$~\cite{Plasmaphysik}.
Magnetic confinement  uses this effect to trap plasma constituents on magnetic field lines.
In tokamaks, the plasma is confined in a torus shaped volume and a large toroidal magnetic field of several T is applied by external coils. 
Additionally, a toroidal current (up to 15 MA) is driven inside to produce a poloidal magnetic field, and a vertical magnetic field of several T is generated by external coils.
The sum of these fields results in a wreath-like helical total magnetic field \cite{Tokamaks}.\\

\noindent
For the discussion of disruptions in tokamaks, first we introduce the coordinate system used to describe a tokamak plasma.
Due to the axisymmetric setup, the coordinate system consists of the toroidal angle $\varphi$, the poloidal angle $\theta$ and the generalized radial coordinate $r$.
Together with the major and minor plasma radius $R_0$ and $a$, as illustrated in fig.~\ref{f:coord}, this system is used for the description of the tokamak equilibrium.
It can be shown that in an axisymmetric tokamak, magnetic field lines form nested toroidal flux surfaces, where the flux of the magnetic field through the surface is constant \cite{Tokamaks}.
The pressure on a flux surface is also constant because transport along flux surfaces occurs orders of magnitude faster compared to transport orthogonal to flux surfaces \cite{Plasmaphysik}.
The generalized radial coordinate $r$ is often used to label the various toroidal surfaces \cite{Review_Boozer}.
The outermost closed flux surface is also called the separatrix, as shown in fig.~\ref{f:coord}.
The twist of the magnetic field lines depends on the radial position $r$ and is often expressed by the safety factor $q$. It is defined as the number of toroidal circumferences per poloidal circumference of the field lines \cite{Tokamaks} and is related to the quantity $\iota$ in fig.~\ref{f:coord} by $q = 2\pi / \iota$.

\FloatBarrier
\section*{Disruptions in Tokamaks}

A disruption is defined as an instability that leads to an uncontrolled loss of plasma current and terminates the discharge \cite{MHD_Zohm}.
These instabilities usually occur through failure of technical subsystems or when a tokamak reactor is driven close to so-called operational limits, which promotes the growth of plasma instabilities \cite{hender07mhd}.
Details on the operational boundaries are summarized by, for example, Koslowski \cite{Koslowski}.
The disruption can be divided into two phases.
First, the thermal energy of the plasma is lost to the vessel wall on a millisecond (ms) timescale.
The mean electron temperature is found to drop rapidly.
This phase is called thermal quench (TQ).
As shown by Cohen, Spitzer and Routly \cite{Spitzer_resistivity}, plasma resistivity $\eta$ rises with decreasing temperature, as $\eta \propto T_e^{-3/2}$.
This resistivity rise causes the plasma current to decay, initiating the second phase of the disruption, called the current quench (CQ) \cite{hender07mhd}.

\noindent
The effects of disruptions can be divided into thermal loads, electromagnetic loads (e.g. halo currents and eddy currents) and so-called runaway electrons~(REs).
During the TQ, high thermal loads are deposited locally on the divertor and on plasma-facing components (PFCs) on a timescale of approx. 0.1 - 1 ms, often exceeding melting/vaporization thresholds of the affected material \cite{hender07mhd}.\par \noindent
During the fast decay of the plasma current in the CQ, eddy currents are produced in PFCs by induction. 
The shorter the CQ duration, the higher the eddy currents.
For ITER, CQ durations below 50 ms lead to undesirably large eddy current forces on components \cite{hollmann2015status}. 
Also, vertical displacement events (VDEs) of the plasma can cause a poloidal current to partially flow through vessel components.
These currents, called halo currents, become more problematic for longer CQ durations, which is why an upper boundary of 150 ms for the acceptable CQ time was chosen for ITER \cite{hollmann2015status}. 

\subsection*{Runaway electrons}
\begin{figure}[h]
     \centering
     \includegraphics[clip, trim = 0 0 0 0,width=0.7 \textwidth]{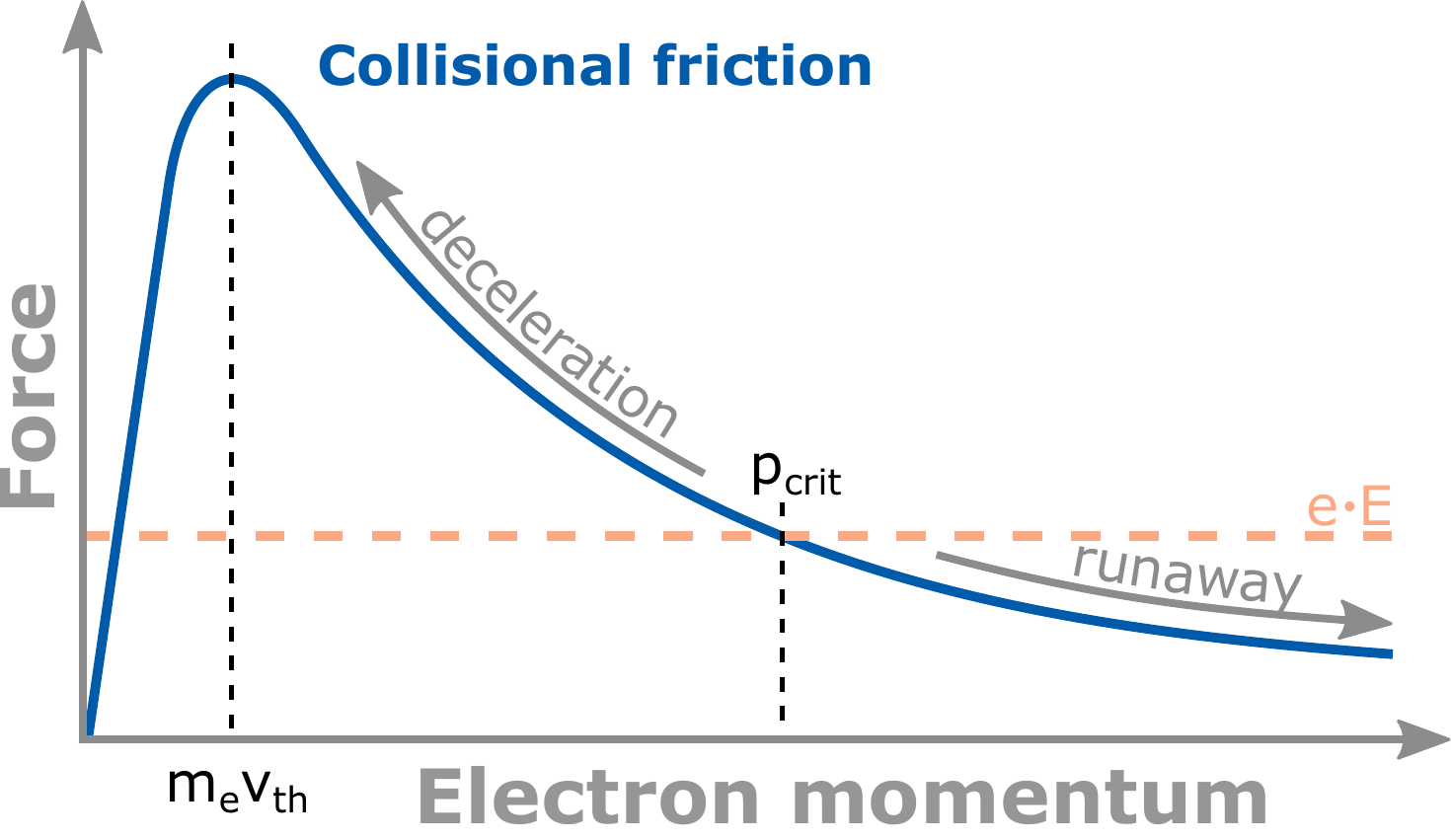}
     \captionsetup{width=0.7\textwidth}
     \caption{Non-monotonic collisional friction inside a plasma as a function of electron momentum. Above a critical momentum, electrons are accelerated to relativistic velocities. Adapted from Stroth~\cite{Plasmaphysik}. }
     \label{f:friction}
\end{figure}

Figure \ref{f:friction} shows the collisional friction force acting on electrons in a plasma as a function of their momentum.
Given an external force parallel to the magnetic field lines, electrons above a certain critical momentum, indicated as $p_\text{crit}$, are accelerated to relativistic energies, hence the name runaway electrons (REs) or runaways.
Several different sources of RE formation exist \cite{breizman2019physics}, but only the three most prominent will be discussed in this thesis.
During a rapid TQ, the (approximately) Maxwellian electron velocity distribution undergoes incomplete thermalization, as the collisional frequency is much lower for more energetic particles \cite{hot_tail}. 
During the CQ, electric fields can then become large enough that the remaining ``hot tail'' of the velocity distribution is in the runaway regime.
This so called ``hot tail'' mechanism is a primary RE generation process and is predicted to be the prominent source of runaways in ITER \cite{breizman2019physics}.
Additionally, the so-called Dreicer generation is a primary generation mechanism where long-range Coulomb collisions can lead to velocity-space diffusion of electrons into the runaway region \cite{hot_tail}.
The avalanche mechanism \cite{Rosenbluth_1997} is a secondary RE generation mechanism, where Coulomb collisions between runaways and electrons that are not yet above the critical energy cause an avalanche multiplication of runaways.
Relativistic electrons often form beams, which impact onto the PFCs when the magnetic equilibrium collapses \cite{hender07mhd}.
This is particularly dangerous because these beams penetrate wall material deeply and deposit their energy locally on a ms timescale \cite{lehnen2015disruptions}.
\\

\noindent
Disruptions can cause serious damage to reactor components and potentially put plasma operations on hold for several days or even weeks because of immediate maintenance needs \cite{hender07mhd}.
Therefore, they must be avoided or must be mitigated to minimize the damage on reactor components.

\newpage

\section{Material Injection for Disruption Mitigation}

The need for effective methods to mitigate disruptions is recognized in the scientific community \cite{hollmann2015status,MHD_Zohm,hender07mhd,Plasmaphysik,lehnen2015disruptions,Disruptions_Boozer,Baylor_2019}.
The goal is to evenly distribute thermal loads during the TQ, minimize electromagnetic loads of both halo currents and eddy currents during the CQ, prevent the generation of runaway electrons and dissipate already existing runaway beams.\\

\noindent
The most prominent approach in disruption mitigation is massive material injection, where a large amount of impurity atoms is introduced into the plasma.
Noble gases and H$_2$ are preferred due to their benign interaction with vessel components and the relative ease of extraction after mitigation \cite{hender07mhd}.
When injected, the impurities radiate the plasma thermal energy towards the vessel wall in an (ideally) isotropic fashion, reducing the heat load during wall impact considerably.
Indeed, it has been found that the injection of high pressure high-Z gas achieves the radiation of more than 90\% of the initial thermal energy in most present tokamaks \cite{hollmann2015status}.
This cooling mechanism can also be used to control the CQ duration and therefore balance forces originating from eddy currents and halo currents.
However, recent simulation results and theoretical work by Hesslow \textit{et al.} \cite{Hesslow2019InfluenceOM} indicate that, under certain circumstances, the avalanche growth can even be increased by the injection of material, as there are more electrons available for the avalanche mechanism.
The challenge of effective runaway electron mitigation is therefore of special relevance for the lifetime of future tokamaks.\\

\noindent
Several methods for the delivery of the impurities exist.
Both the injection of cryogenic impurity pellets and of massive amounts of impurity gas have been extensively studied \cite{Killer_pellet_DIII-D,MGIJET}.
Unfortunately, pellet injection is prone to causing runaway generation due to high electric fields during pellet ablation \cite{hender07mhd,PelletOverview}.
Gas injection, on the other hand, achieves excellent results in medium-sized tokamaks but suffers from comparatively slow delivery time and shallow penetration depth of the impurities in machines of the size of ITER \cite{hender07mhd}.
In ITER, impurities will be delivered in the form of cryogenic fragments via shattered pellet injection. 
Impurity pellets are shot towards the plasma at high speeds and broken up shortly before entering the plasma.
This is done both to increase the pellet ablation inside the plasma and to protect wall tiles in the center of the machine from direct pellet impact.
The penetration depth of the fragments is large enough to allow impurity deposition in the center of the plasma and the material is delivered after 10-20 ms, even in large machines \cite{breizman2019physics}.
There is an argument to be made that starting conditions would more well-defined if, instead of freezing one large pellet and shattering it, one would freeze many small pellets and fire them simultaneously.
However, as I will explain in the next section, the pellet formation and firing process is not trivial, making SPI with a single pellet much more feasible.

\section{Shattered Pellet Injection}

\begin{figure}[h]
     \centering
     \includegraphics[clip, trim = 0 160 0 130,width=0.8 \textwidth]{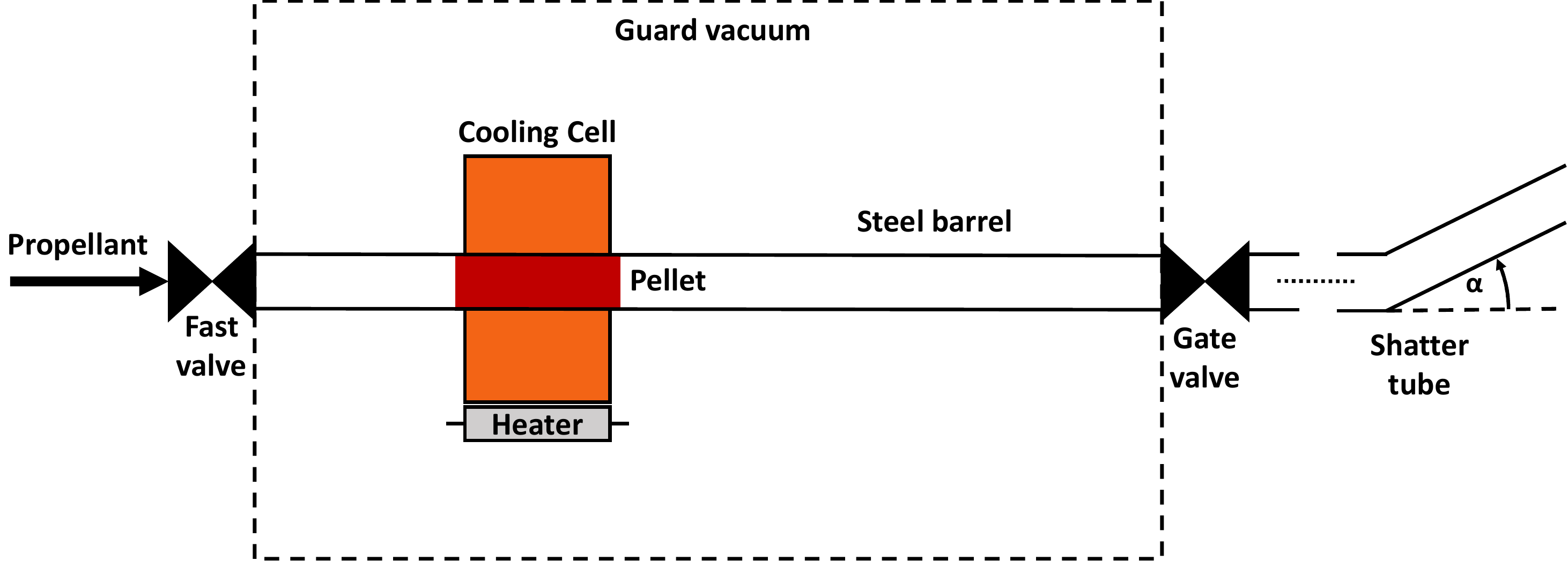}
     \captionsetup{width=0.8\textwidth}
     \caption{Sketch of a simple pipe gun SPI system, with a bent tube (shatter tube) as the shattering unit. $\alpha$ denotes the tube bend angle. Adapded from Combs and Baylor \cite{PelletOverview}.}
     \label{f:SPI_principle}
\end{figure}

A basic SPI system consists of a pellet generator, a launching system and a shattering unit shortly before the plasma.
To better understand steps in the development of present-day SPI systems, it is instructive to first look at the history of SPI.
Key developments in this disruption mitigation method have been made by Oak Ridge National Laboratories, who first demonstrated the use of a pipe gun injector system with a bent tube as the shattering unit in the DIII-D tokamak \cite{PelletOverview}.
A sketch of such a system is shown in fig.~\ref{f:SPI_principle}.

\noindent
Pellets were frozen inside a barrel using a liquid He cooling cycle and fired down a steel barrel using a H$_2$ or He gas pulse.
Pellets were either frozen from pure D$_2$ or Ne and accelerated to speeds of 600 and 300 m/s respectively.
It was found that Ne pellets could not be fired directly at the temperature they were frozen at (approx. 10 K) and instead had to be heated up several degrees to allow dislodging of the pellet.
To overcome this issue, Ne pellets with a thin D$_2$ shell were developed.
Furthermore, the use of D$_2$-Ne mixture pellets was investigated and a mechanical punch mechanism was developed to dislodge and fire slow Ne and Ar pellets.
Now, the performance of SPI as a disruption mitigation tool has been investigated on multiple tokamaks already \cite{KSTAR_SPI,JET_SPI}, but comparatively little effort has gone into the investigation of the pellet fracture process and the resulting fragment sizes.
Simulations with the 3D magnetohydrodynamics code NIMROD \cite{NIMROD_SPI} show a strong dependence of impurity assimilation efficiency on the fragment sizes. 
Breizman \textit{et al.} \cite{breizman2019physics} further explain that  impurity assimilation characteristics and penetration depth of SPI depend also on fragment velocity and the ablation rate. \\

\noindent
In the following section, the research on the shattering of cryogenic pellets, carried out by Oak Ridge National Laboratories, will be summarized shortly.
Then, a theoretical model for pellet fragmentation will be discussed in detail.

\section{Previous studies of SPI fragmentation}

Until 2020, most experimental studies on the fragmentation of SPI pellets have been published by scientists at Oak Ridge National Laboratories.
The results presented here were obtained in a dedicated SPI testing environment, where the shattered pellets are fired into a test tank.\par \noindent
In 2019, Baylor \textit{et al.} \cite{Baylor_2019} presented a three-barrel SPI system that uses a resonant microwave cavity along the flight path to measure pellet integrity and speed and images the fragment spray with a fast camera or witness foils.
They mention that fast camera imaging is challenging when a large number of small fragments is present and that witness foils are compromised when multiple fragments puncture the same location in the foil.
A single histogram of the fragment diameter distribution for an Ar pellet that is 8 mm in diameter, hitting a 20\degree~shatter tube at 180 m/s, is shown in figure 8 in their paper.\par \noindent
A publication by Gebhart \textit{et al.} \cite{Gebhart} focuses on the shattering analysis of pure Ar, Ne and D$_2$ pellets.
Here, threshold impact velocities for pellet breakage were recorded experimentally and embedded in a theoretical model, that will be discussed in the next section.
These threshold velocities were found to be independent of pellet diameter.
Additionally, in figure 8 in their paper, fragment size distributions for 8.5 mm Ar and 12.5 mm Ar, Ne and D$_2$ pellets at different perpendicular impact velocities are shown.
This perpendicular impact velocity is given by

\begin{equation}
     \label{e:vperp}
     v_\text{perp} = v_\text{pellet} \, \cos(\alpha) \, ,
\end{equation}

\noindent
where $v_\text{pellet}$ is the initial pellet speed and $\alpha$ is the tube bend angle. 
Good agreement with the statistical fragmentation model was found.\par \noindent
Afterwards, a similar paper, focusing on the characteristics of D$_2$-Ne mixture pellets, was published by Gebhart \textit{et al.} \cite{DeNeGebhart}, providing size distributions of 26\% and 59\% Ne pellets (partial pressure) in figure 5.
Again, good qualitative agreement with the theoretical model was reported.\par \noindent
Finally, another paper by Gebhart \textit{et al.} \cite{Shatter_tube_ORNL} investigates the impact of a 20\degree~mitre bend and 20\degree~modest S-bend shatter tube (as installed in JET) on the fragment plume.
Also, the effect of entrained propellant gas on plume properties is explored.
The report shows size distributions of 12.5 mm pellets for 5\%, 20\% and 100\% Ne pellets for different combinations of shatter tube designs and propellant gas handling.
These distributions were measured from fast camera videos depicting the fragments exiting the shatter tube using a particle tracking code based on frame subtraction and edge detection.
Furthermore, fragment velocity distributions were measured for these cases by manually picking 20 fragments in each video and following them for a number of frames.
This way, it was found that the front of the plume generally moves faster than the bulk, which in turn moves faster than the rear of the plume. 
This effect is more pronounced when propellant gas is removed before shattering.
Lastly, they present temporal mass evolutions of these pellets, measured 15 cm from the end of the shatter tube in intervals of 0.1 ms.\\

\noindent
Although results for particular combinations of material, pellet speed, shatter angle and geometry exist in the literature, a thorough scan of the parameter space has yet to be provided to conclude relationships between these parameters and plume characteristics.
Furthermore, details about the methods of obtaining fragment size from the fast camera videos are not publicly available.

\newpage

\section{Theoretical Pellet Fragmentation Models}
\label{s:theoretical_model}

To control the material deposition with SPI, it is important to understand the influence of different firing parameters on the resulting fragments.
The fragment size distribution is of considerable interest, because it strongly affects the penetration depth of the material inside the plasma.
Three different theoretical pellet fragmentation models are briefly presented by Parks \cite{Parks}.
The first is based on the fragmentation of exploding munitions, originally formulated by Mott and Linfoot \cite{Mott}. The second is an energy-based model relating fragmentation to the surface energy density of individual shards.
A third model uses the principle of entropy maximization to formulate a fragment size relation based on work by Grady and Kipp~\cite{Max_Ent}.
These models were compared with experimental data by scientists at Oak Ridge National Laboratories (ORNL) and it was concluded that the first model describes their experimental findings best.
This statistical model was then further modified by Gebhart \textit{et al.}~\cite{Gebhart} by substituting constants previously used for normalization with functions of pellet material and normal impact velocity.
As it is the only fragmentation model used in the scientific community for the time being, it is discussed in more detail in this section.\\

\noindent
The first fracture model presented by Parks~\cite{Parks} is based on the statistical probability that a fracture will occur on a grain boundary due to the shockwave generated at pellet impact. 
The modified version, introduced by Gebhart \textit{et al.}~\cite{Gebhart}, predicts that the fragment size distribution is dependent on the pellet's composition, its length, diameter and its velocity perpendicular to the shatter plane $v_\text{perp}$.
The model predicts the relative probability $f(d)$ of a certain fragment diameter as

\begin{equation}
\label{e:parks_dens}
f(d) = \alpha d K_0(\beta d) \, ,
\end{equation}

\noindent
where $d$ is the fragment diameter in cm, $\alpha$ and $\beta$ are functions of pellet velocity and shatter angle in cm$^{-1}$  and $K_0$ is the modified Bessel function of the second kind, used to account for the energy propagation through the cylindrical pellet. 
The functions $\alpha$ and $\beta$ can be written as

\begin{align}
\label{e:parks_const}
\alpha &= X_R / D \\
\beta & = \frac{X_R}{L*C}\\
X_R & = \frac{v^2_{\text{perp}}} {v^2_{\text{thr}}} \, ,
\end{align}

\noindent
where $X_R$ is the ratio of normal kinetic energy to threshold kinetic energy, $L$ and $D$ are the pellet length and diameter in cm and $C$ is a material constant \cite{Gebhart}.
The fragmentation model does not account for evaporation of material during pellet impact and does not consider shatter tube geometry.
For any specific mixture of Ne and D$_2$, $v_{\text{thr}}$ is the threshold perpendicular velocity, above which the pellet will begin to fracture.
These velocities were determined experimentally by Gebhart \textit{et al.}~\cite{Gebhart} and were found to be independent of pellet size.
The corresponding values for $v_{\text{thr}}$ and $C$ for Ne and D$_2$ pellets are shown in tab.~\ref{t:constants}.

\begin{table}[!h]
     \centering
     \captionsetup{width=0.7\textwidth}
     \caption{Values for the threshold perpendicular velocity $v_{\text{thr}}$ and $C$ for pure Ne and D$_2$ pellets, as provided by Gebhart \textit{et al.}~\cite{Gebhart}.} 
     \begin{tabular}{|c|c|c|}
     \hline
      & \textbf{Ne} & \textbf{D$_2$} \\
     \hline
     \textbf{$v_{\text{thr}}$} & 8 m/s & 20 m/s\\
     \hline
     \textbf{$C$} & 5 & 2.5 \\
     \hline
     \end{tabular} 
     \label{t:constants}
\end{table}

\begin{figure}[!h]
     \centering
     \includegraphics[clip, trim = 0 0 0 0,width=0.7\textwidth]{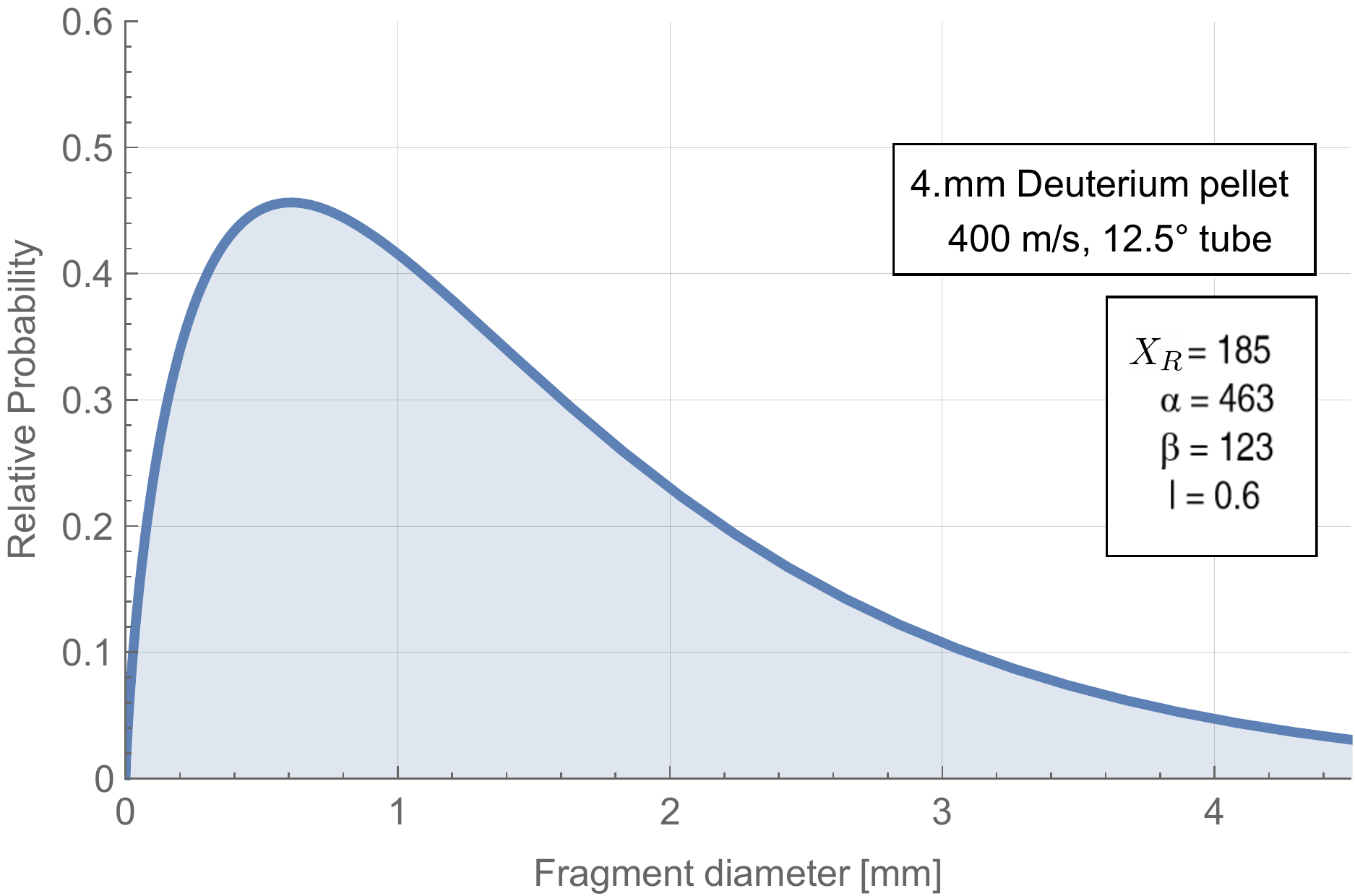}  
     \caption{Example of a theoretical fragment size probability density for a pure D$_2$ pellet, flying at 400 m/s through a 12.5\degree~shatter tube.}
  \label{f:example_parks}
\end{figure}

\noindent
For D$_2$-Ne mixture pellets, $v_{\text{thr}}$ can be obtained from figure 1 by Gebhart \textit{et al.} \cite{DeNeGebhart} and the material constant $C$ can be taken as

\begin{equation}
\label{e:parks_c}
C = 2.5(1.5 + \omega_{\text{Ne}}) \, ,
\end{equation}

\noindent
where $\omega_{\text{Ne}}$ is the neon mass fraction in the pellet \cite{DeNeGebhart}.
An example of the theoretical fragment size probability density is presented in fig.~\ref{f:example_parks} for pure D$_2$ pellet with a diameter $d =$ 0.4 cm, $l =$ 0.6 cm and $v_{\text{perp}} =$ 86.5 m/s. 
The theory predicts a clear peak at low fragment sizes, in this case at fragment diameters of 0.6 mm.
Since the theoretical model gives a probability density for different fragment sizes, any discrete realization of this model will differ from the distribution. 
Pellet fragmentation is treated as a statistical process, so a range of different fragment size distributions must be expected for one set of input parameters.

\subsection*{Generation of discrete realizations}
\label{s:model_generation}

It is important to study not only the probability distribution of the model formulated by Parks but also discrete realizations of it, since, in reality, we will always observe a finite number of fragments.
The variability between these different realizations should be taken into account when comparing predictions with the fragmentation model.
Simulating these discrete fragmentations can be done by repeatedly sampling the probability density until some stopping condition is reached. 
In this section, I will outline a simple numerical random sampling method and briefly explain its application to generate simulated fragment sprays.\\

\noindent
Numerically, one commonly used method for random sampling from a distribution is rejection sampling (also known as accept-reject method), a basic technique used to generate observations from a distribution that does not require the probability density to be invertible \cite{MonteCarlo}. 
Given a probability density $v(x)$ of some distribution in an interval $A \leq x \leq B$ with a maximum value of $v_{\text{max}}$ in this interval, two independent, uniformly distributed random numbers $R_1$ and $R_2$ are generated in the interval $(0, R_{\text{max}})$. 
$R_1$ is mapped onto the interval $(A,  B)$, $R_2$ is mapped onto $(0, v_{\text{max}})$.
Then, the value of $v(R_1)$ is compared to the value of $R_2$. If $R_2 < R_1$, $R_2$ is accepted as an observation.
Otherwise, it is rejected. This process is repeated until the desired number of observations is reached.\\

\begin{figure}[!h]
     \centering
     \captionsetup{width=0.7\textwidth}
     \includegraphics[clip, trim = 0 0 0 0,width=0.7\textwidth]{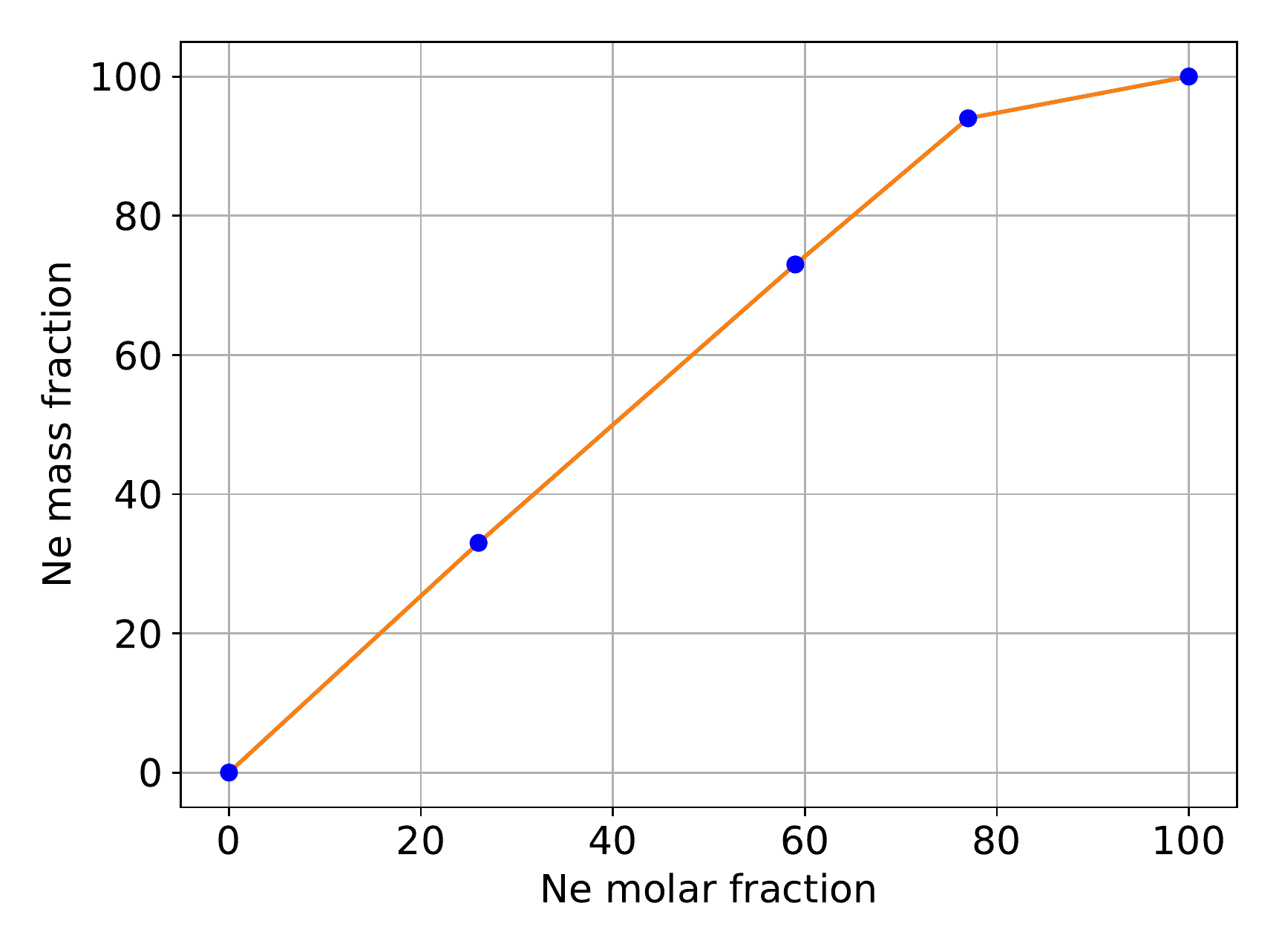}
     \caption{Linear interpolation of Ne molar fraction and mass fraction pairs presented by Gebhart \textit{et al.} \cite{DeNeGebhart}.}
     \label{f:molarmass}
\end{figure}

\noindent
For the evaluation of the theoretical model, neon mass fraction, shatter angle, pellet length, diameter and speed need to be known.
Usually, only the molar neon fraction is available. 
The mass fraction can not be easily computed from knowledge of the molar fraction, since different crystal structures emerge within solid D$_2$-Ne mixtures depending on the mole percentage of Ne \cite{DeNeGebhart}.
However, a rough estimate of the mass fraction of Ne can be inferred for mixture pellets by linearly interpolating molar fraction-mass fraction pairs given by Gebhart \textit{et al.} \cite{DeNeGebhart}. 
These are presented in fig.~\ref{f:molarmass}. 
For example, Ne mass fractions of 6.5\%, 25\% and 50\% correspond to molar fractions of 5\%, 20\% and 40\% respectively.
Threshold velocities $v_{\text{th}}$ for different pellet mixtures can be interpolated from figure 1 in the paper by Gebhart \textit{et al.}~\cite{DeNeGebhart}. 
Using the same Ne molar fractions of 5\%, 20\% and 40\%, velocities of 19 m/s, 15.5 m/s and 12.5 m/s were found. \\

\noindent
So far, experimental validation of the pellet fragmentation model formulated by Parks was only shown qualitatively for a small set of input parameters by Gebhart \textit{et al.}~\cite{Gebhart}.
Since the model is already being used to generate initial conditions for simulations \cite{JOREK_Parks}, it needs to be validated across a larger parameter space. 
In this thesis, I present experimental fragment size distributions of SPI pellets, frozen from different mixtures of D$_2$ and Ne and fired at various speeds. 
For each analyzed pellet, a comparison to discrete realizations of the theoretical model is shown.
Additionally, the impact of different shattering geometries on the spatial fragment distribution and the fragment sizes is discussed.

\chapter{Methods and Experimental Setup}

In this section, the testing setup for the ASDEX-Upgrade shattered pellet injection system will be presented in detail.
The optical design for video diagnostics will be elaborated and I will explain the fragmentation analysis and pellet integrity algorithms.
Also, a method of simulating fragment size distributions from a given probability density is shown.

\section{ASDEX-Upgrade pellet injector}
\label{s:ASDEX-Upgrade pellet injector}

For SPI experiments at ASDEX-Upgrade, a pellet injection system was designed and assembled at IPP, Garching, Germany.
The core of this system, the pellet injector, was designed and constructed by Pelin LLC., Saint Petersburg, Russia.
It is an injection system with 3 flight channels that can produce and fire cryogenic pellets with diameters of 1, 2, 4 and 8 mm made of H$_2$, D$_2$, Ne, Ar and various mixtures of D$_2$/H$_2$ and Ne.

\begin{figure}[!h]
     \centering
     \captionsetup{width=0.7\textwidth}
     \includegraphics[clip, trim = 0 0 0 0,width=0.5\textwidth]{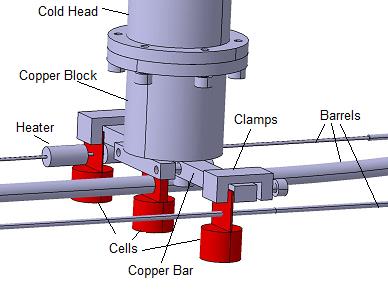}
     \caption{CAD image of the pellet formation stage. Courtesy of I.V.Vinyar, Pelin LLC.}
     \label{f:cooling_stage}
\end{figure}

\noindent
Figure~\ref{f:cooling_stage} shows an annotated CAD image of the pellet formation region inside the injector.
To form a pellet, Ne/D$_2$/mixture gas is filled into a barrel at pressures in the range of 80-250 mbar (depending on pellet material).
Pressure is regulated during the pellet formation process. 
A cryocompressor lowers the temperature of the cooling cells, indicated in red in fig.~\ref{f:cooling_stage}.
Gas freezes on the barrel wall adjacent to the cell and grows inward, forming a cylindrical ice pellet \cite{SPI_manual}.
To control the pellet length, heaters at the front and back side of the cooling cell can be employed to prevent desublimation at a given position.
After enough material is deposited (usually after 200-400 s for 4 mm diameter pellets), the remaining pellet gas is pumped off. \par \noindent

\begin{figure}[!h]
     \centering
     \captionsetup{width=0.9\textwidth}
     \includegraphics[clip, trim = 0 0 0 0,width=0.6\textwidth]{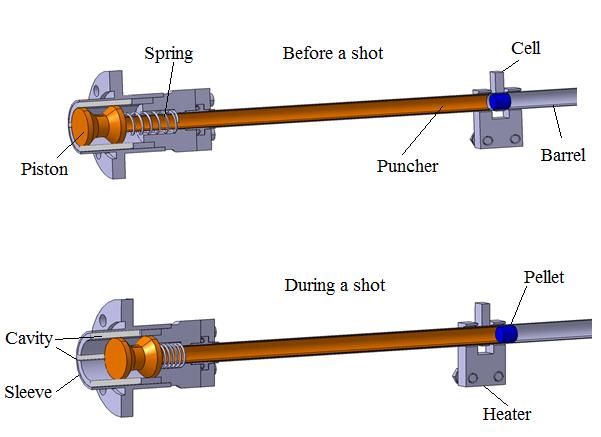}
     \caption{CAD image of the hollow punch before and during firing \cite{SPI_manual}. Courtesy of I.V.Vinyar, Pelin LLC.}
     \label{f:punch_CAD}
\end{figure}

\noindent
Before firing, the gate valves in the injection line are re-opened. 
After a trigger signal, the hollow punch mechanism depicted in fig.~\ref{f:punch_CAD} is activated with propellant gas to launch the pellet.
A fast valve on one end of the barrel opens, introducing He or D$_2$ propellant gas, which moves the piston forward, causing the puncher to scrape the pellet from the barrel wall. 
After the piston has moved a certain distance towards the pellet, the propellant can move around the piston trough cavities in the sleeve and flow through the hollow puncher to propel the pellet.
Propellant pressures in the range of 0.5~-~50~bar can be used.
For Ne pellets, it is necessary to heat the tubes containing the pellets shortly before firing to prevent the pellet from being stuck and to prevent damage on the puncher. 
This is also necessary for firing slow D$_2$ / mixture pellets with low propellant pressure, as there, the punch mechanism is sometimes not able to knock the pellet loose.
Based on the system's laboratory characterization, 4 mm pellets can be fired at 150-750 m/s for H$_2$/D$_2$ and 80-440 m/s for Ne and Ar pellets.
During pellet freezing, the length-to-diameter-ratio of the pellet can be varied between 0.5 - 1.5 by localized heating of the barrels.

\newpage

\section{ASDEX-Upgrade SPI testing setup}
\label{s:SPI_setup}

\begin{figure}[h]
     \centering
     \includegraphics[clip, trim = 0 170 0 120,width=\textwidth]{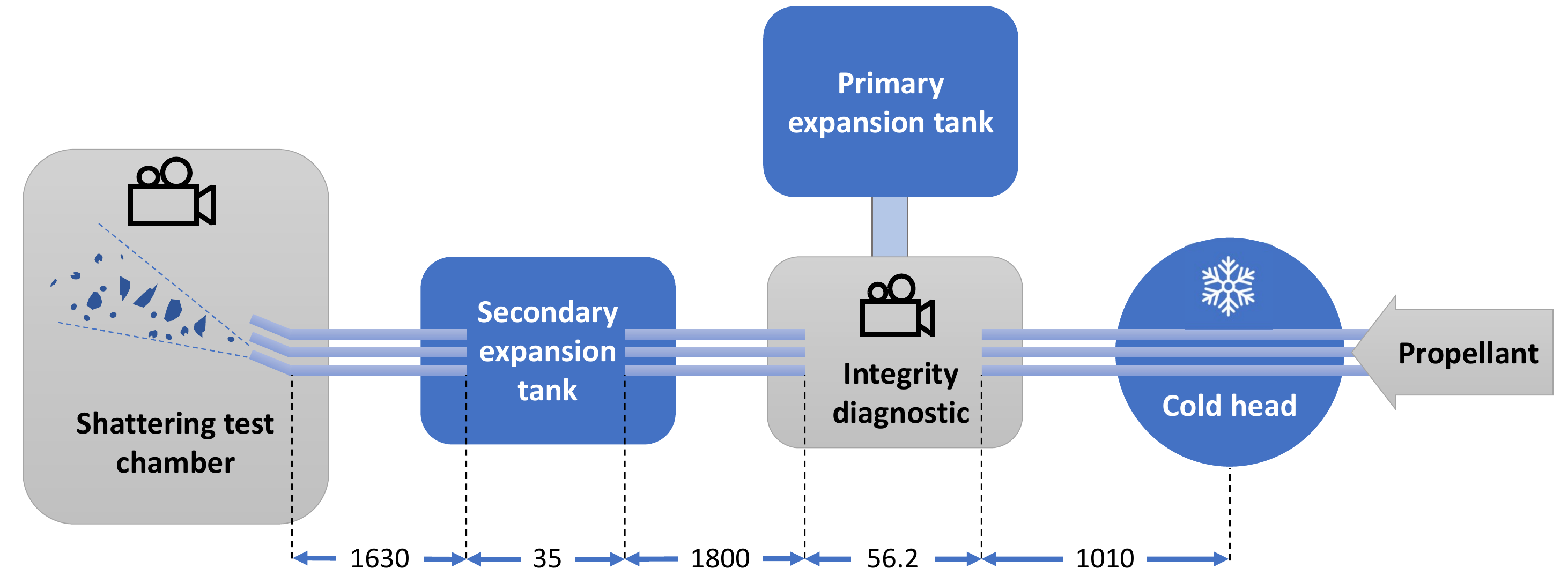}  
     \caption{Schematic diagram of the SPI test setup, with flight distances noted in mm. Horizontal pipes represent the three barrels. 
     Pellets are desublimated at the cold head.
     They are propelled towards the integrity diagnostic, where pellet intactness, size, tilt and speed are measured with a fast camera. 
     Propellant is removed into the first and second expansion tank.
     Inside the shattering test chamber, pellets encounter a shatter head. 
     The resulting fragment spray is recorded with a second fast camera.}
  \label{f:schematic_setup}
\end{figure}

To study the capabilities of the SPI system before installing it to ASDEX Upgrade, a testing setup was designed and constructed around the injector at IPP. 
A schematic image of this setup is shown in fig.~\ref{f:schematic_setup}.

\begin{figure}[h]
      	\centering
      	\includegraphics[clip, trim = 40 100 40 50,width=0.8\textwidth]{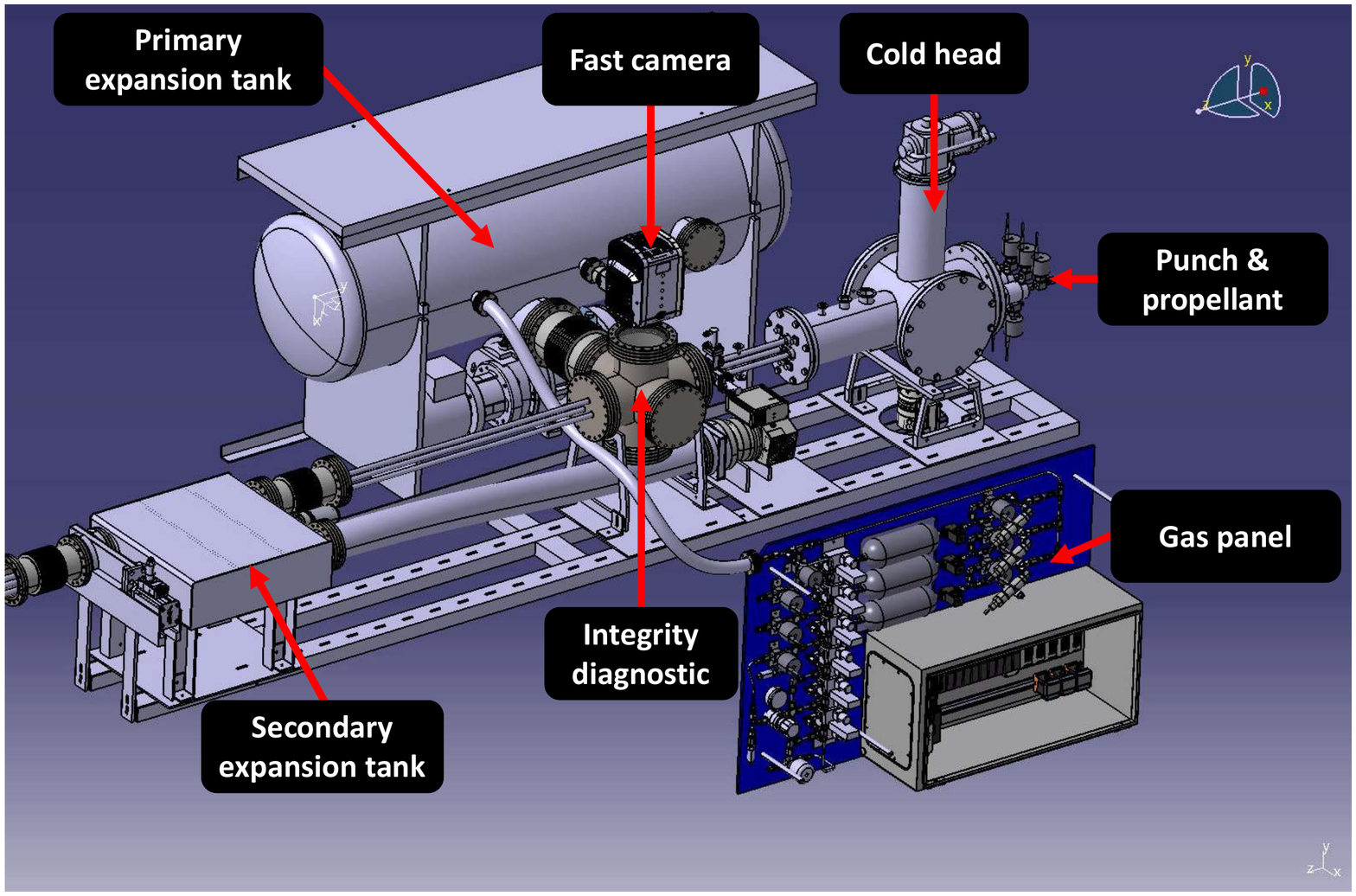}  
      	\caption{CAD image of the testing setup without the shattering test chamber.}
        \label{f:cad_SPI}
\end{figure}

\noindent
Pellets are frozen in the cold head and fired.
They pass three guide tube gaps in their flight paths, one at the barrel gate valve and one at each expansion tank.
The latter two gaps serve as means to remove propellant gas from the system.
As explained by Baylor \textit{et al.}~\cite{Baylor_2019}, propellant gas has a sound speed that is higher than the final pellet velocity and therefore reaches the plasma before the arrival of any pellet material.
This leads to an early thermal collapse of the plasma and reduces the effectiveness of the radiation of plasma kinetic energy.
Also, the injection of propellant gas into a torus causes unwanted stress on the vacuum system and can lead to impurity adsorption in vessel components.
At the integrity diagnostic, a fast camera observes the first pumping gap through a glass window to determine pellet integrity and measures their length, diameter and tilt.
After passing the secondary expansion tank, pellets reach the shattering test chamber, where they fracture upon impact on a shatter tube. 
The resulting fragment sprays are recorded with a second fast camera through a transparent door at the test chamber.
A CAD image of the testing setup without the shattering chamber is presented in fig.~\ref{f:cad_SPI}.
The setup is designed for firing pellets into test chamber pressures of 1 - 10$^{-3}$ mbar \cite{SPI_manual}, with the option of activating turbo pumps to generate pressures similar to a tokamak vacuum vessel.

\newpage

\begin{figure}[!h]
      	\centering
      	\includegraphics[clip, trim = 0 0 50 30,width=0.5\textwidth]{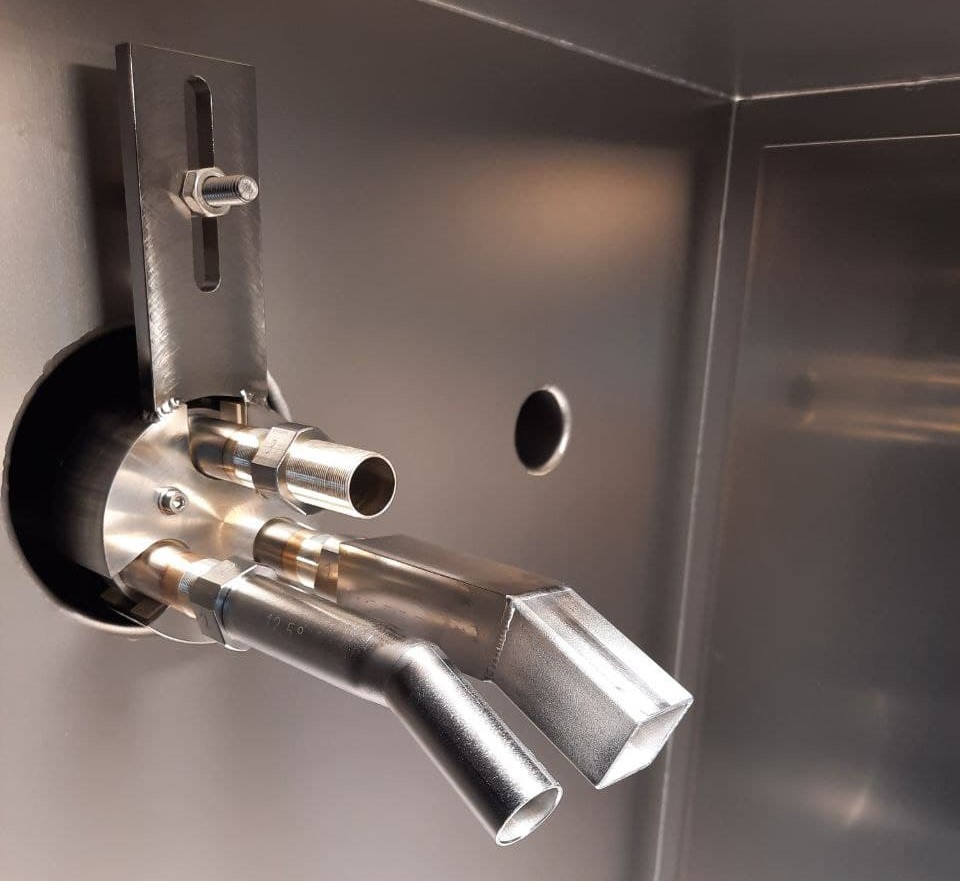}  
      	\caption{Photograph of the shatter heads in the shattering test chamber. At the topmost barrel, no shatter tube is mounted.
      	12.5\degree~mitre-bend shatter tubes with a circular (left) and rectangular (right) cross-section are screwed onto the bottom left/right barrels.}
        \label{f:shatterheads}
\end{figure}

\noindent
In the shattering test chamber, shatter tubes can be connected to each barrel with a silver-coated thread on the outer side of the barrel end.
Figure~\ref{f:shatterheads} shows two shatter tubes/heads, one with a rectangular, one with a circular cross-section. 
On the topmost tube, no shatter head is mounted.
Various shatter geometries, angles and post-shatter lengths were used in the course of the SPI laboratory testing at IPP.
In this thesis, I analyse only a subset of the entire database.
Dimensions and characteristics of shatter heads considered in this thesis are presented in tab.~\ref{t:heads}.
So-called mitre-bends were chosen for every geometry to allow a discrete change of the angle between pellet flight path and tube wall. \\

\begin{figure}[h]
     \centering
     \includegraphics[clip, trim = 0 0 0 0,width=0.65\textwidth]{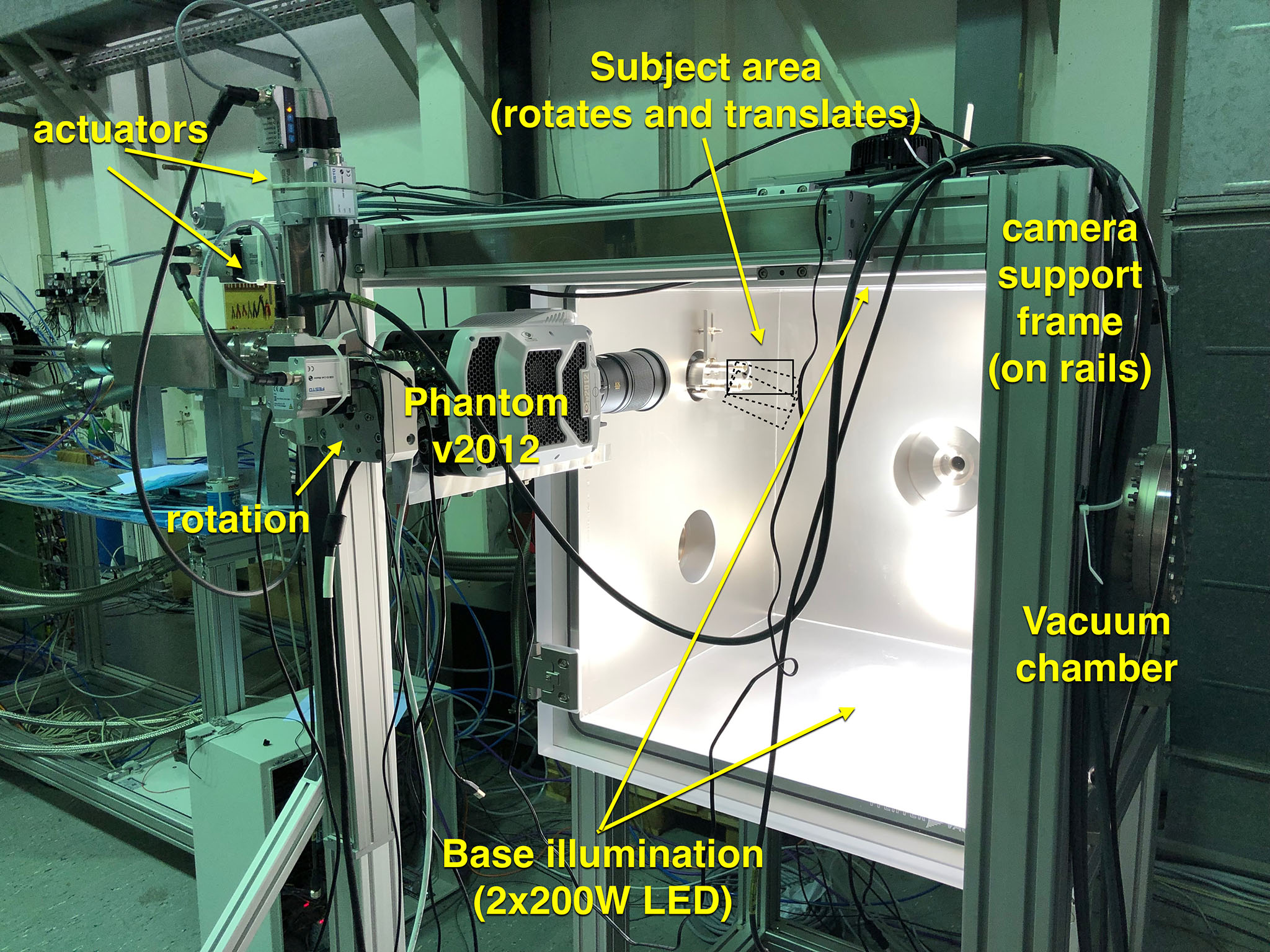}  
     \caption{Photograph of the shattering test chamber and camera mount. The HMI lamp, usually positioned beside the camera on the right (viewed from the back), is not shown here.}
  \label{f:test_annotations}
\end{figure}

\begin{figure}[h]
     \centering
     \includegraphics[clip, trim = 0 0 0 0,width=0.6\textwidth]{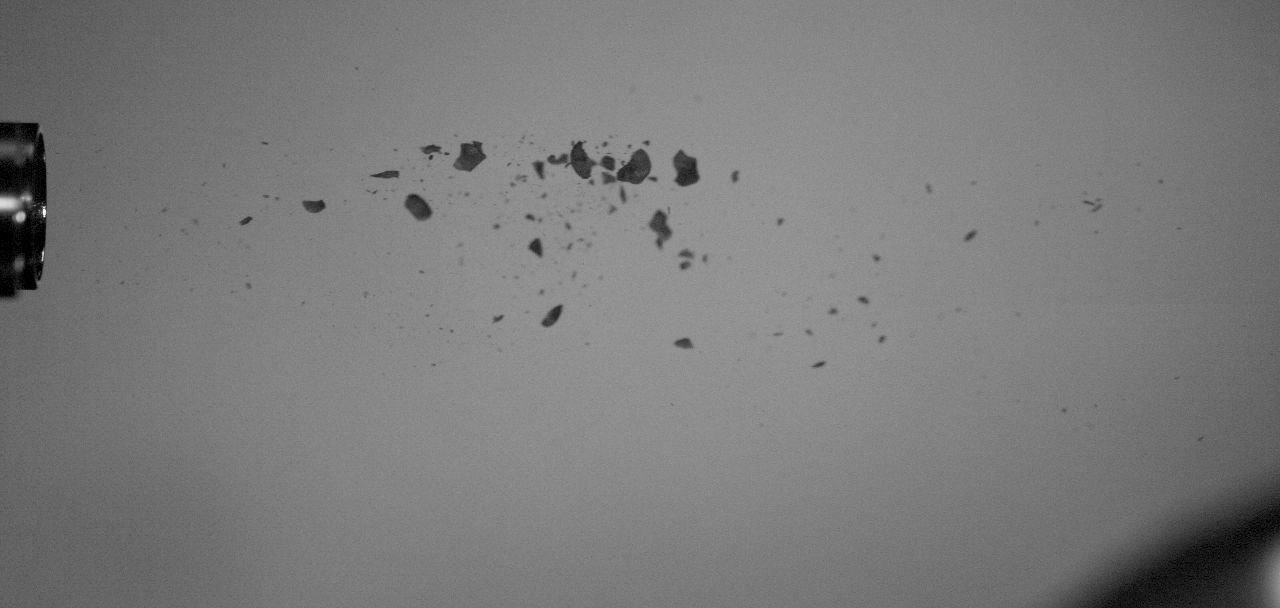}  
     \caption{Example frame of the fast camera FOV at the shattering test chamber.}
  \label{f:shatter_example}
\end{figure}

\noindent
At the shattering test chamber, the fast camera is aligned to look at the shatter heads through the door of the test chamber composed of transparent plexiglas.
The camera mount can be continuously adjusted to move the camera towards and parallel to the door and rotate it around the camera axis.
Lighting is provided by two 200W LED panels from the top and bottom of the chamber and, additionally, by a 400 W Dedolight Daylight 400DT metal-halide lamp with a Fresnel-lens, mounted on a tripod beside the fast camera.
This ensures that sufficient exposure can be achieved even at the shortest exposure time of 1 $\mu$s.
A photograph of the camera mount and shattering test chamber is shown in fig.~\ref{f:test_annotations}.
An example frame of the camera is shown in fig.~\ref{f:shatter_example}.

\begin{figure}[h]
     \centering
     \includegraphics[clip, trim = 0 0 0 0,width=0.6\textwidth]{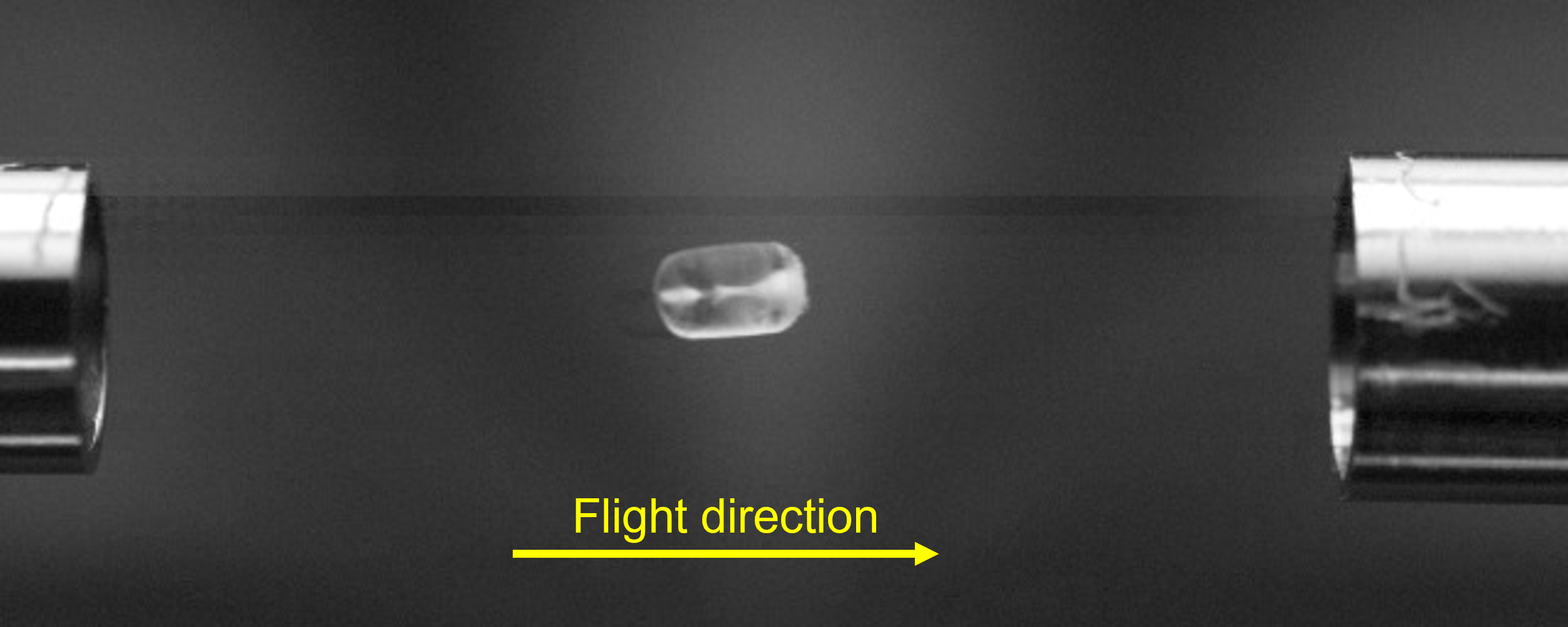}  
     \caption{Example of the FOV of the fast camera at the integrity diagnostic. Pellets fly from left to right.}
  \label{f:integrity}
\end{figure}

\noindent
At the integrity diagnostic, a fast camera observes the first pumping gap through a glass window.
A 200 W LED light illuminates the gap.
Figure~\ref{f:integrity} shows an example frame of a video recorded at the integrity diagnostic.
\newpage

\noindent
Both fast cameras are Phantom UHS-v2012 cameras, were manufactured by Vision Research Inc., New Jersey, USA, and are capable of recording at up to 622k frames per second (fps).
For the videos analyzed in this thesis, the fast cameras were recording at a resolution of 640$\times$256 pixel with a framerate of 110k fps at the integrity diagnostic, and 1280$\times$608 pixels and 29676 fps at the shattering test chamber.
The exposure time was set to 1 $\mu$s for both cameras to minimize motion blur.

\begin{table}[!h]
     \centering
     \captionsetup{width=0.82\textwidth}
     \caption{Characteristics of shatter heads considered in this thesis. Post-shatter length denotes the length after the tube bend, the cross-section states the dimensions inside the tube.} 
     \begin{tabular}{|c|c|c|c|}
     \hline
     \textbf{Angle} & \textbf{Shape} & \textbf{Post-shatter length} & \textbf{Cross-section}\\
     \hline
     12.5 \degree & circular & 78 mm & 16 mm\\
     \hline
     15 \degree& circular & 78 mm& 16 mm\\
     \hline
     25 \degree& circular & 78 mm& 16 mm\\
     \hline
     25 \degree& rectangular & 78 mm& 21x21 mm\\
     \hline
     30 \degree& circular & 78 mm& 16 mm\\
     \hline
     \end{tabular} 
     \label{t:heads}
     \end{table}

\clearpage

\section{Test chamber optical design}
\label{s:Test chamber optical design}

\begin{figure}[h]
      	\centering
      	\includegraphics[clip, trim = 0 0 0 0,width=\textwidth]{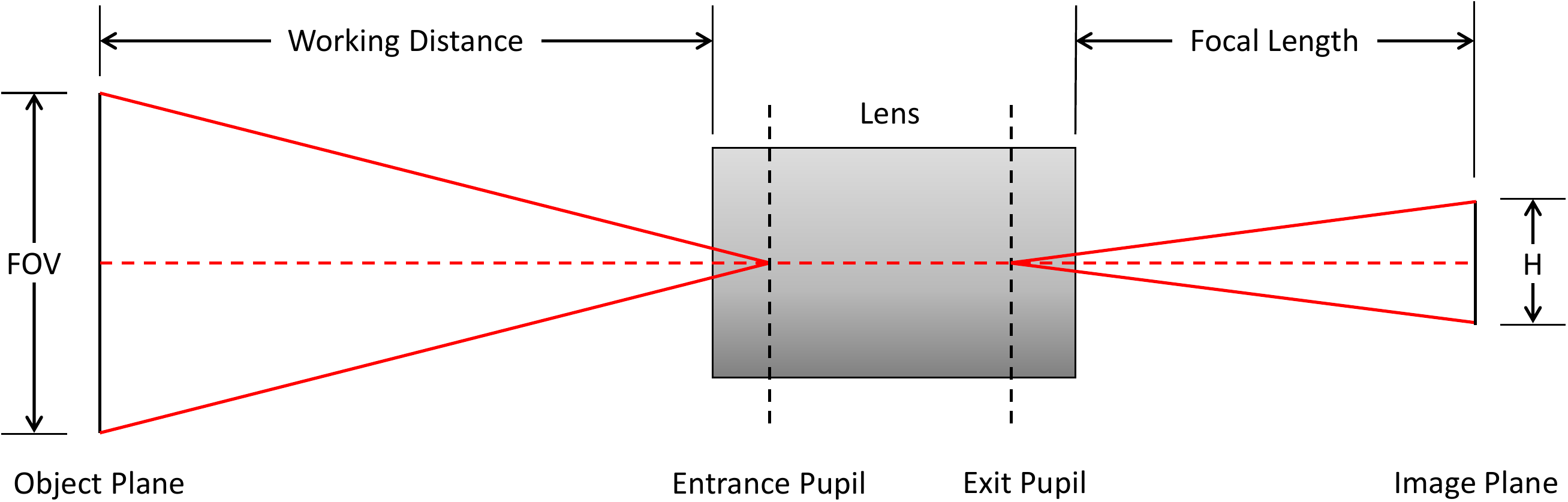}  
      	\caption{Schematic optical system with single lens. In the thin lens approximation, the distance between entance and exit pupil is assumed to be zero.}
        \label{f:optics_schematic}
\end{figure}

The optical system in the test tank must be able to image the fragments in a sufficiently broad and wide field of view (FOV) and provide enough depth of field (DOF) that only a negligible amount of fragment material is blurred during the plume expansion.
For this, an appropriate lens system and camera setup had to be found.
A two-dimensional schematic image of a simple optical system is depicted in fig.~\ref{f:optics_schematic}. 
For the calculations presented in this section, the distance between entrance pupil and exit pupil was assumed to be zero (thin lens approximation).

\noindent
Here, $H$ is sensor size of the camera. 
Given a specific sensor size $H$, the FOV can be varied by changing the working distance WD or the focal length $f$.
In the thin lens approximation, this relation is described in the equation for the magnification $m$ of a lens, given by

\begin{equation}\label{e:magn}
m = \frac{H}{\text{FOV}} = \frac{f}{\text{WD}}\,.
\end{equation}

\noindent
In the shattering test chamber, the working distance WD is limited to a minimum of 470 mm by the distance of the shatter tubes to the glass door. 
The horizontal and vertical sensor size $H$ of our fast cameras is 35.8 mm x 22.4 mm. \par \noindent
To  have the fragment plume in focus during expansion, a deep enough DOF is required.
Gebhart \textit{et al.}\cite{Gebhart} found that the shatter plume disperses at less than 20$\degree$~half-angle from the center line of the shatter tube. 
Therefore, a (conservative) minimum DOF of $2\times \sin{(20\degree)} \times \text{FOV}$ is required to have the whole plume in focus. 
According to Hecht~\cite{Hecht}, the DOF can be approximately expressed by 

\begin{equation}\label{e:DOF}
     \text{DOF} = \frac{2 N C (\text{WD})^2}{f^2}\,.
\end {equation}

\begin{figure}[h]
      	\centering
      	\includegraphics[clip, trim = 50 130 20 130,width=0.8\textwidth]{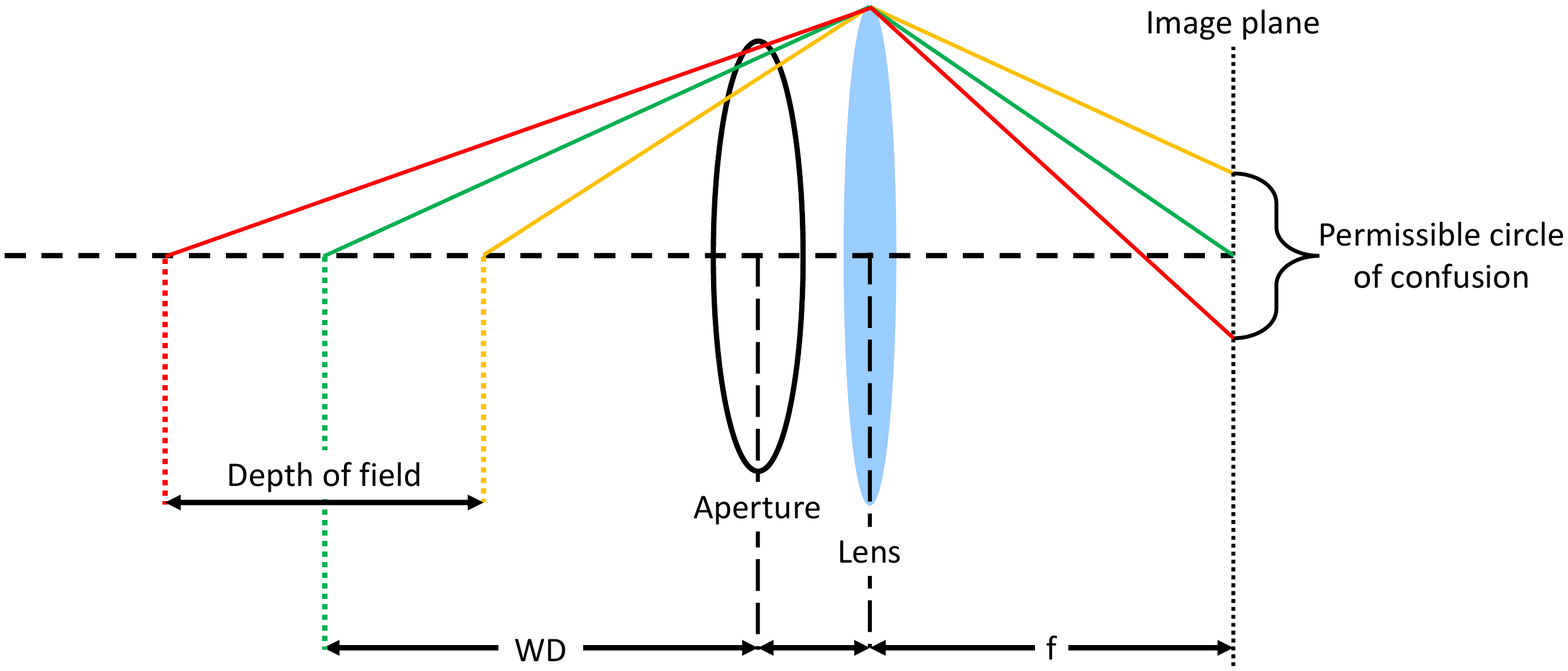}  
      	\caption{Schematic optical system, depicting the DOF for a given circle of confusion.}
        \label{f:DOF_schematic}
\end{figure}

\noindent
Here, WD is the working distance, $f$ is the focal length, $C$ is the circle of confusion and $N$ is the f-number of the lens, which is defined as the ratio of the focal length $f$ to the entrance pupil diameter. 
Typical f-numbers $N$ range from 0.5 to 16 and determine the amount of light that can pass the lens system.
Figure~\ref{f:DOF_schematic} illustrates the relation of DOF and circle of confusion.
We define a maximum circle of confusion $C$ of 28 $\mu$m for both cameras (pixel size on the sensor).

\noindent
By solving eq.~\eqref{e:magn} for $f$ and inserting in eq.~\eqref{e:DOF}, we find

\begin{equation}\label{e:DOF2}
DOF = \frac{2 N C (\textit{FOV})^2}{H^2} = \frac{2 N C }{m^2} \,.
\end {equation}

\noindent 
Therefore, DOF increases quadratically with the FOV.
On the other hand, it is desirable to have a small enough FOV to achieve higher mm/pixel resolution (= higher magnification). 
Assuming 1280 pixels as the horizontal sensor resolution, the effective mm/pixel resolution is given by $FOV/1280$.
We want to have a sufficiently wide FOV to keep most of the fragments inside the DOF, while keeping the mm/pixel resolution as small as needed to resolve fragments. \\

\begin{figure}[!h]
     \centering
     \begin{subfigure}[b]{0.49\textwidth}
         \centering
         \includegraphics[clip, trim = 0 0 0 0, width=\textwidth]{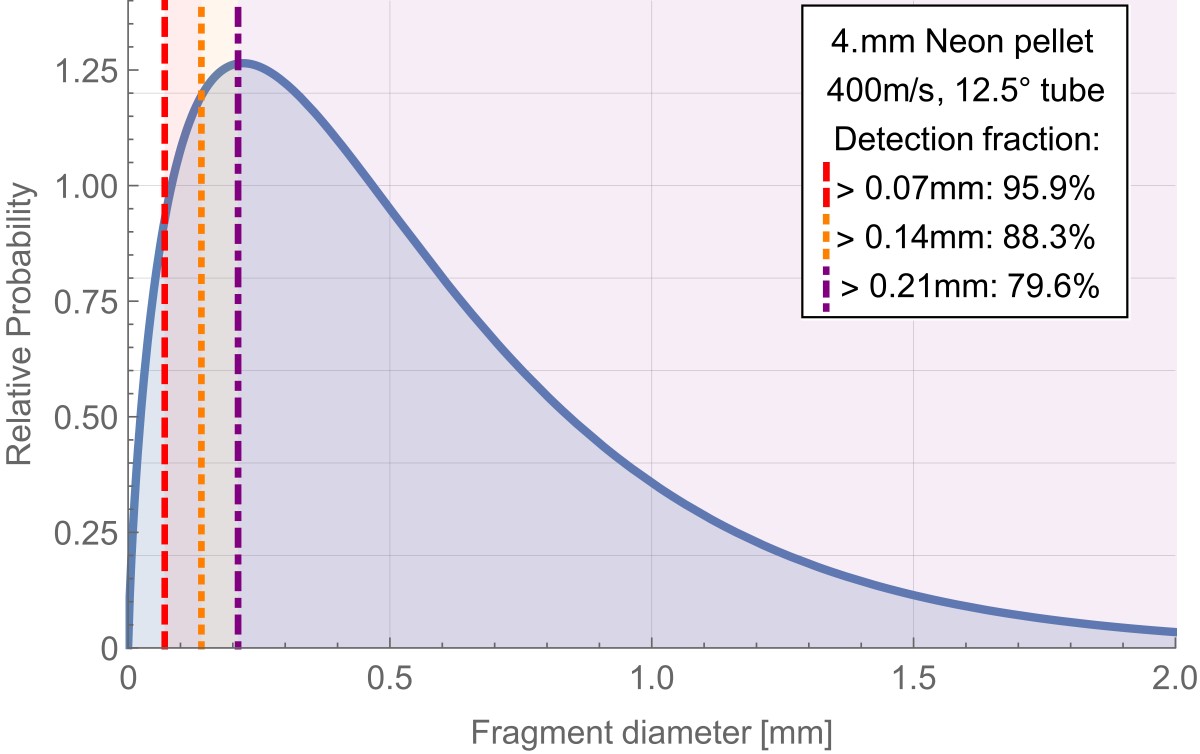}
         \caption{}
         \label{f:res_limit1}
     \end{subfigure}
     \hfill
     \begin{subfigure}[b]{0.49\textwidth}
         \centering
         \includegraphics[clip, trim = 0 0 0 0, width=\textwidth]{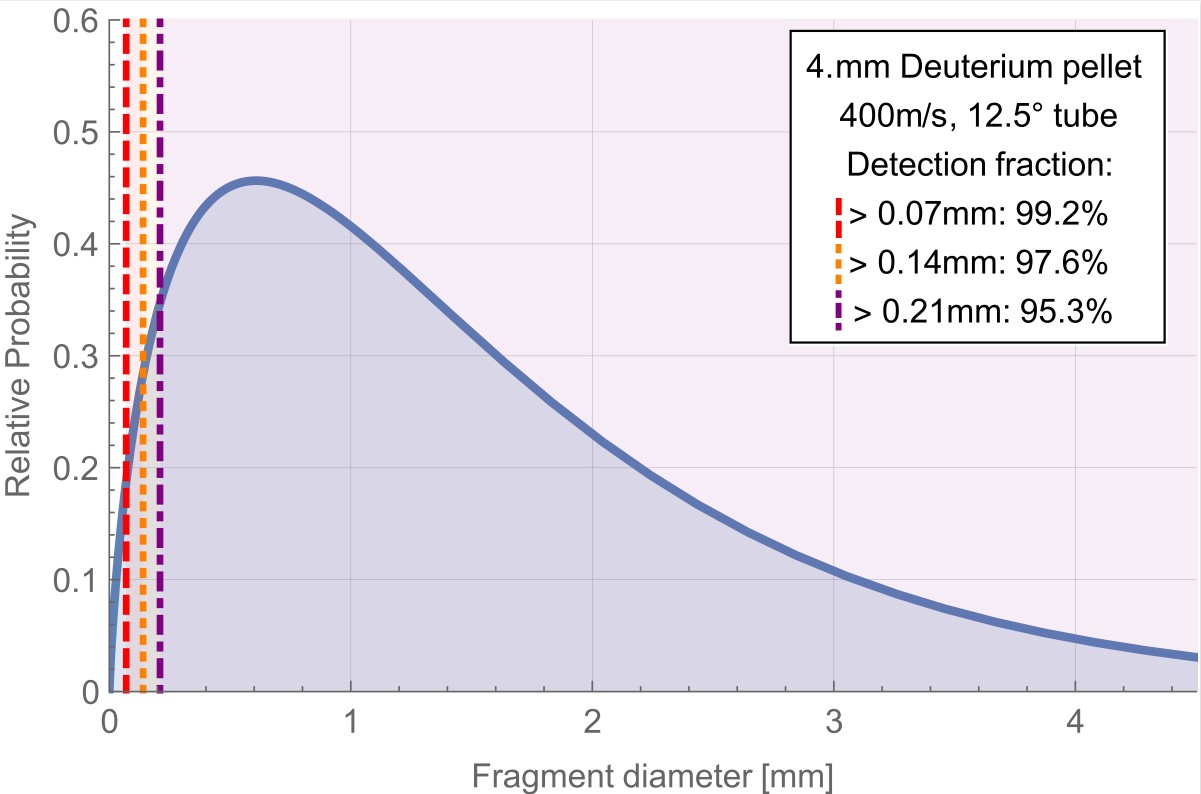}
         \caption{}
         \label{f:res_limit2}
     \end{subfigure}
     \vskip\baselineskip
     \begin{subfigure}[b]{0.49\textwidth}
         \centering
         \includegraphics[width=\textwidth]{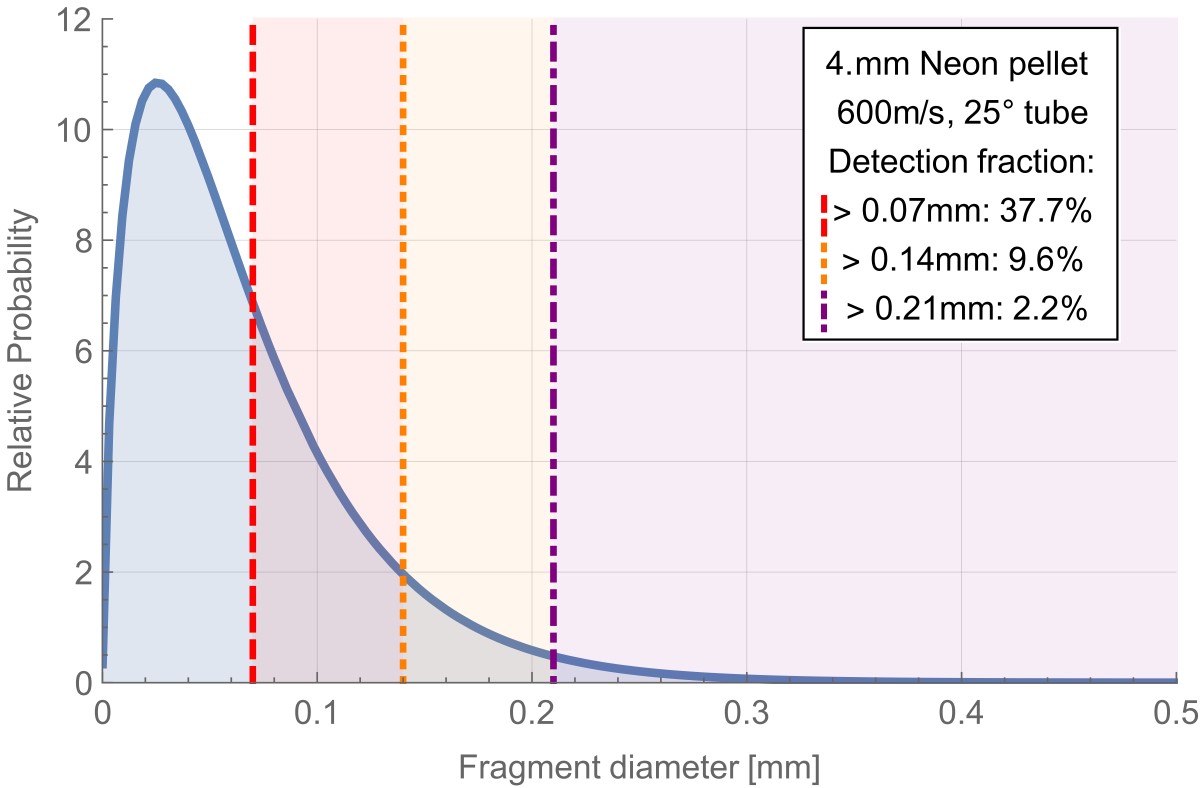}
         \caption{}
         \label{f:res_limit3}
     \end{subfigure}
	\hfill
     \begin{subfigure}[b]{0.49\textwidth}
         \centering
         \includegraphics[clip, trim = 0 0 0 0, width=\textwidth]{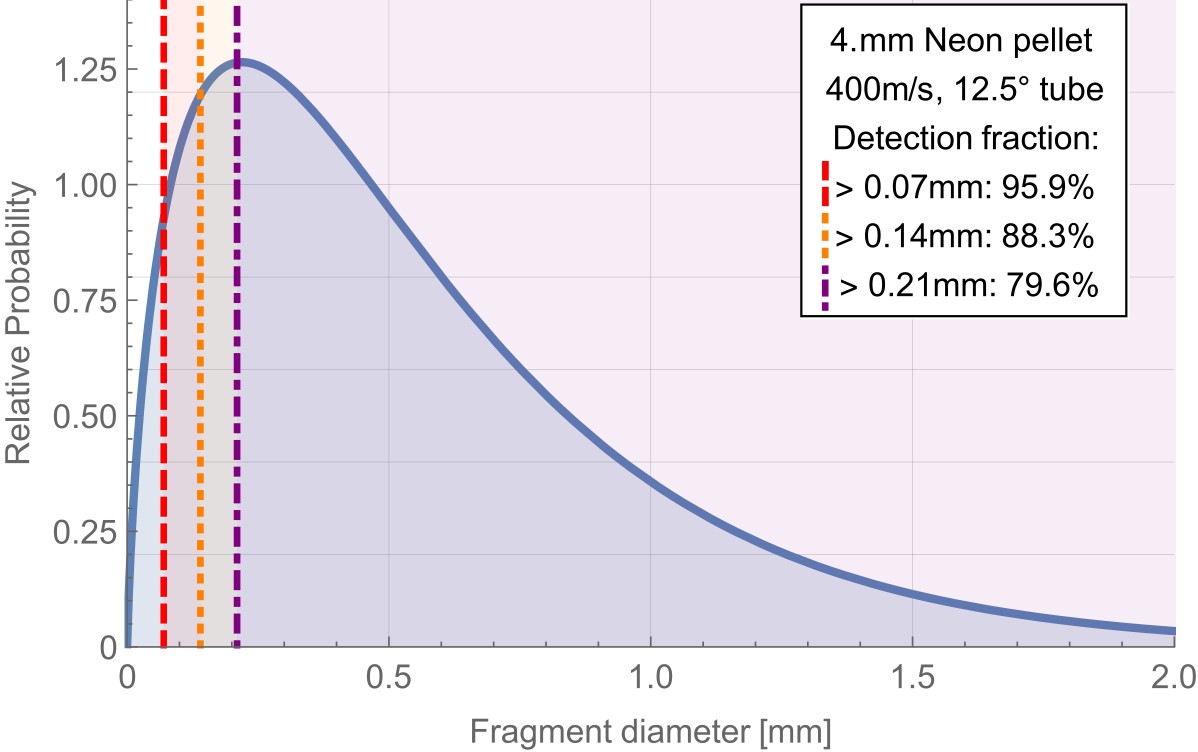}
         \caption{}
         \label{f:res_limit4}
     \end{subfigure}
     
        \caption{Theoretical fragment size distribution for 4 mm Ne and D$_2$ pellets at different perpendicular velocities. 
        Plot a) and b) show theoretical size distributions for D$_2$ pellets, fired through 25\degree~/~12.5\degree~tubes at 600 m/s. 
        In c) and d), distributions for Ne pellets, fired through 25\degree~/~12.5\degree~tubes at 600 m/s~/~400 m/s are shown. 
        Dashed lines in orange, yellow and purple indicate the resolution in mm/pixel with a focal length of 60, 100 and 200 mm for a fixed working distance of 470 mm.
        The resolvable fractions of all fragments is stated in the upper right of the plots.
        }
     \label{f:res_limit}
\end{figure}

\noindent
To get an estimate for the required resolution to measure the fragmentation of pellets with 4 mm diameter, the fragmentation model by Parks \cite{Parks}, explained in section~\ref{s:theoretical_model}, was evaluated for a range of pellet materials and speeds. 
The maximum speed for pellets was estimated to be 600 m/s, material constants were taken from Gebhart \textit{et al.}\cite{Gebhart}. 
Also, the fraction of resolvable fragments was computed for resolutions of 0.21, 0.14 and 0.07 mm/pixel, roughly corresponding to focal lengths of 60, 100 and 200 mm.
These resolutions were computed by inserting in eq.~\eqref{e:magn}, assuming a WD of 470 mm and a sensor resolution of 1280 pixels.
Theoretical fragment size distributions are plotted in fig.~\ref{f:res_limit}, with vertical lines showing the fraction of resolvable pellet material produced by shattering for different focal lengths.\\

\noindent
Figure~\ref{f:res_limit} indicates that high focal lengths (dashed red lines) are advisable to be able to resolve a significant portion of the fragment plume.
This insight led to the choice of a 180mm f/2.8 full frame macro lens for the shattering test chamber.
It is a lens with a fixed focal length, with minimal image distortion.
With a WD of 470 mm, the resulting FOV is approx. 94.5 mm wide.
By inserting into eq.~\eqref{e:DOF2}, we find that the DOF is now only dependent on the f-number of the aperture $N$ and can be expressed approximately as

\begin{equation}\label{e:DOF3}
DOF \approx 0.4 \times N  \, \, [mm] \, .
\end {equation}

\noindent 
Now, the DOF can be increased by closing down the aperture. 
The possible f-numbers for the Sigma lens range from f/2.8 to f22. 
Evaluation of the plume expansion estimate yields a required DOF of approx. 64 mm.
However, one can argue that this is a rather conservative estimate and that only the smallest fragments, which carry a negligible amount of material, disperse far from the cylinder axis of the shatter tube.

\newpage

\noindent
The further the aperture is closed down, the less light reaches the sensor, increasing the need for powerful lighting to have acceptable levels of contrast.
Therefore, the powerful HMI lamp, mentioned in ch.~\ref{s:SPI_setup}, was added to the existing LED panels.
As a rule of thumb, diffraction of visible light (wavelengths of 0.5 $\mu$m) occurs when it passes a slit with a diameter $D<$~10 $\mu$m \cite{Demmi2}.
Since the entrance pupil diameter is approx.~8 mm wide at the smallest setting (f22), diffraction is negligible for our optical setup.\\

\noindent
For the integrity diagnostic, a 60mm f/2.8 macro lens was used, producing a FOV of approximately 70 mm at a WD of approx. 118 mm.
Here, DOF is not an issue, as the intact pellet is not expected to travel out of the focal plane.
The processing methods and analyses used for the videos recorded with both cameras will be discussed in the following section.

\section{Computer Vision}
\label{s:computer_vision}
At the integrity diagnostic, the pellet length, diameter and speed are measured using a (semi)automatic computer vision algorithm.
These values are then stored in an .xlsx file, including the corresponding ID, neon percentage, firing pressure, shatter angle and many other parameters for each fired pellet.
This file will be referred to as the pellet database.
The measured pellet speeds are then used as initial conditions for the fragmentation analysis.
Multiple computer vision concepts and image processing methods were used.
These will be summarized shortly before showing their application in the analysis algorithms.

\subsection*{Image representation}

Digital images are generally represented by matrices, where every entry corresponds to one pixel in the image and the shape of the matrix corresponds to the pixel resolution (e.~g. 640$\times$480).
For a black and white image, the matrix values encode the brightness of the respective pixel.
In colored images, three values are used to represent one pixel, the exact encoding depends on the colorspace that is used.
While the default representation of color is usually done using red, green and blue saturation values, working in different colorspaces can make image processing tasks easier.
Videos (without sound) are treated as an ordered list or sequence of image matrices.

\subsection*{Pre-processing}
\label{s:preprocessing}

\begin{figure}[!h]
     \centering
     \includegraphics[clip, trim = 0 0 0 0,width=\textwidth]{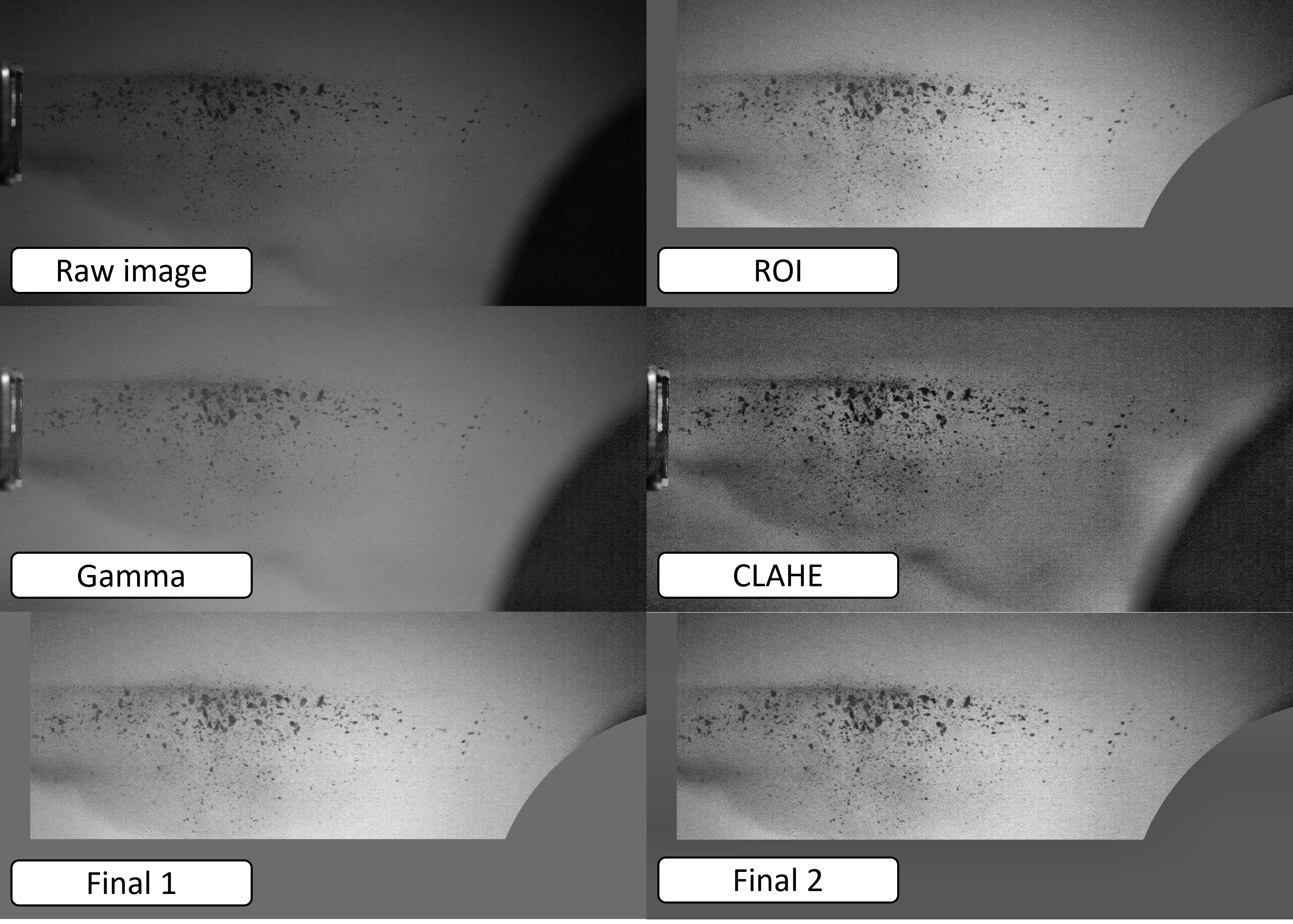}
     \caption{Different pre-processing methods and their effects on the video frames.}
     \label{f:preprocess_methods}
\end{figure}

\noindent
For image or video manipulation, performance can often be improved considerably by using pre-processing methods.
I tested the capabilities of several pre-processing algorithms and will present the most successful ones shortly.
For a more thorough review, I refer to Chaki and Dey~\cite{Preprocessing}.
Pre-processing methods can be roughly divided into contrast modification, denoising and FOV augmentation.
Contrast modification is used to alter the visibility of objects by making differences in pixel intensity more apparent.
Denoising tries to eliminate different noise contributions (e.g. Gaussian noise from a camera sensor) to achieve a sharper image.
By further augmenting the FOV (e.g. cropping noisy parts and bright/darks spots), the performance of the first two methods can be improved.
In fig.~\ref{f:preprocess_methods}, the effects of some example methods on a raw video frame are shown.
On the top left, a raw video frame with integer pixel values ranging from 0 to 255 is shown. 
The top right, pixels outside a specific region of interest (ROI) are painted gray, thereby removing the darkest parts of the image.
Pixel values are then re-scaled so they again span the maximum possible range, which leads to a significant increase of contrast.
On the middle left, a gamma transformation using  $\gamma > 1$ was applied, making the background appear brighter while keeping very dark pixels almost equally dark (e.g. black spot on the right).
Here, the pixels are scaled down to the interval [0,1], before undergoing the transformation

\begin{equation}
     I_{\text{out}} = I_{\text{in}}^{1/\gamma} \, .
     \label{e:gamma}
\end{equation}

\noindent
Values of $\gamma > 1$ ($< 1$) lead to a shift to the brighter (darker) side of the spectrum~\cite{OpenCV}.
Since gamma transformations are a non-linear transformation, they can be used to highlight a specific end of the brightness spectrum.
On the middle right, Contrast Limited Adaptive Histogram Equalization (CLAHE) was performed. 
CLAHE can be used to locally increase the contrast of an image and can help with the segmentation of objects that are e.g. partially in the shade.
The method operates on tiles of several pixels in the image and increases the contrast inside them until a threshold value is reached.
Then, the tile boundaries are smoothened by using bilinear interpolation \cite{OpenCV}.
This method is powerful, but usually needs to be used together with a denoising algorithm because it can also amplify noise.
The bottom images in fig.~\ref{f:preprocess_methods} were created by combining several of these methods.
The choice of these specific algorithms, their parameters and their order of execution of these methods is nontrivial and required several weeks of trial-and-error learning.
On the lower left (``Final 1"), the FOV was first cropped to an optimal ROI, then pixel values were rescaled to span the maximum possible range and finally, gamma transformation ($\gamma > 1$) was used to further highlight the fragments.
The image on the lower right (``Final 2") was produced by then applying a denoising algorithm implemented in OpenCV and subsequent histogram equalization using CLAHE.
In both images, noise is kept at an acceptable level to avoid false segmentation in the following steps while contrast is significantly increased.
For the segmentation of videos, this procedure is applied to each frame individually using the same parameters.

\subsection*{Background subtraction}
\label{s:background_subtract_concept}

Background subtraction is a method that benefits strongly from pre-processing and that is widely employed in computer vision applications such as traffic surveillance or motion detection.
The goal of this method is to separate a video into a moving foreground and a static background using the information contained in the similarity of subsequent images.
Conceptually, the process of background subtraction can be divided into three steps \cite{Sobral}:

\begin{itemize}
\item Model initialization: A background model is built based on a number of frames.
\item Foreground detection: The model is compared to the next frame. Pixels that are not well-described by the model are considered foreground.
\item Model maintenance: The background model is updated with the frame it was just compared to. In this way, the model can adapt to permanent changes in the background.
\end{itemize}

\noindent
The most simple background subtraction method is static background subtraction. 
Here, the background is assumed to never change and is simply represented by a single video frame, ideally containing only the background of a scene.
This frame could show an empty road in the context of traffic surveillance or an empty test chamber in the case of pellet shattering tests. 
Every pixel in subsequent frames whose brightness/color is not identical to its respective pixel in the static background model is considered foreground. 
While being easy to implement, this method rarely leads to satisfying results, as it is not stable under illumination changes and as even statistical fluctuations in pixel intensities can lead to false positives.\\

\noindent
A more sophisticated method is the single Gaussian model \cite{Pfinder}. 
Here, each pixel is assigned a Gaussian distribution centered around the mean value of that pixel in previous frames. 
If a future pixel value is close to this mean, it is considered background.
However, pixel values often have complicated distributions that require even more elaborate models.
For example, a tree background of a video should ideally be segmented as background even if it is swaying in the wind slightly.
Gaussian mixture models, as first proposed by Friedmann and Russell \cite{GMM_first}, use multiple Gaussian distributions to describe a single pixel. 
This allows better modeling of more complex conditions.
Zivkovic has further improved this approach, as described in his 2004 paper \cite{MOG2_2004}, to adapt the number of Gaussian distributions continuously for each pixel. 
This specific background subtraction method is implemented in OpenCV under the name \texttt{MOG2}. After testing the capabilities of different background subtraction algorithms implemented in OpenCV \cite{Projektarbeit}, I decided to use this algorithm both for the integrity measurements and the fragment analysis.
Furthermore, the pre-processing and background subtraction workflow developed in the course of this thesis was also used successfully in multi-perspective SPI fragmentation studies performed by Chongsathapornpong \cite{Jade_presentation} at ITER. \\

\begin{figure}[!h]
     \centering
     \captionsetup{width=\textwidth}
     \includegraphics[clip, trim = 0 0 0 0,width=0.55\textwidth]{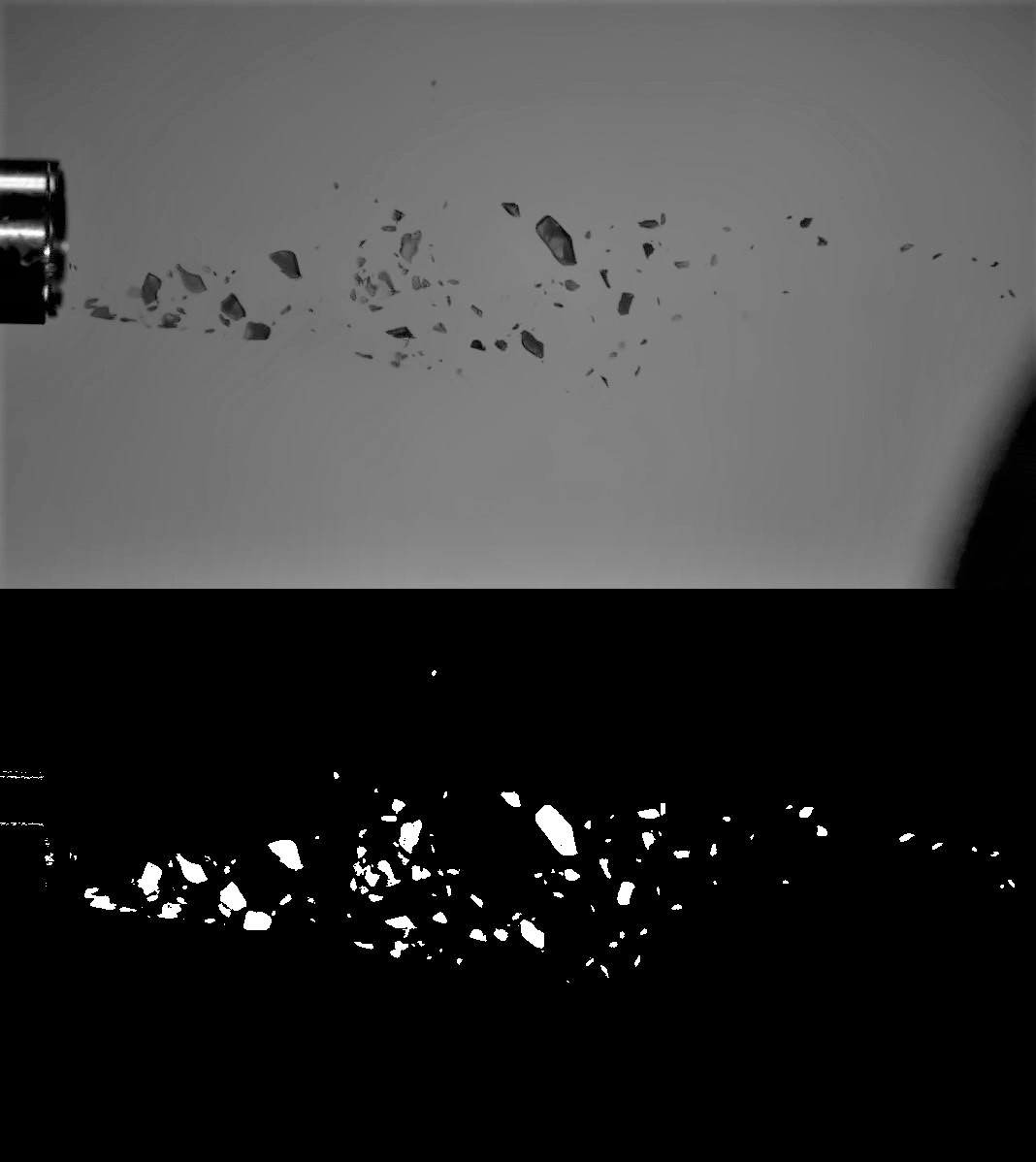}
     \caption{Example image with the corresponding foreground mask, generated by applying the MOG2 background subtractor. Moving objects are marked in white, background in black.}
     \label{f:bgsub_example}
\end{figure}

\noindent
For better understanding of this tool, let us consider one single image pixel over multiple video frames.
The model assumes that if this pixel is part of the video background, its intensity $I$ can be described by a Gaussian.
Its mean $\mu$ is computed from the intensities $I$ of a fixed number of previous frames (determined by learning rate).
The variation of $I$ over the considered frames is used to construct $\sigma$ such that a high variation corresponds to a high $\sigma$ (broad Gaussian) and vice versa.
If, for a new frame, the gaussian at the new pixel value is above a pre-defined sensitivity threshold, the pixel is considered background, as it is well described by the Gaussian model.
To be able to account for changes in the background throughout the video, the number of Gaussian components is adapted dynamically throughout the video.
This process is done for every pixel in a frame to produce a segmented image (also called foreground mask), with pixel values of 1~(white) for every foreground pixel and 0~(black) for every background pixel, as shown in fig.~\ref{f:bgsub_example}.
When executed for every frame in the video, one obtains a full background segmentation of the video.

\begin{figure}[!h]
     \centering
     \includegraphics[clip, trim = 70 50 70 60,width=0.8\textwidth]{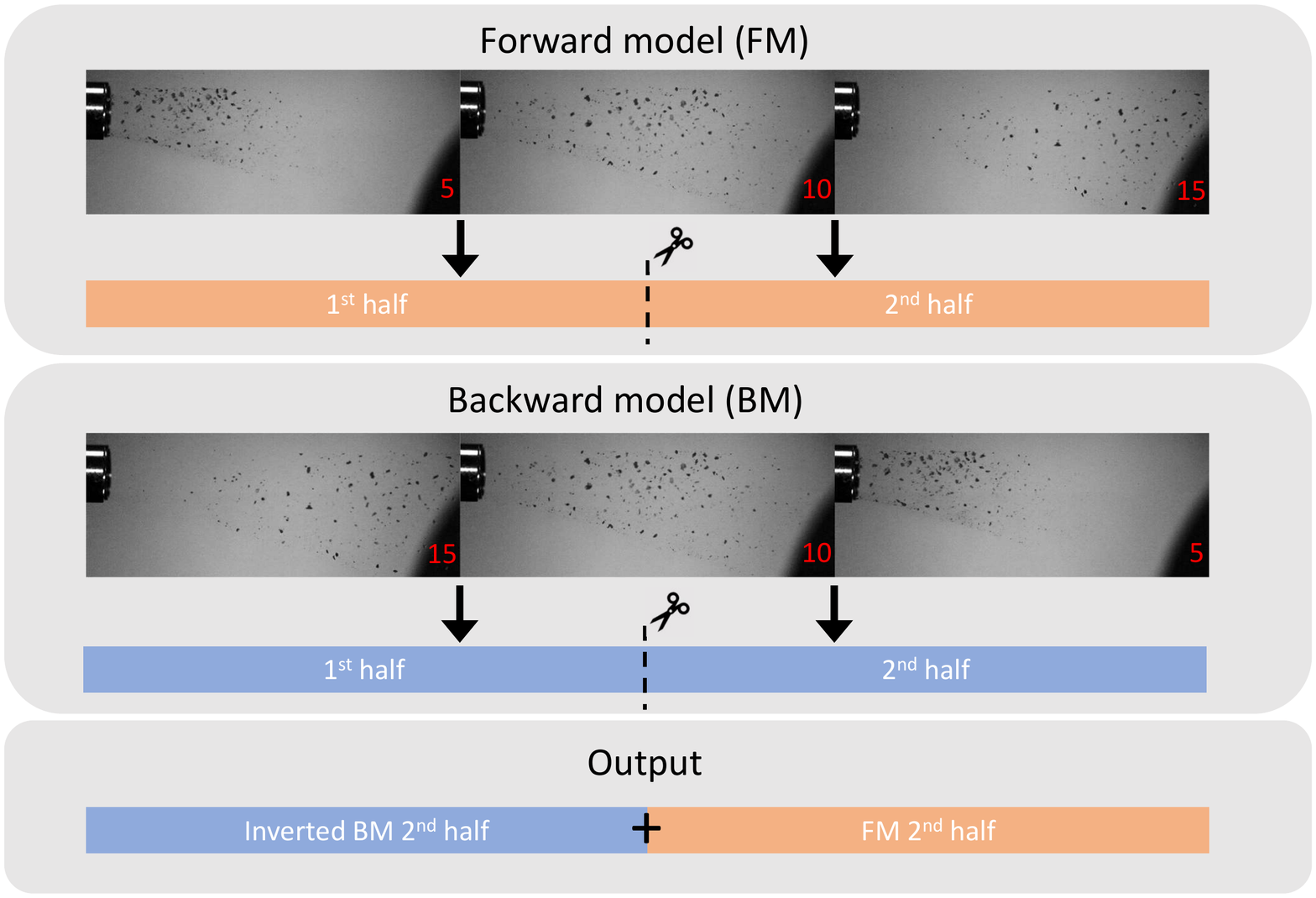}
     \caption{Visualization of the custom background subtraction scheme. Frame numbers are shown on the lower right. 
     The video is fed to two background subtractor models, once in forward direction (FM), once in backward direction (BM).
     Each one produces a foreground mask for each video frame.
     These masks are halved and a complete foreground mask of the video is generated by joining the inverted second half of BM with the second half of FM.}
     \label{f:subtractDouble}
\end{figure}

\noindent
Since I found that, when applying background subtraction, the later half of the segmentation was almost always better than the first, I decided to use a custom technique for the analysis.
It is illustrated in fig.~\ref{f:subtractDouble}.
In this method, two background models with identical initial settings are trained simultaneously.
To one of them, the video is fed in forward direction, the other one gets the time-reversed version (order of images is reversed).
The resulting foreground masks are separated into two halves with the same length. 
The second half of the time-reversed background subtraction is again time-reversed and joined with the second half of the forward background subtraction.
By doing so, a complete foreground mask with enhanced quality is generated.

\subsection*{Contours and Image Moments}
\label{s:contour_detection}

So-called contour detection is a computer vision method based on edge detection that works well with binary images (e.g. foreground masks)
Contour detection searches for closed boundaries in the image can be used to obtain separate objects from the black-and-white foreground masks.
For each visible object, a list of pixel coordinates along the object boundary is produced.
Each contour is assigned an ID and is unique to one object in the foreground mask.\\
 

\noindent
Now that single objects have been identified, we can proceed to measure position and size of these objects.
To do this, each contour is re-drawn in white~(1) onto a black~(0) image and the pixel values inside it are set to white.
Position and size are then computed using image moments $M_{ij}$, which are defined by 

\begin{equation}
     M_{ij} = \sum_x \sum_y x^i y^j I(x,y) \, ,
\end{equation}

\noindent
where $x$ and $y$ are the horizontal and vertical pixel coordinate and $I(x, y)$ is the corresponding pixel value.
In binary images, the object area $A$ and the centroid coordinates $\bar{x}$ and $\bar{y}$ are given by

\begin{align}
     A & = M_{00}\\
     \bar{x} & = \frac{M_{10}}{A}\\
     \bar{y} & = \frac{M_{01}}{A} \, .
\label{e:moments}
\end{align}

\section{Fragmentation Video Analysis}
\label{s:analysis_software}

With the tools outlined in the previous sections, I constructed an algorithm that can automatically identify, follow and measure fragments in SPI laboratory test videos. 
A detailed description of this algorithm will be given in this section.

\subsection*{Pellet data acquisition and calibration}

During the SPI laboratory testing, a .xlsx file was used as a database to store information for every fired pellet. 
This database contains, among other things, the unique pellet ID and the pellet velocity, measured by its time of flight from the integrity diagnostic to the test chamber.
This time of flight is inferred from the frame number at which the bulk of fragments becomes visible in the test chamber video, which is recorded manually right after firing the pellet.
These measurements were largely carried out by my colleagues at IPP Paul Heinrich, Gergely Papp, Pascal de Marné and Mathias Dibon.
Before analyzing the fragmentation of any pellet, the corresponding entry is automatically searched in the pellet database to obtain the initial pellet speed and the frame number in which the main plume enters the FOV are taken from there.
This frame number is used to skip to the first frames in the video where fragments become visible, decreasing computation time and required memory.
The pellet speed (time of flight) is used as a prior for the individual fragment velocity.
This approach was found to be more robust than automated tracking of the center of mass, since it does not rely on the visibility of the full fragment plume for multiple frames.\\

\begin{figure}[!h]
     \centering
     \includegraphics[clip, trim = 50 80 50 80,width=0.5\textwidth]{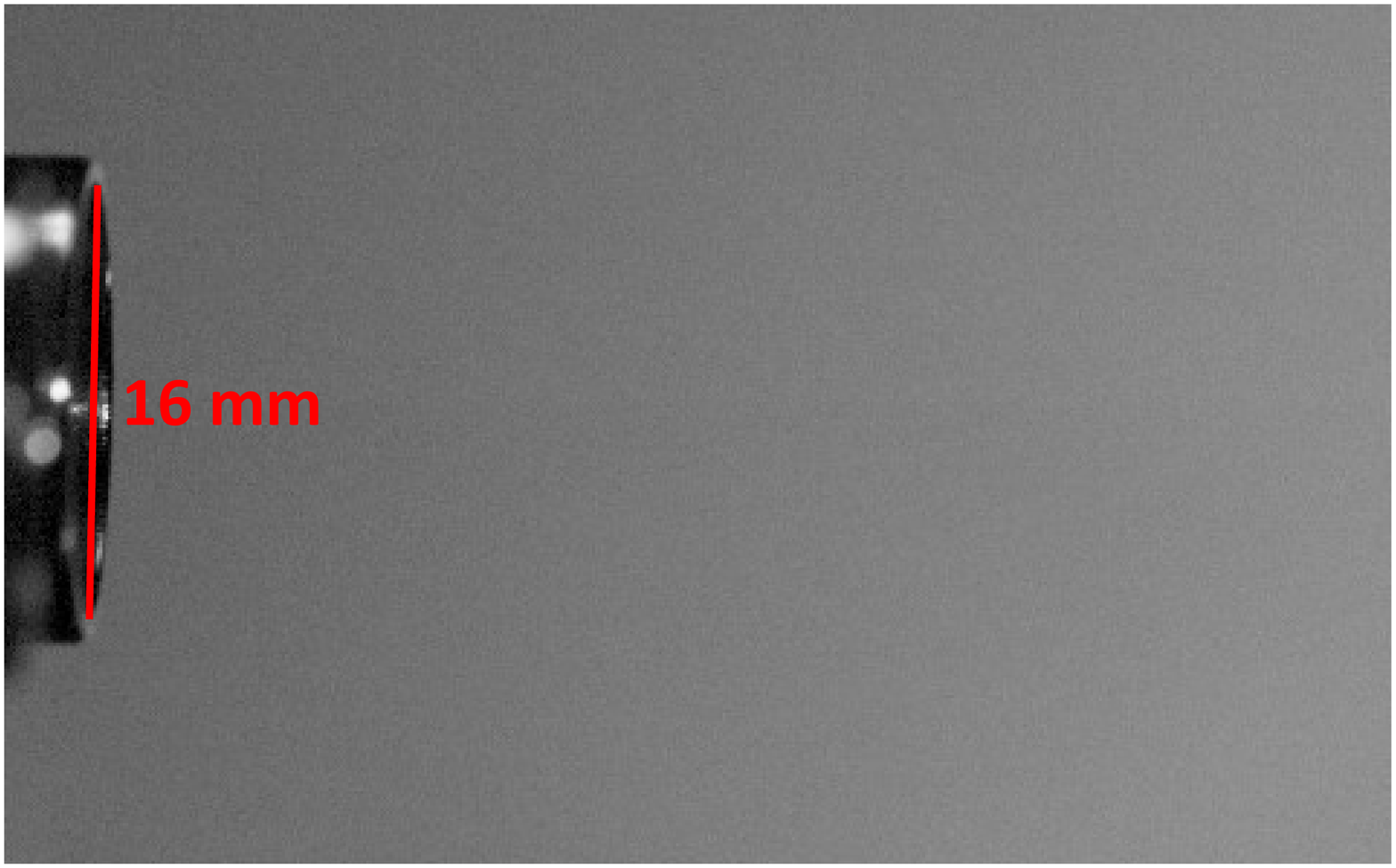}
     \caption{Calibration using the inner diameter of a circular shatter tube.}
     \label{f:calibration}
\end{figure}

\noindent
To convert measured distances from pixels to mm, videos need to be calibrated to a known distance. 
As shown in fig.~\ref{f:calibration}, the inner diameter of the shatter tubes (16 mm) is used for this.
Depending on the exact camera position and the injection line used, this conversion factor was found to be within 0.1 and 0.113 mm/pixels.
However, since calibration is necessary for every video and has to be done manually, a constant mm/pixel factor of 0.111 was assumed to automate the analysis procedure.

\subsection*{Pre-processing and Background subtraction}

During the experiments, the camera was tilted so that the fragments are always recorded flying in horizontal direction in order to image the largest portion of the plume.
Therefore, the camera was repositioned every time a new shatter angle was tested.
Furthermore, several alterations and improvements to the recording setup (e.g. stronger lighting, background plate) were made in the course of several months of recording pellets.
Due to these changes in lighting conditions and camera perspective, pre-processing routines and the variance threshold used in background subtraction had to be manually tailored towards multiple camera and illumination settings instead of using one setting for all of them. 
The settings used for different experiment days are stored on a GitHub repository with the analysis code itself.\\

\begin{figure}[!h]
     \centering
     \includegraphics[clip, trim = 0 50 0 30,width=0.6\textwidth]{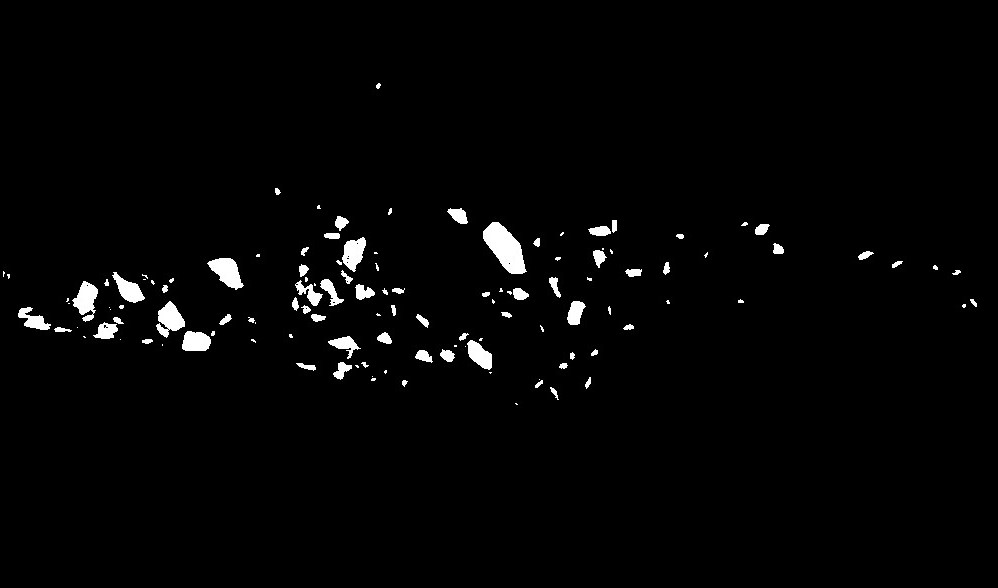}
     \caption{Example frame from the foreground mask for pellet \#434.}
     \label{f:fgmask}
\end{figure}

\noindent
After pre-processing, I used the custom background subtraction scheme outlined in section~\ref{s:background_subtract_concept} to obtain a binary foreground mask for each video frame.
For the analyses presented in this thesis, MOG2 sensitivity thresholds between 40 and 250 were used, depending on the amount of contrast modification done in the previous step.
The learning rate of the background model was set to automatic.
An example foreground mask image is shown in fig.~\ref{f:fgmask}.

\subsection*{Position and Size measurement}

As already mentioned in section~\ref{s:contour_detection}, contour detection is used to identify individual objects in the foreground masks.
For each visibile contour, the area $A$ and centroid coordinates $\bar{x}$ and $\bar{y}$ is calculated using image moments. 
Under the assumption of a spherical fragment volume, which has been tested and found successful compared to other assumptions by Chongsathapornpong \cite{Jade_presentation}, the fragment diameter can be calculated from the visible area by

\begin{equation}
     d = 2\sqrt{\frac{A}{\pi}}\, .
\end{equation}

\subsection*{Fragment tracking}
\label{s:tracking}

Fragment tracking is done by characterizing fragments in one frame and looking for similar fragments in the next frame.
Since each fragment is moving, the corresponding fragment in the next frame will be displaced by shifts $\Delta x_j$ and $\Delta y_j$.
Here, $j$ denotes the fragment index.
In the first frame where a fragment $j$ is detected,  $\Delta x_j$ is set as the distance the initial pellet travels per frame.
This distance is calculated using the initial pellet velocity from the database, the camera framerate and the mm-pixel conversion obtained during calibration.
$\Delta y_j$ is assumed to be zero.
This prior has proven to provide a robust first estimate for fragment displacement.
Then, after each successful tracking step, $\Delta x_j$ and $\Delta y_j$ are set to the mean displacement of the corresponding fragment.
This is motivated by the assumption that individual fragments have a constant velocity during the video.\\

\begin{figure}[!h]
     \centering
     \includegraphics[clip, trim = 0 0 0 0,width=0.6\textwidth]{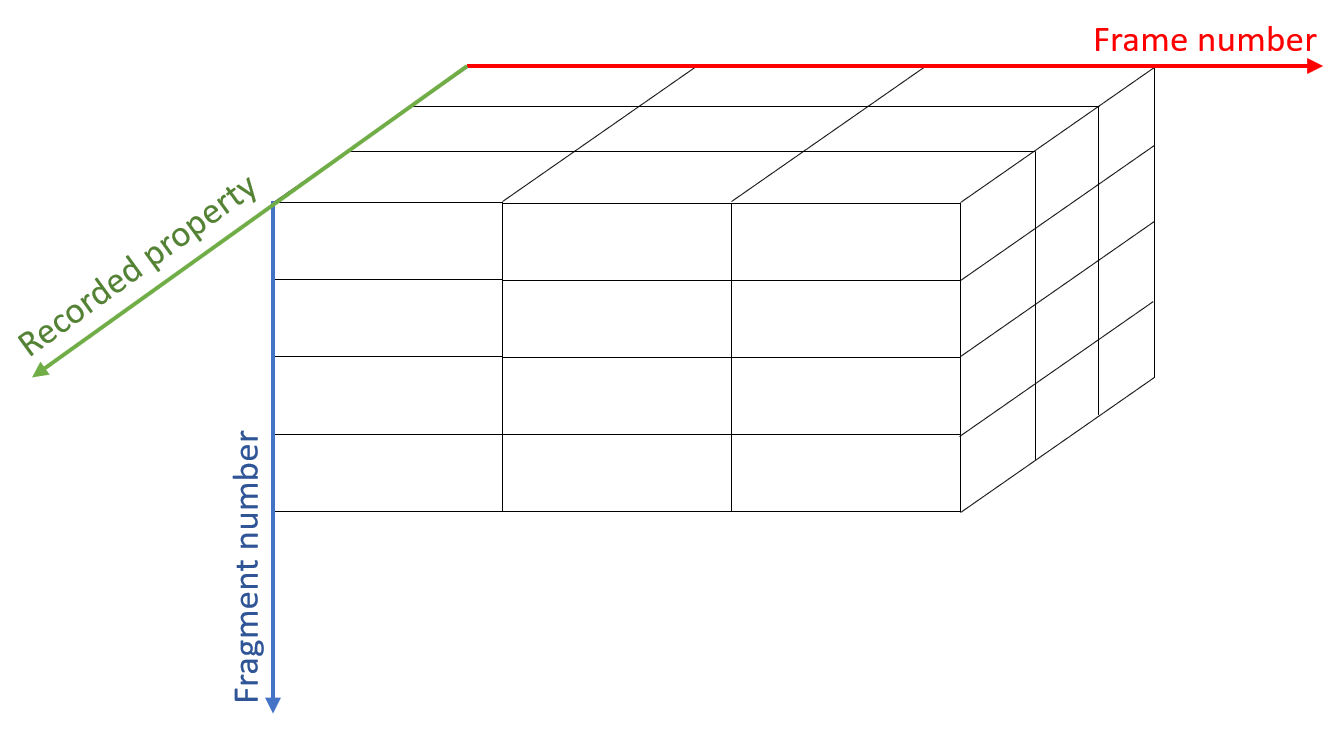}
     \caption{Schematic structure of output array.}
     \label{f:output_array}
\end{figure}

\noindent
The area and centroid coordinates $A$, $\bar{x}$ and $\bar{y}$ are computed for each object in the  current (``cur'') and next (``new'') frame.
In frame \#1, each object is assigned a unique fragment number/ID and the corresponding information ($\bar{x}$, $\bar{y}$, $A$) is entered into the output array.
The output array structure is illustrated fig.~\ref{f:output_array}.

\noindent
To identify the objects in subsequent frames with each other, I defined a similarity cost matrix $C$ as

\begin{equation}
C_{ij}= \sqrt{(\bar{x}_i^\text{new} - \bar{x}_j^\text{cur} + \Delta x_j)^2 + (\bar{y}_i^\text{new} - \bar{y}_j^\text{cur} + \Delta y_j)^2} + (A_i^\text{new} - A_j^\text{cur})^2\, .
\label{e:cost}
\end{equation}

\noindent
This matrix contains a similarity score for each combination of object $i$ in ``new'' and object $j$ in ``cur''.
$C_{ij}$ is minimal when the difference in shifted position and visible area is minimal.\par \noindent
An iterative process is used for matching fragments. 
It is illustrated in eq.~\eqref{e:cost_matrix}. 
First, the minimum of $C$ is computed. 
It is denoted by an underline in eq.~\eqref{e:cost_matrix}.
Let this minimum be at $C_{kl}$.
If it lies below a threshold value $T$ , the object $k$ in ``new'' is identified with object $l$ in ``cur'' and is assigned the same fragment ID as $l$.
A constant value of $T$ was used in the course of this thesis to keep the sensitivity of the tracking algorithm constant for all videos.
After the position and area of object $k$ have been entered into the output array, all values in $C$ along column $k$ and row $l$ are ruled out from the search (as these objects have already been matched).
Then, the search for a minimum begins again.
If, at any point, the minimum of $C$ is above $T$, the search is stopped.
All remaining objects in ``new'' are then considered newly detected fragments and are assigned a new fragment ID. 
This happens when new fragments exit the shatter tube, but also when fragments break apart or when occluded/overlapping fragments can become fully visible while flying through the FOV.
Remaining objects in ``cur'' could not be tracked successfully, so no entry in the output array will be made for the next frame and the trace of those objects is lost.
This, in turn, happens when a fragment exits the FOV or when fragments start occluding/overlapping.

\begin{equation}
\begin{pmatrix}
c_{00} & c_{10} & c_{20} \\
c_{01} & \underline{c_{11}} & c_{21} \\
c_{02} & c_{12} & c_{22}  \\
c_{03} & c_{13} & c_{23}
\end{pmatrix}
 \Rightarrow 
 \begin{pmatrix}
c_{00} & \infty & c_{20} \\
\infty & \infty & \infty  \\
c_{02} & \infty & c_{22}   \\
c_{03} & \infty & c_{23}
\end{pmatrix}
 \Rightarrow 
 \begin{pmatrix}
c_{00} & \infty & c_{20} \\
\infty & \infty & \infty  \\
c_{02} & \infty & c_{22}\\
c_{03} & \infty & \underline{c_{23}}
\end{pmatrix}
\Rightarrow ...
\label{e:cost_matrix}
\end{equation}

\noindent
After fragment matching has been done for all frames of a single video, the output array is stored as a .npy file for further processing.
This file can then be used to reconstruct the trace of an individual fragment through the FOV by looking at its centroid coordinates over all recorded frames.

\section{Integrity diagnostic video analysis}
\label{s:integrity_diagnostic}

At the integrity diagnostic, the pellet passes a pumping gap where it is filmed and analyzed to assess the integrity, size and speed of the pellet.
The software presented in this section has been in use in tuning pellet firing recipes for ASDEX-Upgrade SPI experiments as well as to aid tokamak experiment execution.
Measurements are done by applying multiple methods outlined in the previous sections.
During pre-processing, CLAHE is used to increase the pellet visibility. 
The custom background subtraction algorithm, illustrated in fig.~\ref{f:subtractDouble}, is applied to separate the moving pellet from the static background.
A MOG2 sensitivity threshold of 40 and an automatic learning rate were used. \\

\noindent
A constant mm-pixel conversion factor, obtained from calibration to the inner diameter of the visible tubes, is used.
To gain a prior of the pellet displacements $\Delta x$ and $\Delta x$, the position of the (two-dimensional) center of mass is calculated.
To do this, image moments $M_{00}$, $M_{10}$ and $M_{01}$ are computed for the whole resulting foreground masks without using contour detection to seperate objects.
The center of mass is followed over multiple frames to obtain a prior for the velocity.
Then, one of four different measurement modes is chosen:

\begin{enumerate}
     \item ``\texttt{biggest}"
     \item ``\texttt{all}"
     \item ``\texttt{manual}"
     \item ``\texttt{backtilt}"
\end{enumerate}

\noindent
The default mode is of the integrity analysis software is called ``\texttt{biggest}".
It automatically tracks the largest detected object (by area $A$) over multiple frames and measures pellet size, speed and tilt.
These measurements are conducted by using an OpenCV routine to find the smallest enclosing rotated rectangle for a set of (foreground) pixels. 
The size and pellet tilt is then given by the median of the dimensions and tilt of the rectangle.
Speed is calculated by using the tracking algorithm shown in the previous section.

\noindent
In the so-called  ``\texttt{all}"-mode, not only the biggest object, but all visible objects are tracked automatically. 
This setting is useful for the analysis of pellets that are broken when entering the integrity diagnostic.
In both these modes, the measured median size, speed and tilt are written into the console and stored in a database automatically.\par
\begin{figure}[!h]
     \centering
     \begin{subfigure}[b]{0.3\textwidth}
          \centering
          \includegraphics[clip, trim = 0 0 0 0, width=0.8\textwidth]{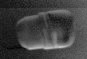}
          \caption{}
          \label{f:all_mode_pellet}
      \end{subfigure}
      \hfill
     \begin{subfigure}[b]{0.3\textwidth}
         \centering
         \includegraphics[clip, trim = 0 0 0 0, width=0.8\textwidth]{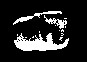}
         \caption{}
         \label{f:all_mode_ex}
     \end{subfigure}
     \hfill
     \begin{subfigure}[b]{0.3\textwidth}
         \centering
         \includegraphics[clip, trim = 0 0 0 0, width=0.8\textwidth]{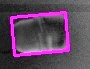}
         \caption{}
         \label{f:all_mode_res}
     \end{subfigure}
     \caption{Exemple screenshots of pellet (a),  foreground mask (b) and detected pellet size (c) using ``\texttt{all}''-mode. 
     Distance-based hierarchical clustering is used to group the separate contours depicted in (b). Then, the smallest enclosing rectangle of the group is used as an estimate for the pellet size in (c).}
        \label{f:all_mode}
\end{figure}

 \noindent
Both measurement modes have also been modified to be able to deal with partially transparent pellets (see fig.~\ref{f:all_mode_pellet}). 
In these, segmentation with contour detection often only segments pellet edges instead of the whole pellet, as depicted in fig.~\ref{f:all_mode_ex}. 
To overcome this problem, a hierarchical, distance-based clustering algorithm is used to nearby group pixels.
This way, even though the pellet has not been fully segmented as foreground, the size of the initial pellet can be estimated by ``stitching together'' detected objects, as shown in fig.~\ref{f:all_mode_res}.
However, clouds of pellet material can pose a problem, as they are often also detected as part of the pellet.

\noindent
This is adressed in one of the two semi-automatic analysis settings.
``\texttt{Manual}" mode is used when automatic size detection fails due to the pellet being obscured by dust. 
The operator is shown a video frame and asked to manually skip forward to a frame that is beneficial for measurement.
By clicking into the video frame in ``\texttt{manual}" mode, the corners of the pellet can be marked, from which the smallest enclosing rectangle is then calculated.

\begin{figure}[!h]
     \centering
     \captionsetup{width=\textwidth}
     \includegraphics[clip, trim = 0 0 0 0,width=0.7\textwidth]{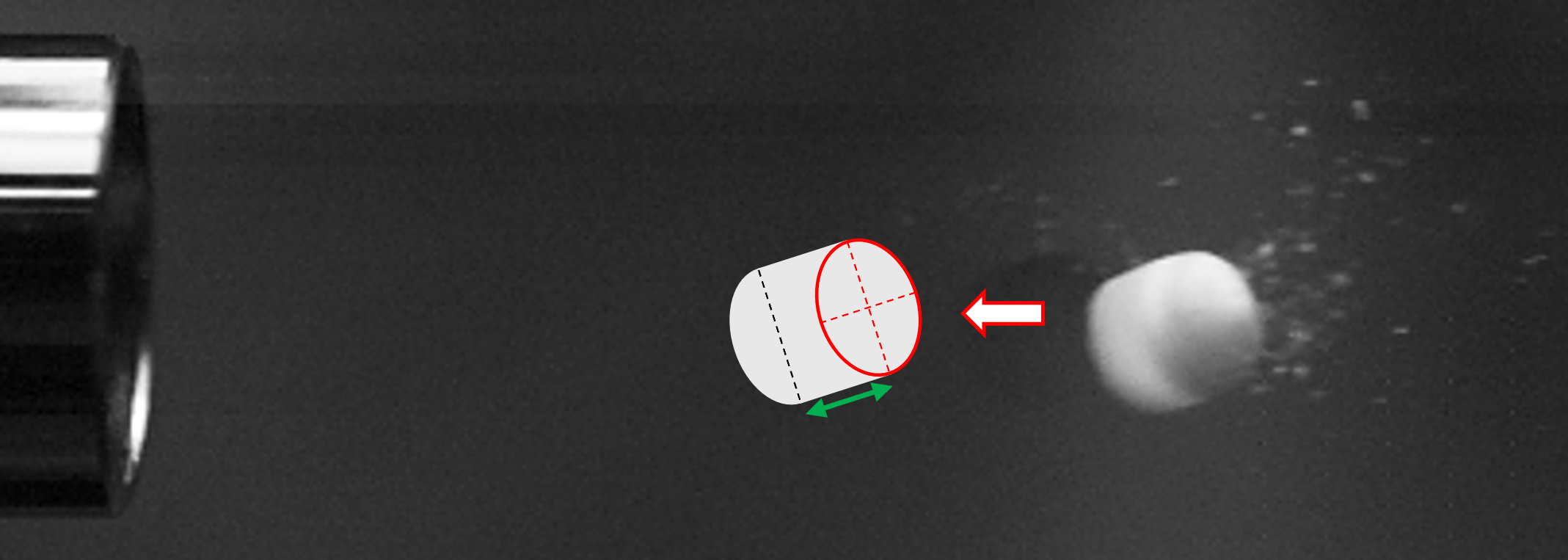}
     \caption{Schematic representation of the measurement of pellet tilt towards the camera. An ellipsis is fitted to the base of the cylindrical pellet to determine the pellet tilt and the real pellet length is calculated from the visible length.}
     \label{f:backtilt}
\end{figure}

\noindent
The ``\texttt{backtilt}" mode is used when the pellet length can not be measured correctly because the pellet is strongly tilted towards the camera.
A schematic image is given in fig.~\ref{f:backtilt}
Again, the operator skips through the frames until the pellet is well visible.
Then, the visible pellet length $l_v$ is marked by right-clicking and dragging.
After this, several points along the  base area edge of the cylindrical pellet are marked by clicking.
An ellipsis is fit to these points and the in-plane tilt $\alpha$ is calculated from the ratio of its major and minor axis.
The real pellet length is then given by $l_v / \sin{\alpha}$.\\

\noindent
The measured pellet speed, diameter, length and tilt are recoreded into the pellet database.
It has to be noted that the pellet speed is not calculated when a semi-automatic analysis mode is used.
This is not a problem however, as the pellet speed is also measured using time-of-flight calculations.
During the experiments, both measurements showed agreement up to $\pm$ 10 m/s.

\chapter{Results}
\label{s:Results}

In this chapter, a detailed description of the analyzed pellets and examples of the performance of the previously discussed computer vision methods are given.
Results obtained by video analysis of cryogenic pellet shattering and from theoretical fragmentation simulations are presented.
These include fragment size distributions obtained experimentally and through simulations, statistical parameters of these distributions.
I present a detailed comparison of theoretical and experimental fragment size distributions and discuss the influence of shatter head geometry on the experimental results.
Furthermore, a spatial spray distribution analysis of different shatter heads is shown.
Finally, a comparison of initial pellet volume and cumulative fragment volume is shown and some examples of experimentally measured fragment velocity distributions are given.
However, before presenting these results, I would like to discuss some limitations of the methods used to obtain them.

\section{Limitations of fragment size measurements}
\label{s:limitations}

Since fragment size measurements were carried out using a single-perspective,  high-speed camera and automated software, there are some limitations to consider when looking at the results.
Firstly, when the 3-dimensional fragment spray is projected onto a single image plane, fragments that are separated in a direction perpendicular to the FOV can appear as if they touch/overlap. 
These will then either be segmented as a single, big chunk rather than several individual fragments or one fragment will occlude another one, reducing the total fragment count. 
This is especially relevant for rectangular shatter geometries that were found to produce a collimated  fragment spray.\\
\noindent
A second problem also arises with the projection onto the image plane: Any depth information of the debris is lost.
In this thesis, I approximated the fragments to be spherical. 
This is reasonable when one knows only the projected area of a fragment rather than its entire shape, but this approximation introduces an amount of uncertainty into the measurements.
It can be argued that the ambiguous three-dimensional rotation of the fragments justifies the assumption that the average fragment is spherical. 
Simultaneous multi-axis camera views are currently being investigated by ITER to overcome these problems \cite{Jade_presentation}.\\
\noindent
Thirdly, pellets with perpendicular velocities above 100 m/s often produce bunches of fragments with diameters smaller than 0.1 mm.
These are often segmented as one single fragment larger than 3 mm in diameter rather than many different ones and can also obscure other fragments inside/behind the cloud.
This effect is sometimes visible in the experimental size distributions at the highest perpendicular velocities and can also distort  derived statistical parameters such as the 20\% mass quantile.\\
\noindent
The last limitation is systematic: 
The camera and the HMI lamp were moved every time the shatter heads were changed to get an optimal image of the shatter spray. 
While we tried to achieve similar lighting conditions, these modifications still required the pre-processing settings to be chosen manually. 
Since these settings were not constant for every measured pellet, the effective sensitivity of fragment detection differs slightly between analyzed fragmentations.
In an effort to keep the last two errors as small as possible, every single analysis result was reviewed manually before using it in this thesis.

\section{Analysis of AUG-SPI pellets}
\label{s:analysis_of_spi_pellets}

With the laboratory settings discussed in ch.~\ref{s:SPI_setup}, approx. 1400 pellets were launched.
The algorithm presented in ch.~\ref{s:analysis_software} was used to analyze all recorded videos of pellets with a diameter of 4 mm fired between August 5$^{\text{th}}$, 2021 and September 1$^{\text{st}}$, 2021.
This pellet size was expected to be the most used in SPI experiments at ASDEX upgrade.
After this time, a campaign with larger pellets (8 mm in diameter) was started.
This data was not included in this thesis, as it would have exceeded the scope the thesis.\\

\noindent
In the analyzed time period, 493 pellets with a diameter of 4 mm were shot.
146 of those were not recorded successfully. 
They were either released prior to the firing trigger through evaporation of the pellet material and self-propelling or arrived at the camera positions too late to be recorded, which might have been due to improper pellet formation.
The analyses of the remaining 347 pellets were evaluated manually by looking at the reconstruction of the fragment plume for each of them.
The following criteria were used to assess analysis quality:

\begin{itemize}
     \item Most visible fragments must be detected as foreground.
     \item False segmentation due to noise/reflections on background must be minimal.
     \item No significant fragment overlap must be present.
\end{itemize}

\noindent
Of the  347 recorded videos, 170 fragmentation analyses were deemed successful, based on the criteria stated above. 
Results for these pellets are presented in this thesis.
63 of them were pure D$_2$ pellets, 56 consisted of pure Ne, the remaining 51 were mixture pellets containing 5\% , 20\% or 40\% Ne (molar fraction).
An overview of all analyzed pellet compositions and tested shatter heads is given in tab.~\ref{t:analyzed_pellets}.
Speed and length of these pellets were varied to cover a larger parameter space.
For each fragmentation, size distribution and velocity distribution were measured with the algorithm presented in ch.~\ref{s:analysis_software}. 
Additionally, a set of discrete realizations of the fragmentation model by Parks, as outlined in section~\ref{s:model_generation}, was created for each analyzed pellet.\\

\begin{table}[!h]
\centering
\captionsetup{width=0.82\textwidth}
\caption{Number of analyzed pellets for different combinations of shatter tube pellet composition in molar \% Ne. $\bigcirc$ and $\square$ denote a circular/rectangular tube cross section.} 
\begin{tabular}{|c||c|c|c|c|c|}
\hline
 & \multicolumn{5}{c|}{\textbf{Neon fraction} } \\
\hline
\textbf{Shatter tube} &\textbf{0\%}& \textbf{5\%} & \textbf{20\%} & \textbf{40\%} & \textbf{100\%} \\
\hline

\textbf{12.5\degree} $\bigcirc$& 9 & 12 & - & - & 10 \\
\hline
\textbf{15\degree} $\bigcirc$& 13 & 7 & - & - & 7 \\
\hline
\textbf{25\degree} $\bigcirc$ & 14 & 9 & 7 & 2 & 23 \\
\hline
\textbf{25\degree} $\square$& 12 & 6 & - & - & 10 \\
\hline
\textbf{30\degree} $\bigcirc$ & 15 & 8 & - & - & 6 \\
\hline
\end{tabular} 
\label{t:analyzed_pellets}
\end{table}

\begin{figure}[!h]
     \centering
     \begin{subfigure}[b]{0.45\textwidth}
         \centering
		\includegraphics[clip, trim = 0 0 0 0,width=\textwidth]{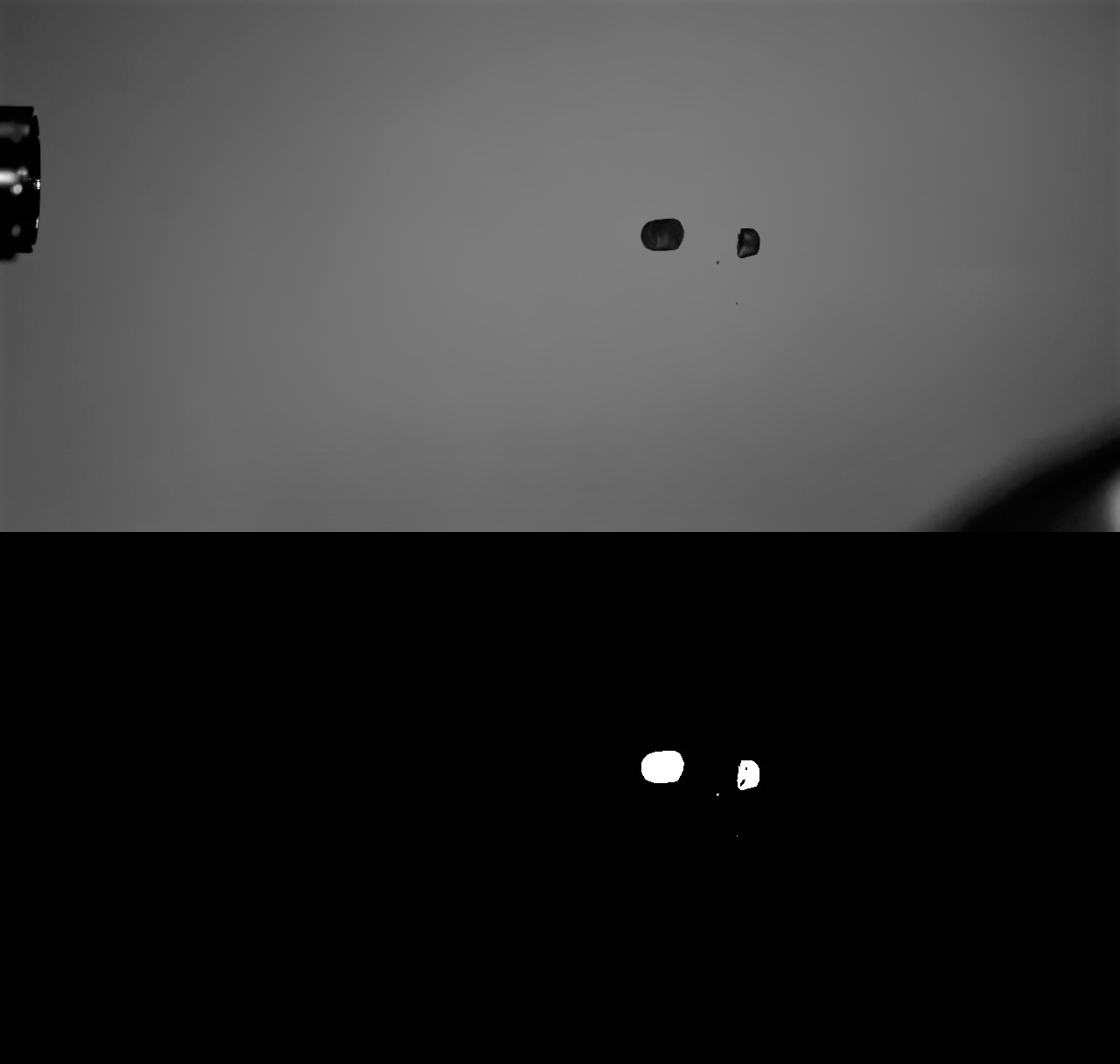}
     		\caption{Pellet \#380, 100\% Ne, 12.5\degree\, shatter tube, 112~m/s pellet speed.}

         \label{f:demo_bgsub1}
     \end{subfigure}
     \hfill
     \begin{subfigure}[b]{0.45\textwidth}
         \centering
       	\includegraphics[clip, trim = 0 0 0 0,width=\textwidth]{434_demo_bgsub.jpg}
     		\caption{Pellet \#434, 100\% Ne, 25\degree\, shatter tube, 140~m/s pellet speed.}
         \label{f:demo_bgsub2}
     \end{subfigure}
     \vskip\baselineskip
     \begin{subfigure}[b]{0.45\textwidth}
         \centering
        	\includegraphics[clip, trim = 0 0 0 0,width=\textwidth]{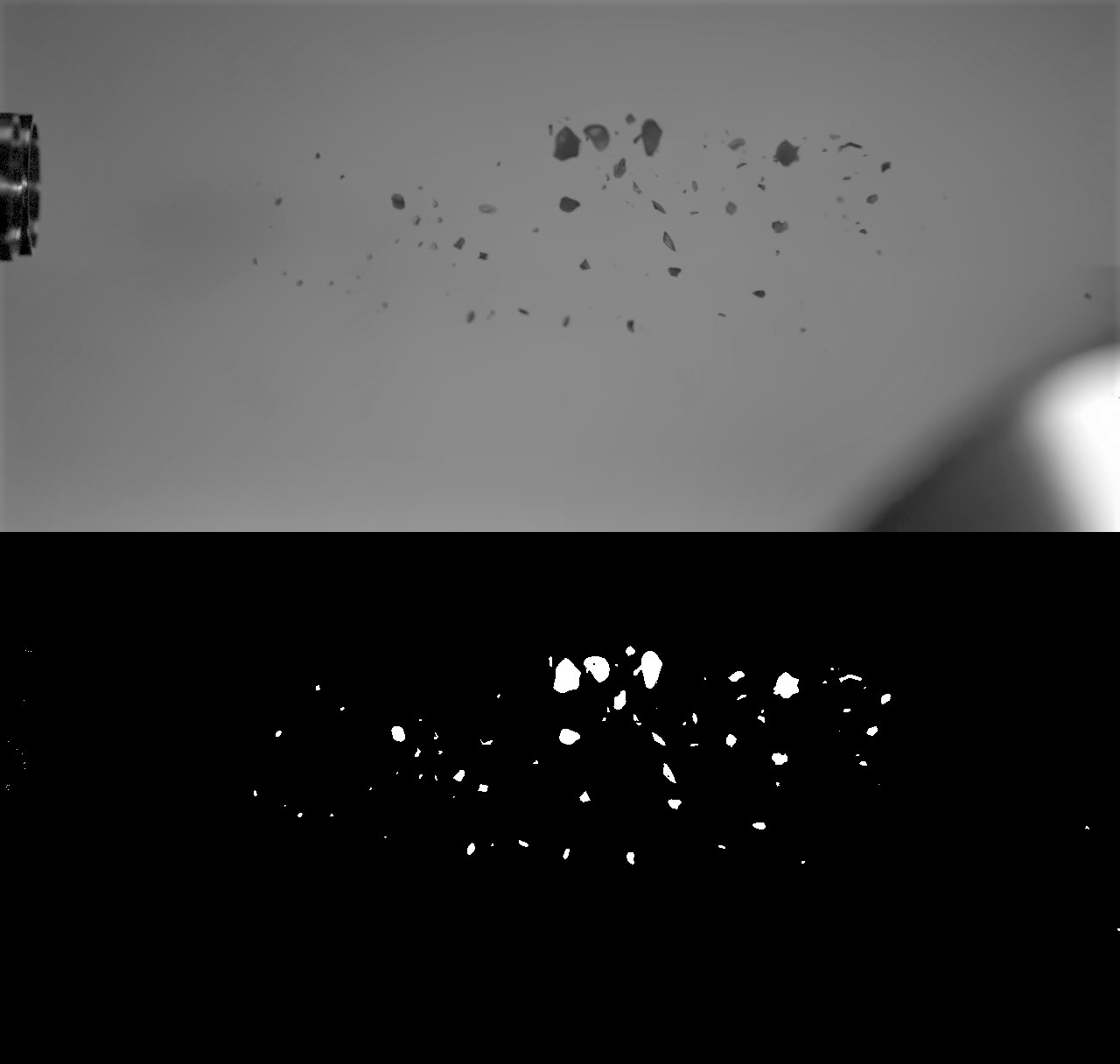}
     		\caption{Pellet \#234, 100\% Ne, 25\degree\, shatter tube, 202~m/s pellet speed.}
         \label{f:demo_bgsub3}
     \end{subfigure}
	\hfill
     \begin{subfigure}[b]{0.45\textwidth}
         \centering
        	\includegraphics[clip, trim = 0 0 0 0,width=\textwidth]{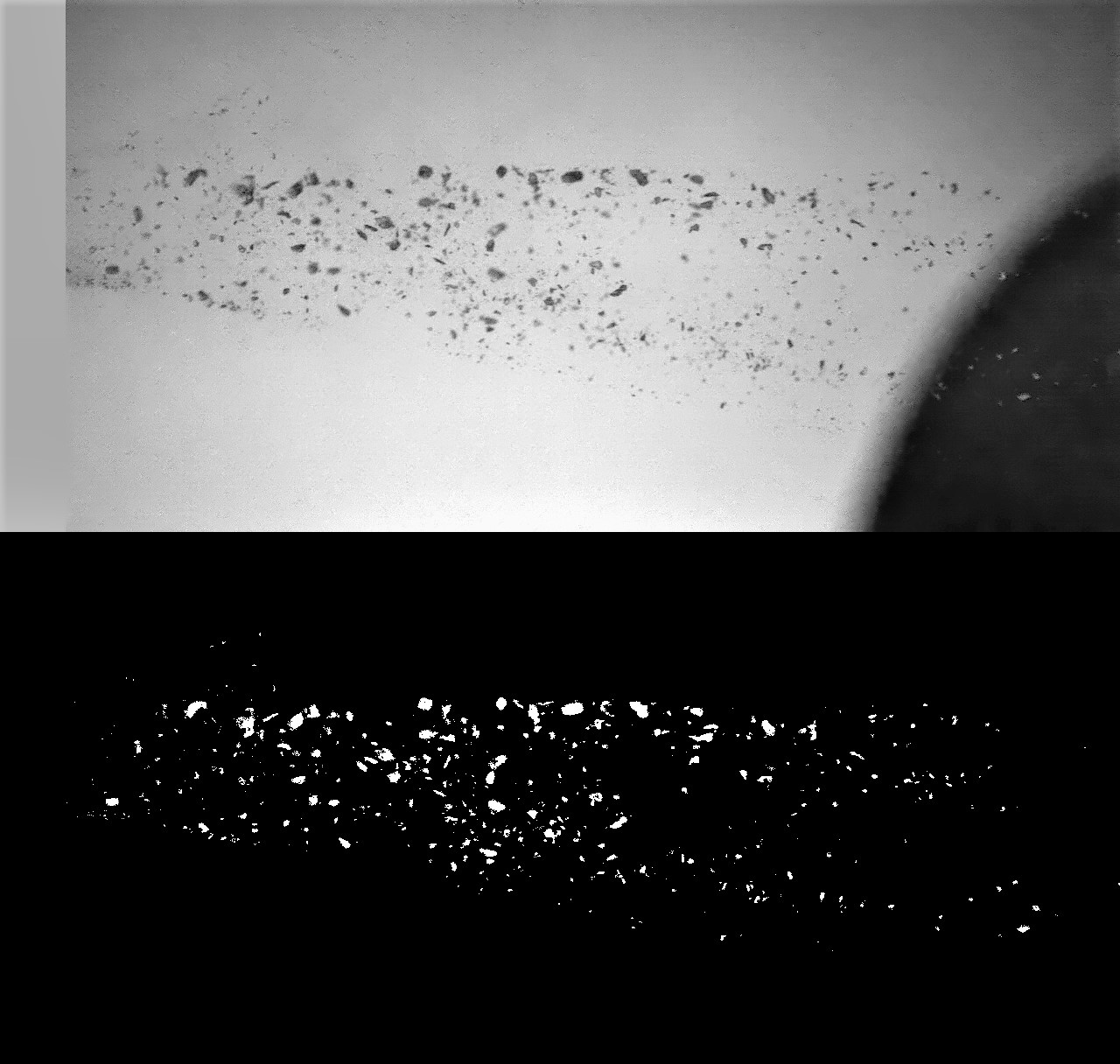}
     		\caption{Pellet \#549, 100\% Ne, 25\degree\, shatter tube, 260~m/s pellet speed.}
         \label{f:demo_bgsub4}
     \end{subfigure}
     
        \caption{Examples of 4 different pellet fragmentations and the corresponding foreground segmentation at different pellet speeds.}
        \label{f:demo_bgsub}
\end{figure}

\noindent
To illustrate the performance of fragment detection, four different pellet fragmentation scenarios along with the resulting foreground segmentation are shown in fig.~\ref{f:demo_bgsub}. 
All shown pellets are pure Ne, fired at different speeds and normal impact velocities. 
Pellet \#380 is shown in fig~\ref{f:demo_bgsub1}, fired at a 12.5\degree~circular shatter head at 112 m/s. 
This corresponds to a normal velocity of approximately 24 m/s. 
The pellet broke into two large chunks and 2-3 shrapnels.
It has to be noted that, although the foreground segmentation is not influenced by any noise, the transparency of the pellet material leads to two voids inside one of the large fragments.\par\noindent 
In fig.~\ref{f:demo_bgsub2}, pellet \#434 is shown.
It was fired at a circular 25\degree~shatter tube at 140 m/s, which corresponds to a normal velocity of about 59 m/s.
Some comparatively small fragments fly out of the DOF and are therefore not segmented as foreground.
Furthermore, reflections on the shatter tube (left) introduce false positives.
\par\noindent 
Figure~\ref{f:demo_bgsub3} shows pellet \# 234 after hitting a 25\degree~circular shatter tube at 202 m/s, corresponding to a normal velocity of 85 m/s.
2-3 fragments seem to fly out of focus at the end of the plume and are not detected. \par\noindent 
Pellet \#549 is depicted in fig.~\ref{f:demo_bgsub4}.
It was fired at a 25\degree~circular shatter tube at 260 m/s, producing an impact with a normal velocity of 109 m/s. 
This pellet broke into two pieces during firing, producing an elongated fragment spray.

\clearpage

\section{Experimental and simulated fragment size distributions}
\label{s:size_dist_res}

\begin{figure}[!h]
     \centering
     \includegraphics[clip, trim = 120 30 100 10,width=\textwidth]{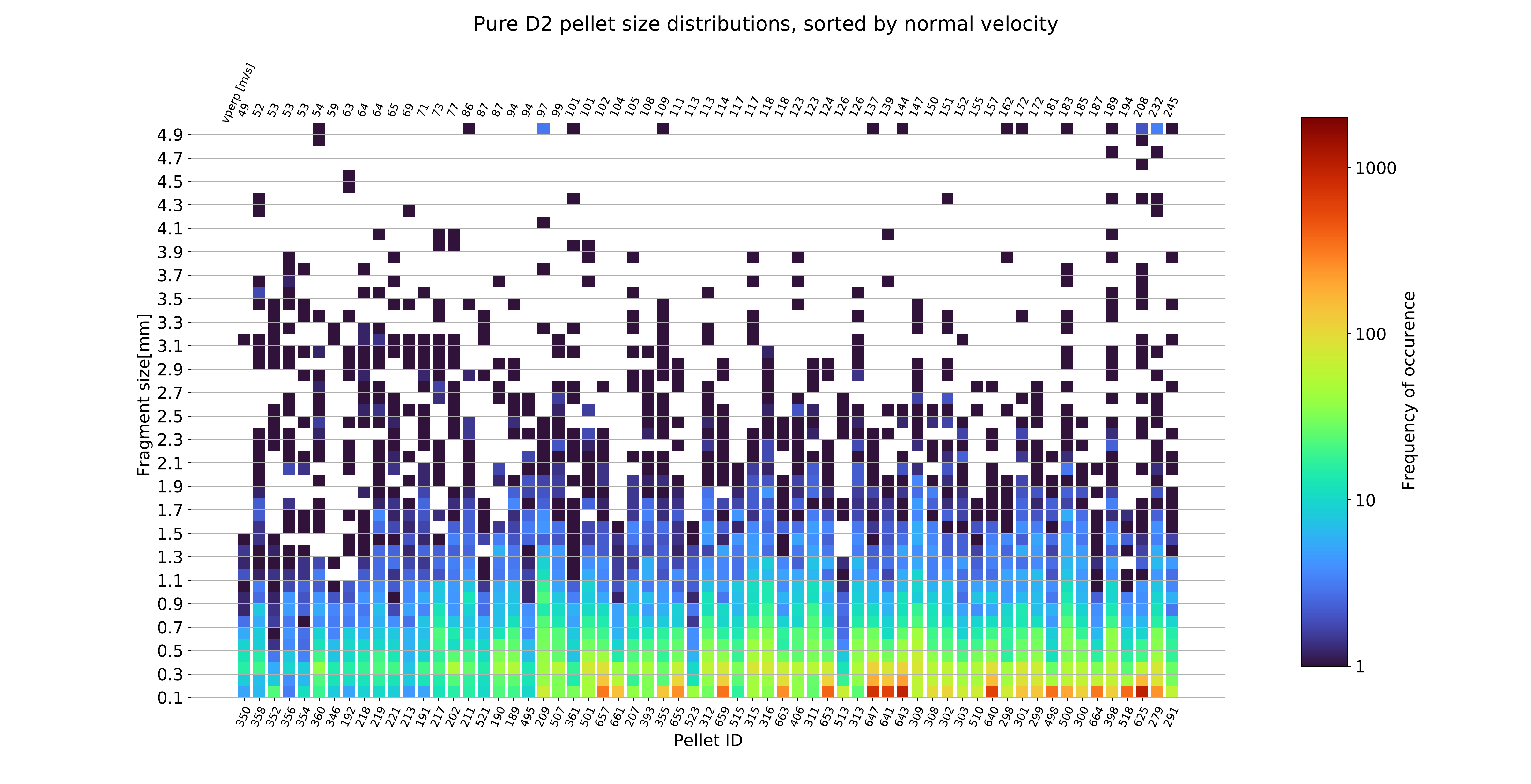}
     \caption{Experimental size distribution for pure D$_2$ pellets, sorted by normal velocity (top). On the bottom, the corresponding pellet ID is given.}
     \label{f:de_exp_vperp}
\end{figure}

As explained in detail in ch.~\ref{s:analysis_software}, the fragment diameter is computed from the visible shard area by assuming spherical fragments. 
In the course of this thesis, these diameters will simply be called fragment sizes.
Fragment size distributions for pure D$_2$ and Ne pellets are displayed in fig~\ref{f:de_exp_vperp} and fig.~\ref{f:ne_exp_vperp}. 
Distributions for 5\%, 20\% and 40\% Ne pellets are shown in the appendix.
The vertical axis denotes fragment size bins, ranging from 0.1 mm to 4.9 mm, with a bin width of 0.1 mm.
In the uppermost bin, fragments with a diameter larger than 4.9 mm are counted.
Note that this scale goes up to 4.9 mm in diameter, as an intact pellet with 4$\times$6 mm (side view) would be detected as a fragment with a diameter of approx.~5.5 mm.
In horizontal direction, the pellet ID is shown on the bottom. 
On top, the normal velocity during tube impact is stated as \textit{vperp}. 
Pellets are sorted from left to right by ascending \textit{vperp}.
The color of each histogram entry encodes the mean frequency of occurrence of the corresponding fragment size in a shot on a logarithmic color scale.
Entries are left blank if the mean fragment count for the corresponding size is smaller than 0.25. 
It is important to note that the difference in normal velocity between neighboring pellet IDs is not constant.
Normal velocities of down to 24 m/s were reached with pure Ne pellets, the highest normal velocity of 245 m/s was achieved with pure D$_2$. \\

\noindent
For each analyzed pellet, 1000 discrete fragment size distributions were generated using the methods explained in section~\ref{s:model_generation}.
For the evaluation of the theoretical model, molar Ne fraction, shatter angle, pellet length, diameter and speed were taken from the .xlsx database containing information on all fired pellets.
The mass fraction and threshold velocity were inferred from the molar fraction as explained in section~\ref{s:model_generation}.\\
In each discrete realization, samples were drawn from the probability density until the same cumulative fragment volume as in the experimental case was reached.
Fragment size histograms, using the same bin width and bin edges as in the experimental case, were calculated for each realization. 
Then the expectation value and statistical error of the predicted fragment count per measured fragment size bin was calculated.\\

\begin{figure}[!h]
     \centering
     \includegraphics[clip, trim = 120 30 100 10,width=\textwidth]{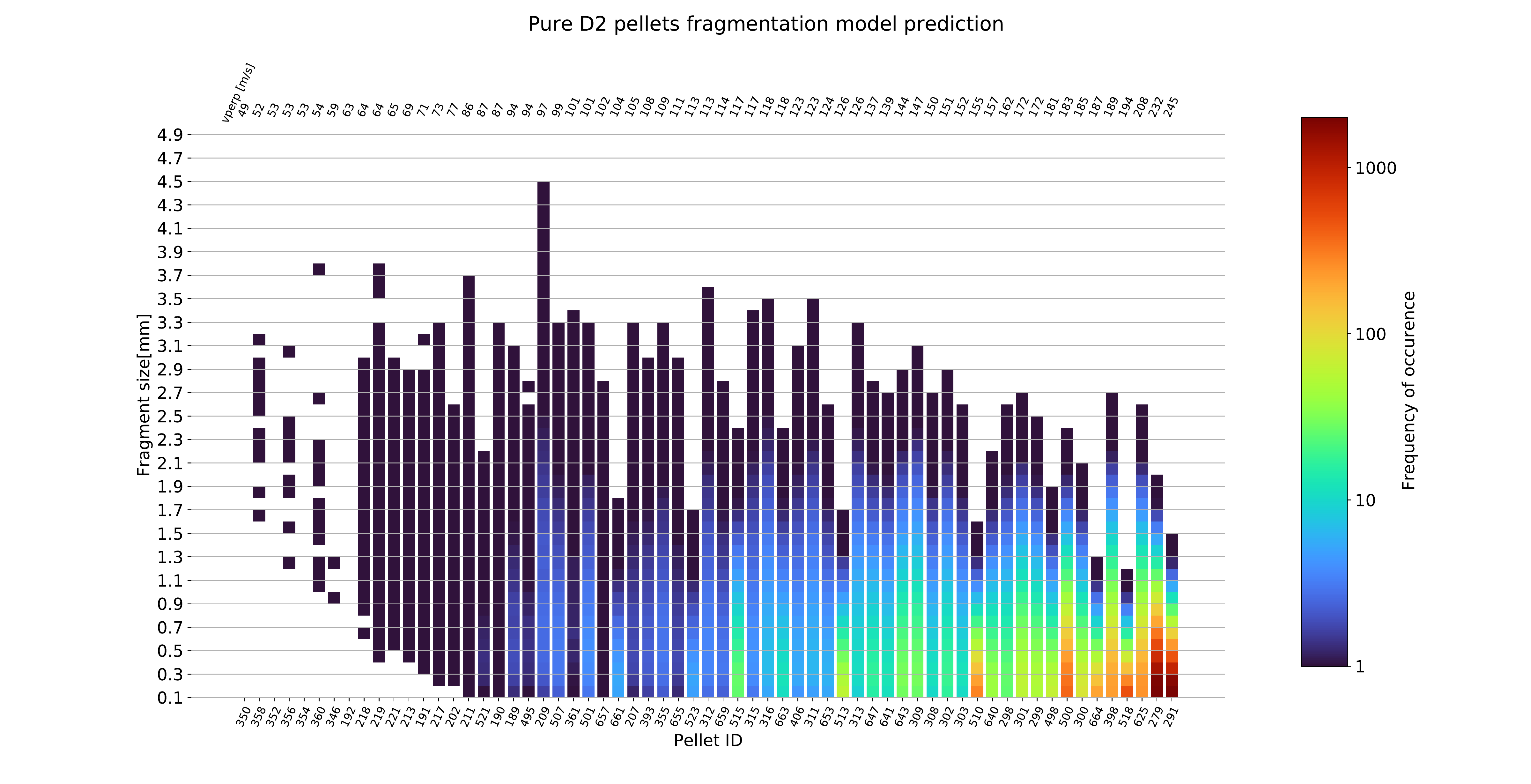}
     \caption{Modeled size distribution for pure D$_2$ pellets, sorted by normal velocity (top). On the bottom, the corresponding pellet ID is given.}
     \label{f:de_model_vperp}
\end{figure}

\noindent
The resulting mean distributions for D$_2$ and Ne pellets are shown in fig.~\ref{f:de_model_vperp} and fig.~\ref{f:ne_model_vperp}. 
Simulation results for 5\%, 20\% and 40\% Ne pellets are presented in the appendix.
The same color encoding, number of bins and bin locations as in the experimental distributions were used.
Experimental result and simulation result are shown in alternating order to simplify comparison.
In some simulation results, especially for pellets below 40 m/s, every bin is left blank. 
This tends to happen for long, slow pellets, as their theoretical size distribution is comparatively flat and has a maximum at fragment sizes above 2 mm.
Therefore, it is likely that only few fragments are drawn from the distribution until the stopping condition is met.
This way, the mean fragment count can fall below the threshold of 0.25 counts/bin for all size bins.

\newpage

\begin{figure}[!h]
     \centering
     \includegraphics[clip, trim = 120 30 100 10,width=\textwidth]{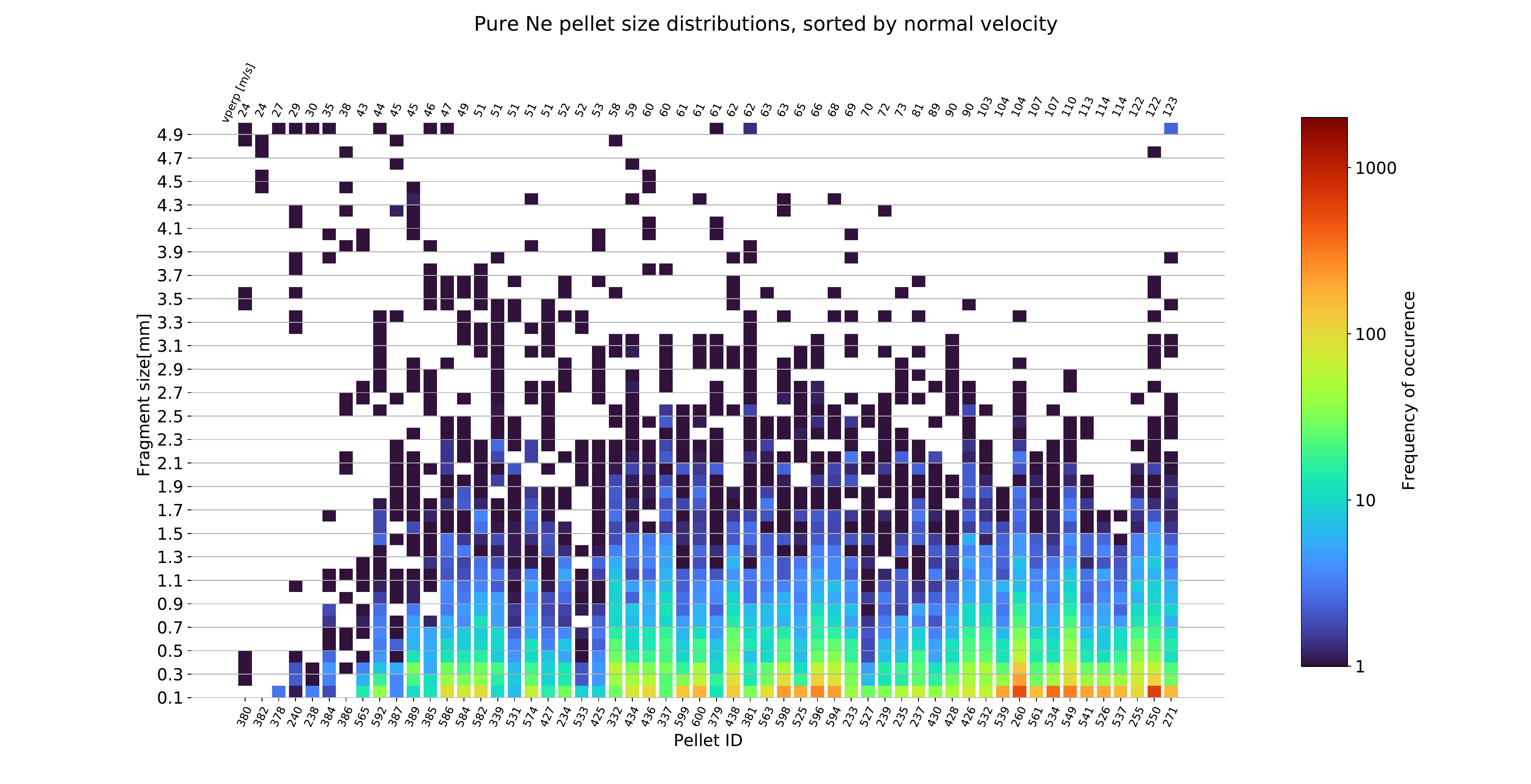}
     \caption{Experimental size distribution for pure Ne pellets, sorted by normal velocity (top). On the bottom, the corresponding pellet ID is given.}
     \label{f:ne_exp_vperp}
\end{figure}

\begin{figure}[!h]
     \centering
     \includegraphics[clip, trim = 120 30 100 10,width=\textwidth]{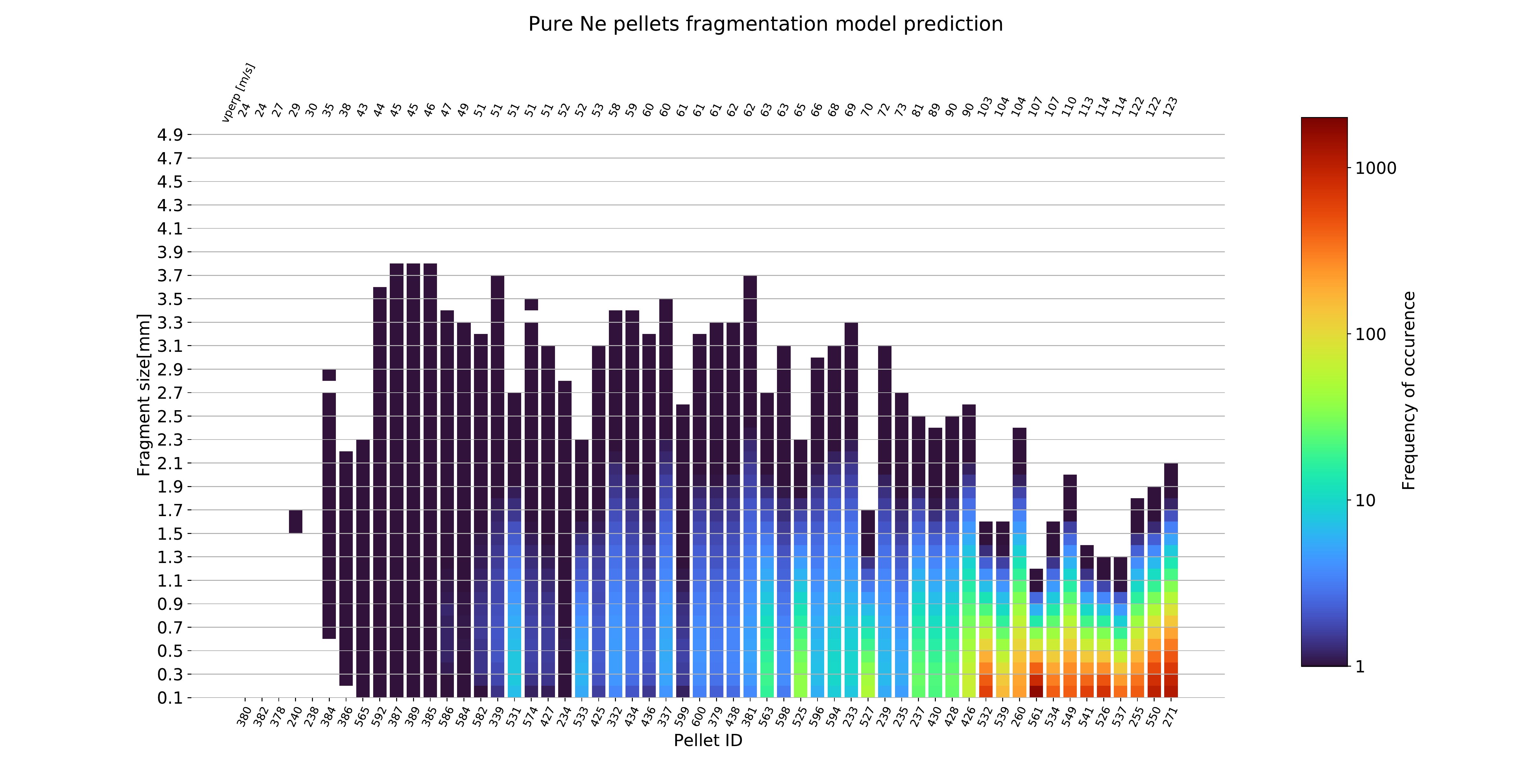}
     \caption{Modeled size distribution for pure Ne pellets, sorted by normal velocity (top). On the bottom, the corresponding pellet ID is given.}
     \label{f:ne_model_vperp}
\end{figure}

\clearpage

\section{Statistical fragment size parameters}
\label{s:Statistical fragment size parameters}

To further examine the fragment size distributions presented in the previous chapter, it is instructive to look at statistical parameters like mean fragment size, its standard deviation, the number of observed fragments and the 20\% mass quantile.
The location of the 20\% mass quantile indicates that 80\% of the total fragment mass are contained in fragments larger than this value.
It is a useful indicator for the distribution of mass among fragments.
Fragment volume $V$ and mass were calculated from the fragment diameter $d$ by assuming a spherical volume:

\begin{equation}
V = \frac{4r^3\pi}{3} = \frac{d^3\pi}{6} \, .
\end{equation}

\begin{figure}[!h]
     \centering
     \includegraphics[clip, trim = 110 30 90 10,width=0.98\textwidth]{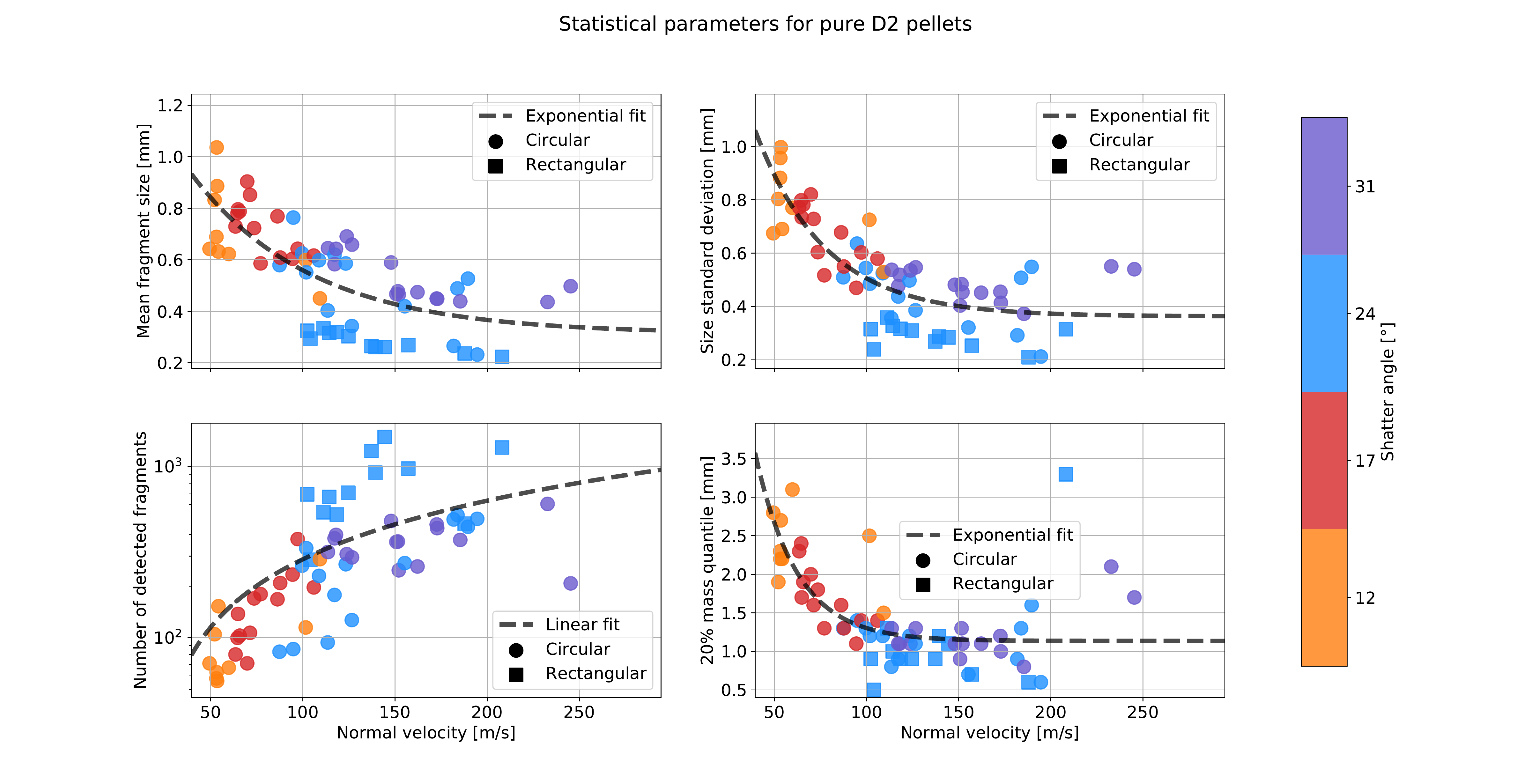}
     \caption{Statistical parameters of experimental D$_2$ pellet size distributions, fitted with linear/exponential functions.}
     \label{f:de_exp_fit}
\end{figure}

\begin{figure}[!h]
     \centering
     \includegraphics[clip, trim =  110 30 90 10,width=0.98\textwidth]{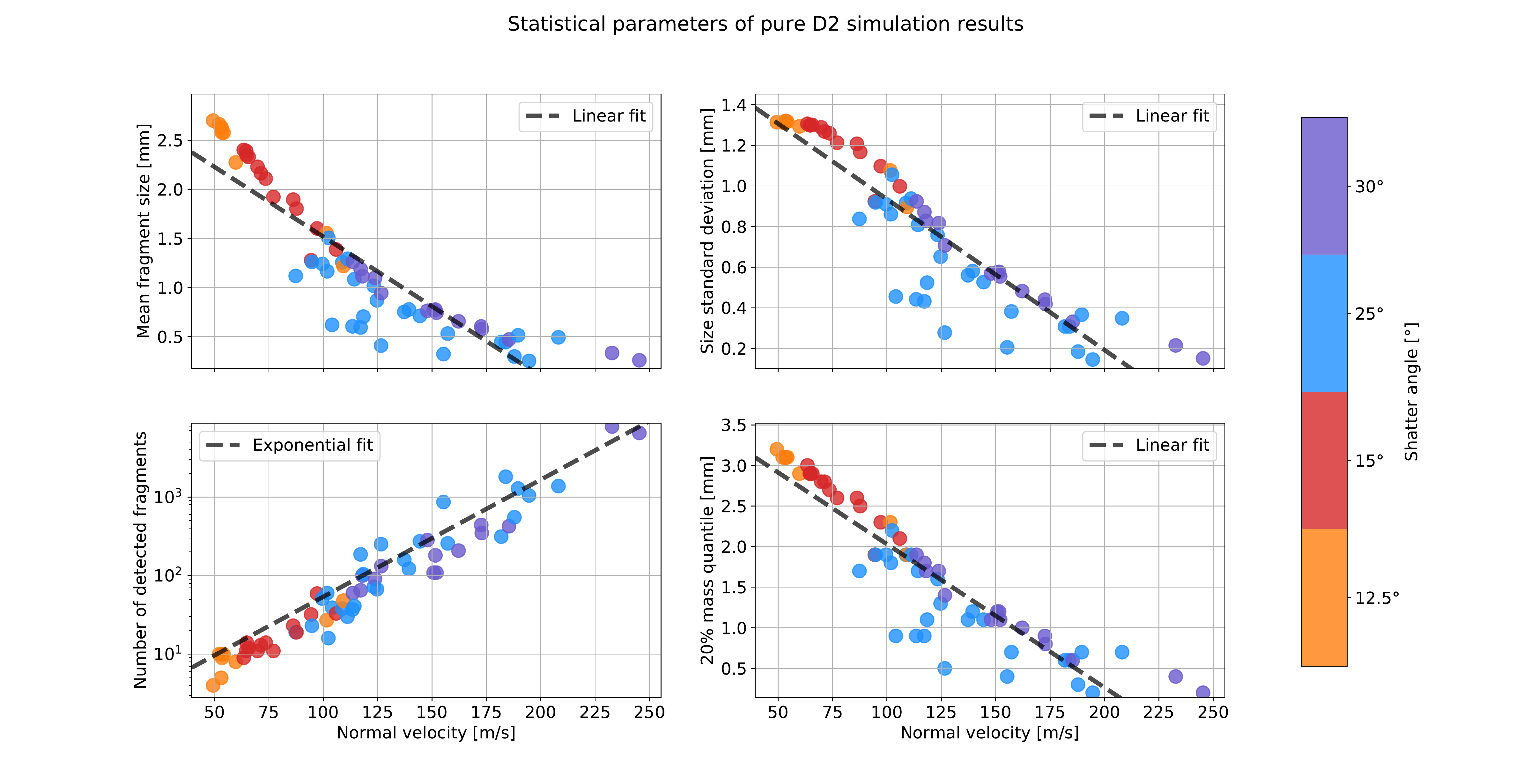}
     \caption{Statistical parameters of simulated D$_2$ pellet size distributions, fitted with linear/exponential functions.}
     \label{f:de_model_fit}
\end{figure}

\begin{figure}[!h]
     \centering
     \includegraphics[clip, trim = 110 30 90 10,width=0.98\textwidth]{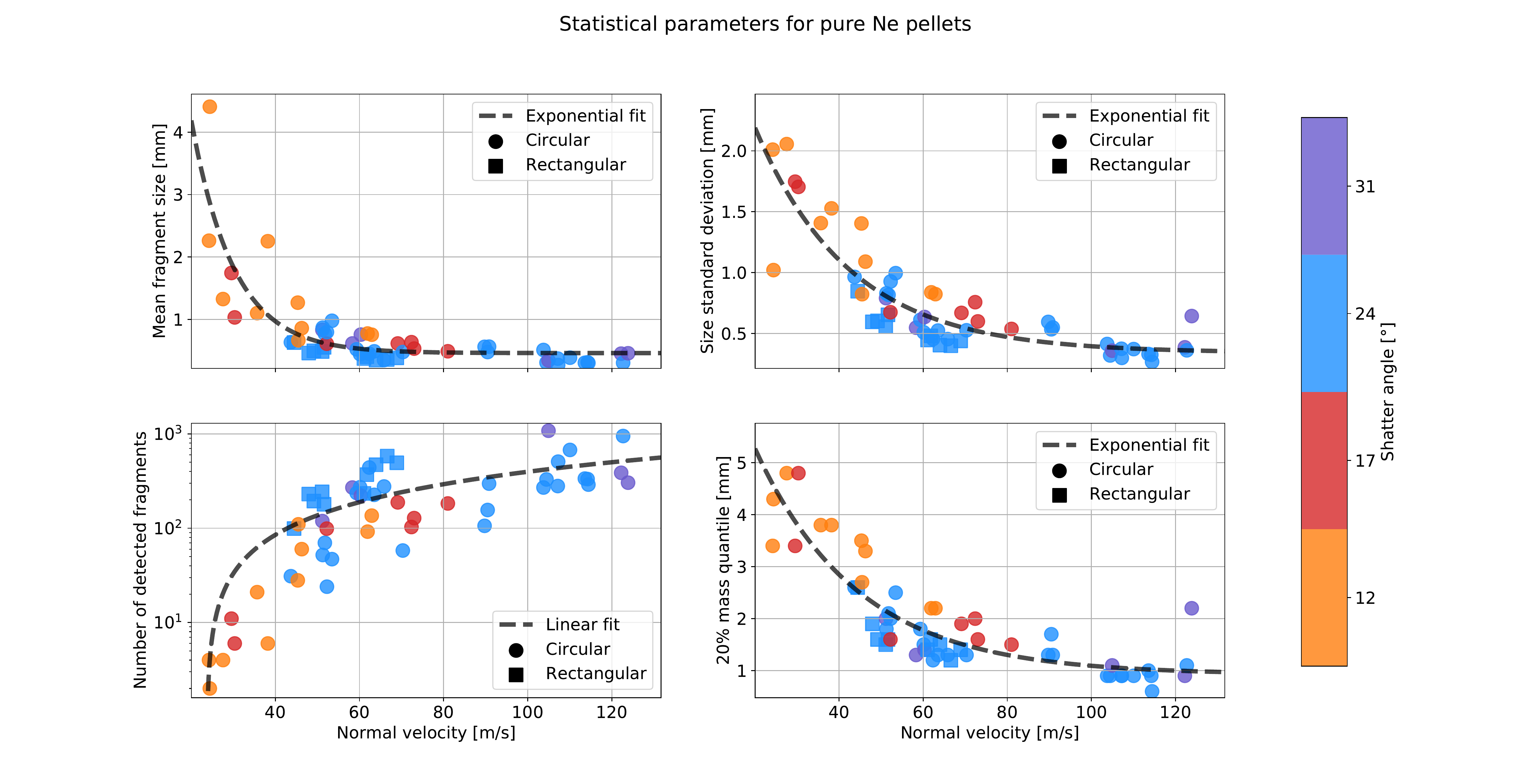}
     \caption{Statistical parameters of experimental Ne pellet size distributions, fitted with linear/exponential functions.}
     \label{f:ne_exp_fit}
\end{figure}

\begin{figure}[!h]
     \centering
     \includegraphics[clip, trim =  110 30 90 10,width=0.98\textwidth]{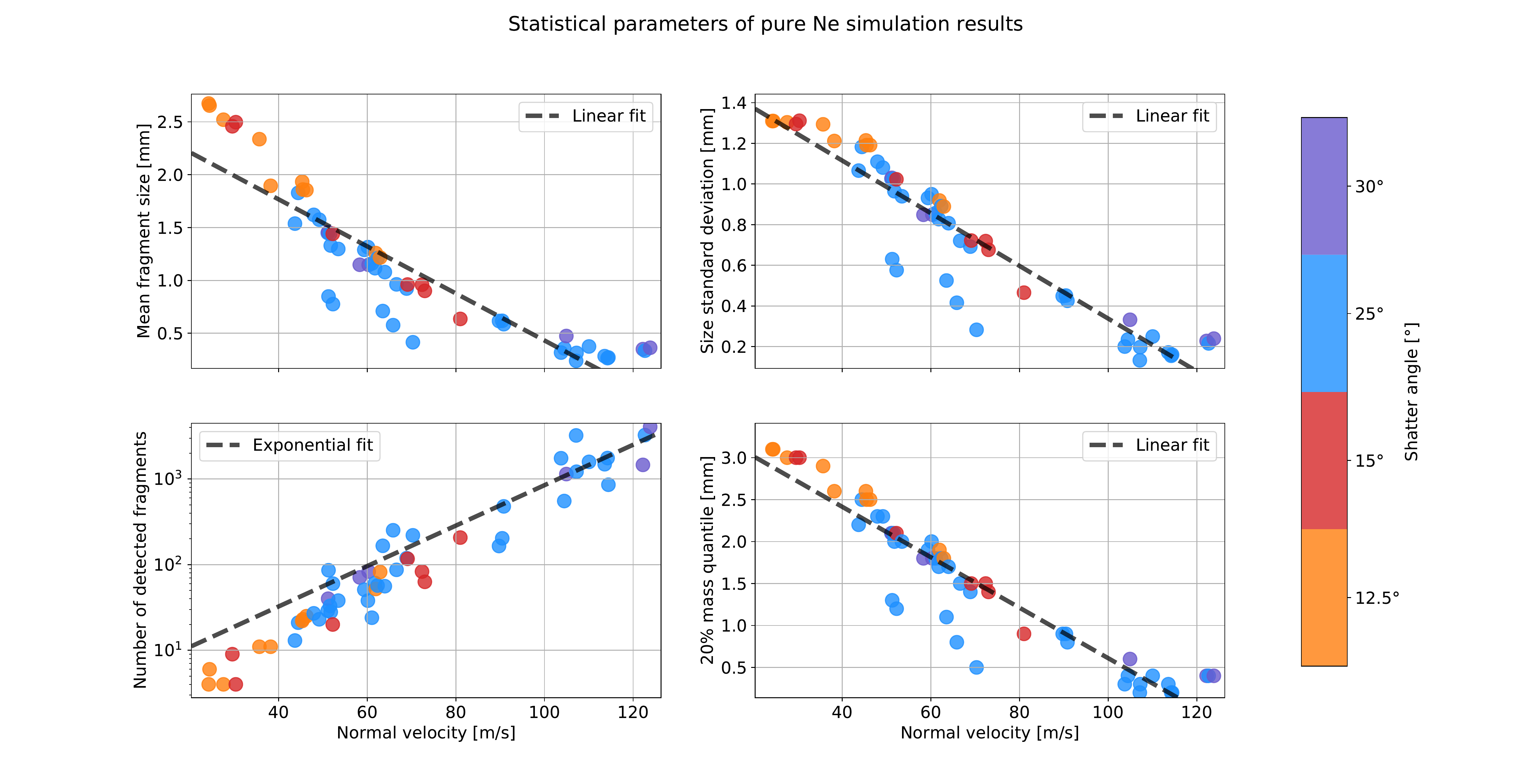}
     \caption{Statistical parameters of simulated Ne pellet size distributions, fitted with linear/exponential functions.}
     \label{f:ne_model_fit}
\end{figure}

\noindent
Figures~\ref{f:de_exp_fit}~-~\ref{f:ne_model_fit} show these statistical parameters, plotted against the normal velocity for both experiment and simulation of each pellet composition. 
In plots for experimental distributions, circles indicate circular shatter heads, rectangles represent the rectangular one.
Note that pellet length can differ between the respective shots and has not been accounted for in these diagrams.

\noindent
While clear dependences are hard to recognize in the size distributions shown in fig.~\ref{f:de_exp_vperp}-~\ref{f:ne_model_vperp} due to the non-linear horizontal axis, all investigated statistical parameters were found to be dependent on perpendicular impact velocity.
In an effort to visualize trends, the observed parameters were fitted with exponential/linear functions, shown in fig.~\ref{f:de_exp_fit}~-~fig.~\ref{f:ne_exp_fit}.
The linear function was defined as

\begin{equation}
f_\text{lin}(x) =  a' x + b' \, .
\end{equation}

\noindent
The exponential fit function used is defined by

\begin{equation}
f_\text{exp}(x) = a e^{-b x} + c \, 
\end{equation}

\noindent
with the fit parameters $a'$, $b'$ and $a$, $b$, $c$.
Values for these parameters resulting from the fit are given in tab.~\ref{t:fit_params_ne} and tab.~\ref{t:fit_params_de} for pure Ne and pure D$_2$.

\begin{table}[!h]
     \centering
     \captionsetup{width=0.82\textwidth}
     \caption{Fit parameters $a'$ and $b'$ / $a$, $b$ and $c$ for linear/exponential fits in fig.~\ref{f:ne_exp_fit} and \ref{f:ne_model_fit} of statistics for pure Ne pellets.} 
     \begin{tabular}{|c|c|c|c|c|}
          \hline
          & \textbf{Function} & $a$ / $a'$ & $b$ / $b'$ & $c$ \\
          \hline
          Experiment mean & $f_{exp}$ & 27.43 & 0.1 & 0.46 \\
          \hline
          Experiment std & $f_{exp}$ & 4.49 & 0.044 & 0.34 \\
          \hline
          Experiment num & $f_{lin}$ & 5.2 & 123 & - \\
          \hline
          Experiment m$_{20}$ & $f_{exp}$ & 9.85 & 0.041 & 0.9261 \\
          \hline
          Simulation mean  & $f_{lin}$ & -0.022 & 2.66 & - \\
          \hline
          Simulation std  & $f_{lin}$ & -0.013 & 1.63 & - \\
          \hline
          Simulation num  & $f_{exp}$ & 3.685 & 0.054 & 0 \\
          \hline
          Simulation m$_{20}$  & $f_{lin}$ & -0.03 & 3.617 & - \\
          \hline
     \end{tabular} 
     \label{t:fit_params_ne}
\end{table}

\begin{table}[!h]
     \centering
     \captionsetup{width=0.82\textwidth}
     \caption{Fit parameters $a'$ and $b'$ / $a$, $b$ and $c$ for linear/exponential fits in fig.~\ref{f:de_exp_fit} and \ref{f:de_model_fit} of statistics for pure D$_2$ pellets.} 
     \begin{tabular}{|c|c|c|c|c|}
          \hline
          & \textbf{Function} & $a$ / $a'$ & $b$ / $b'$ & $c$ \\
          \hline
          Experiment mean & $f_{exp}$ & 1.13 & 0.015 & 0.31 \\
          \hline
          Experiment std & $f_{exp}$ & 1.97 & 0.026 & 0.36 \\
          \hline
          Experiment num & $f_{lin}$ & 3.44 & 56.8 & - \\
          \hline
          Experiment m$_{20}$ & $f_{exp}$ & 14.08 & 0.044 & 1.14 \\
          \hline
          Simulation mean  & $f_{lin}$ & -0.014 & 2.94 & - \\
          \hline
          Simulation std  & $f_{lin}$ & -0.007 & 1.68 & - \\
          \hline
          Simulation num  & $f_{exp}$ & 1.74 & 0.034 & 0 \\
          \hline
          Simulation m$_{20}$  & $f_{lin}$ & -0.018 & 3.796 & - \\
          \hline
     \end{tabular} 
     \label{t:fit_params_de}
\end{table}
     
\clearpage

\clearpage 

\section{Comparison of experimental and simulated size distributions}
\label{s:Comparison of experimental and simulated size distributions}

The fragmentation model by Parks~\cite{Parks} is used to generate initial conditions for various disruption mitigation simulations \cite{JOREK_Parks}.
It is therefore instructive to compare the predicted fragment sizes to the experimentally observed ones.
A strong difference is reflected in the statistical parameters shown in fig.~\ref{f:de_exp_fit} - fig.~\ref{f:ne_model_fit}.
In both experiment and simulation, the observed parameters clearly follow trends when plotted against the normal velocity. 
Mean fragment size, standard deviation and the 20\% mass quantile seem to follow an exponential-like trend in the experiment, while the dependence is approximately linear for modeled distributions.
Also, the number of fragments seems to be linearly dependent in the experiment and exponentially dependent in the simulations.
Note that the number of fragments is logarithmically scaled.
There is a significant amount of outliers in the simulated statistics (see fig.~\ref{f:ne_model_fit}).
These are caused by the strong dependence of the model on the initial pellet length.

\begin{figure}[!h]
     \centering
     \includegraphics[clip, trim = 30 0 40 20,width=\textwidth]{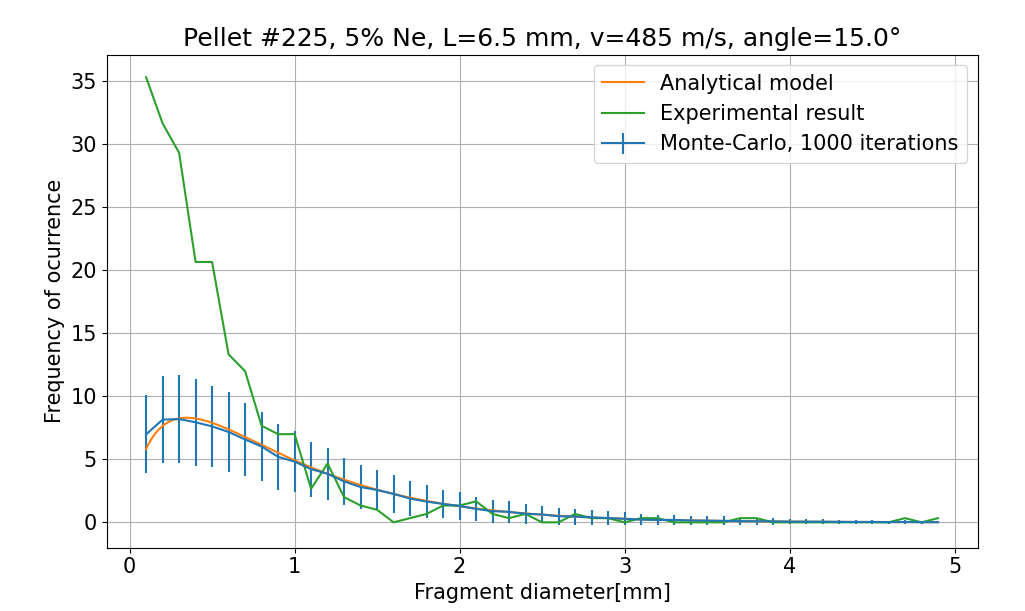}
     \caption{Experimental, analytical and simulated fragment size distribution for pellet \#255, normalized to the same observed fragment volume.}
     \label{f:theory_comp_ex}
\end{figure}

\noindent
This difference is also visible when directly comparing fragment size distributions.
An example comparison is given in fig.~\ref{f:theory_comp_ex}, showing the analytical fragmentation result, the simulation result (standard deviation as error bars) and the experimental result for pellet \#225.
The curves were normalized to contain the same cumulative fragment volume as observed in the experiment.
Especially at fragment sizes below 0.9 mm, the fragmentation model mentioned in ch.~\ref{s:theoretical_model} does not describe the experiment well.
A peak at fragment sizes of approx. 0.3 mm, as visible in the model prediction in fig.~\ref{f:theory_comp_ex}, is also not observed.\\

\begin{figure}[h!]
     \centering
     \includegraphics[clip, trim = 120 30 50 10,width=\textwidth]{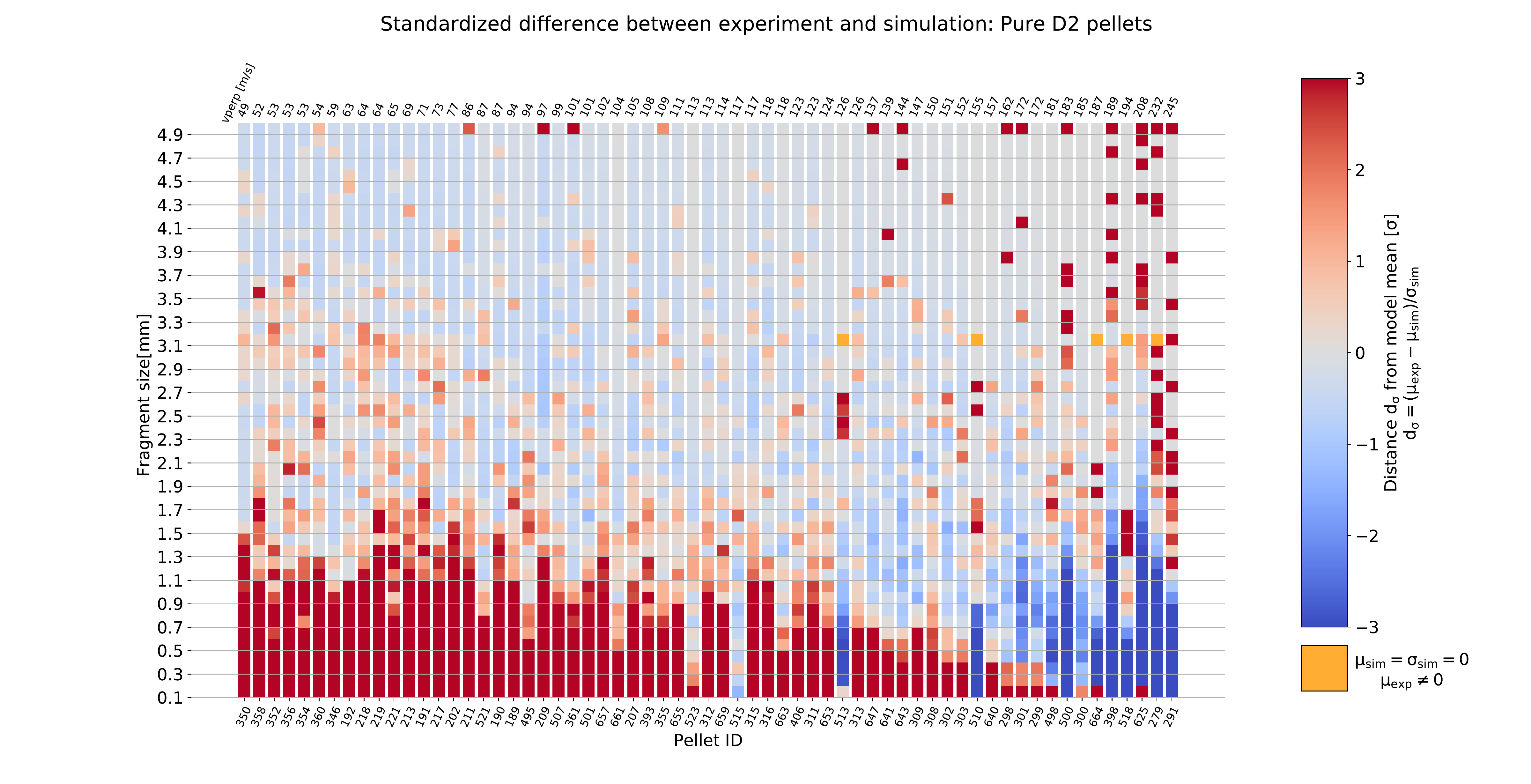}
     \caption{Normalized distance of experimental and simulated size distributions for pure D$_2$ pellets. Red indicates that the model predicts significantly less fragments for a bin, blue indicates that the model predicts significantly more fragments than observed.}
     \label{f:de_comp}
\end{figure}

\begin{figure}[h!]
     \centering
     \includegraphics[clip, trim = 120 30 100 10,width=\textwidth]{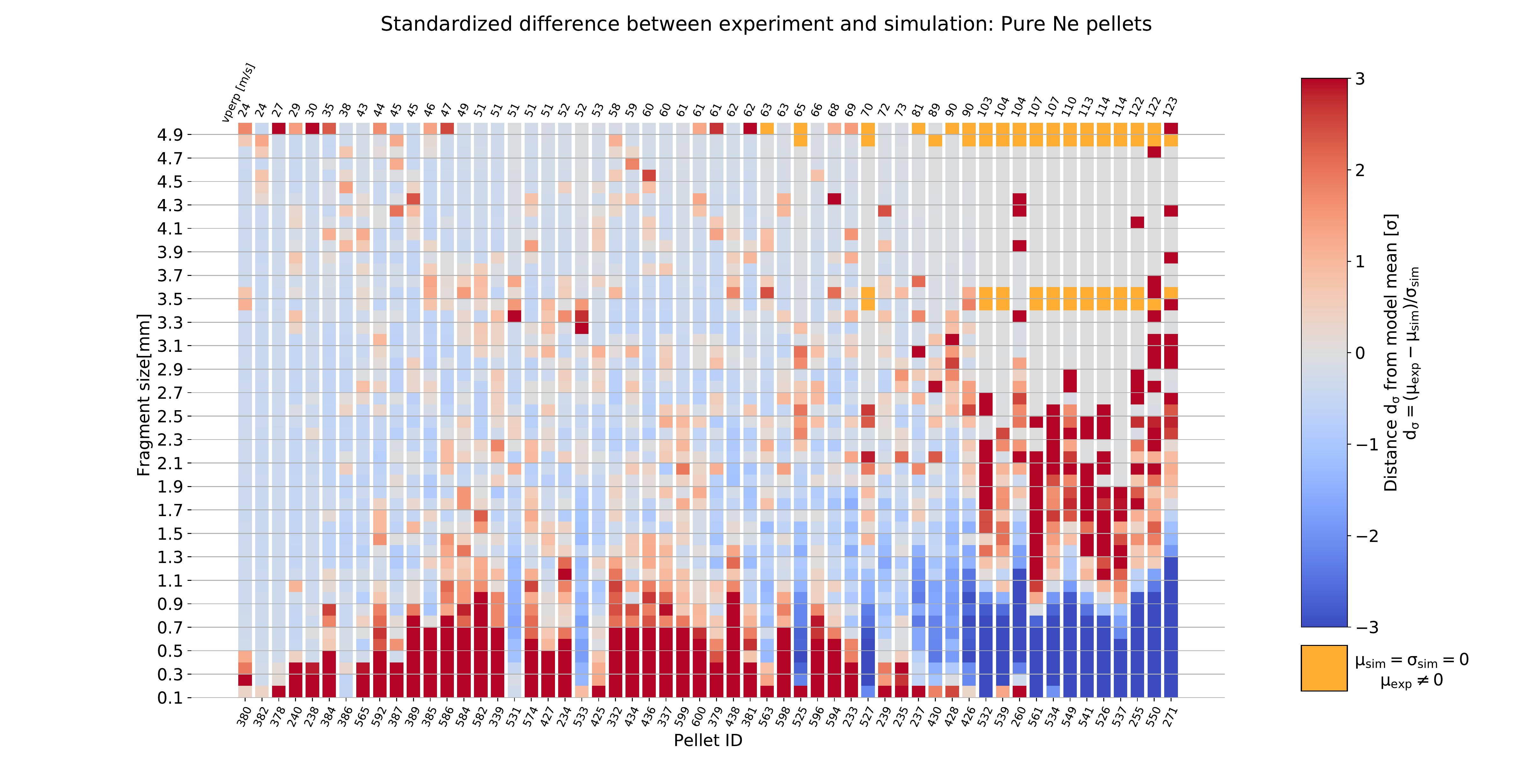}
\caption{Normalized distance of experimental and simulated size distributions for pure Ne pellets. Red indicates that the model predicts significantly less fragments for a bin, blue indicates that the model predicts significantly more fragments than observed.}
     \label{f:ne_comp}
\end{figure}

\noindent
To further compare model and experiment, the normalized distance $d_{\sigma}$ between model and experimental mean fragment count ($\mu_\text{sim}$ and $\mu_\text{exp}$) was computed for each bin and pellet. This distance $d_{\sigma}$ is given by 

\begin{equation}
\label{e:norm_diff}
d_{\sigma} = \frac{\mu_\text{exp} - \mu_\text{sim}}{\sigma_\text{sim}} \, ,
\end{equation}

\noindent
where $\sigma_\text{sim}$ is the standard deviation of the simulated data.
It was obtained from the 1000 randomly sampled fragment distributions generated for each pellet (see ch.~\ref{s:size_dist_res}).
Distance maps, encoding $d_{\sigma}$ as color for each bin and pellet are shown in fig.~\ref{f:ne_comp}, fig.~\ref{f:de_comp} and fig.~\ref{f:mix5_comp}.
Values outside the depicted distance range of (-$3\sigma, 3\sigma$) are set to the nearest boundary.
For some bins, $\sigma_\text{sim}$ is zero, causing $d_{\sigma}$ to go to $\infty$.
These bins were colored yellow.

\FloatBarrier

\noindent
In pure D$_2$ pellets, below 100 m/s normal velocity, the simulation underestimates the amount of fragments smaller than 1.1 mm by more than 3 standard deviations for most pellets.
There are some pellets, where the opposite happens, e.g. pellet \#513 and \#510 (pure D$_2$).
These pellets have a length of 2.7 mm and 2.4 mm, which is rather short compared to other pellets with lengths of 6-7 mm.
Since the theoretical fragmentation model depends heavily on the initial pellet dimensions, a much higher amount of small fragments is predicted for these cases.

\newpage

\section{Spatial spray distribution analysis}
\label{s:spatial}

\begin{figure}[!h]
     \centering
     \includegraphics[clip, trim = 0 135 0 65,width=0.9\textwidth]{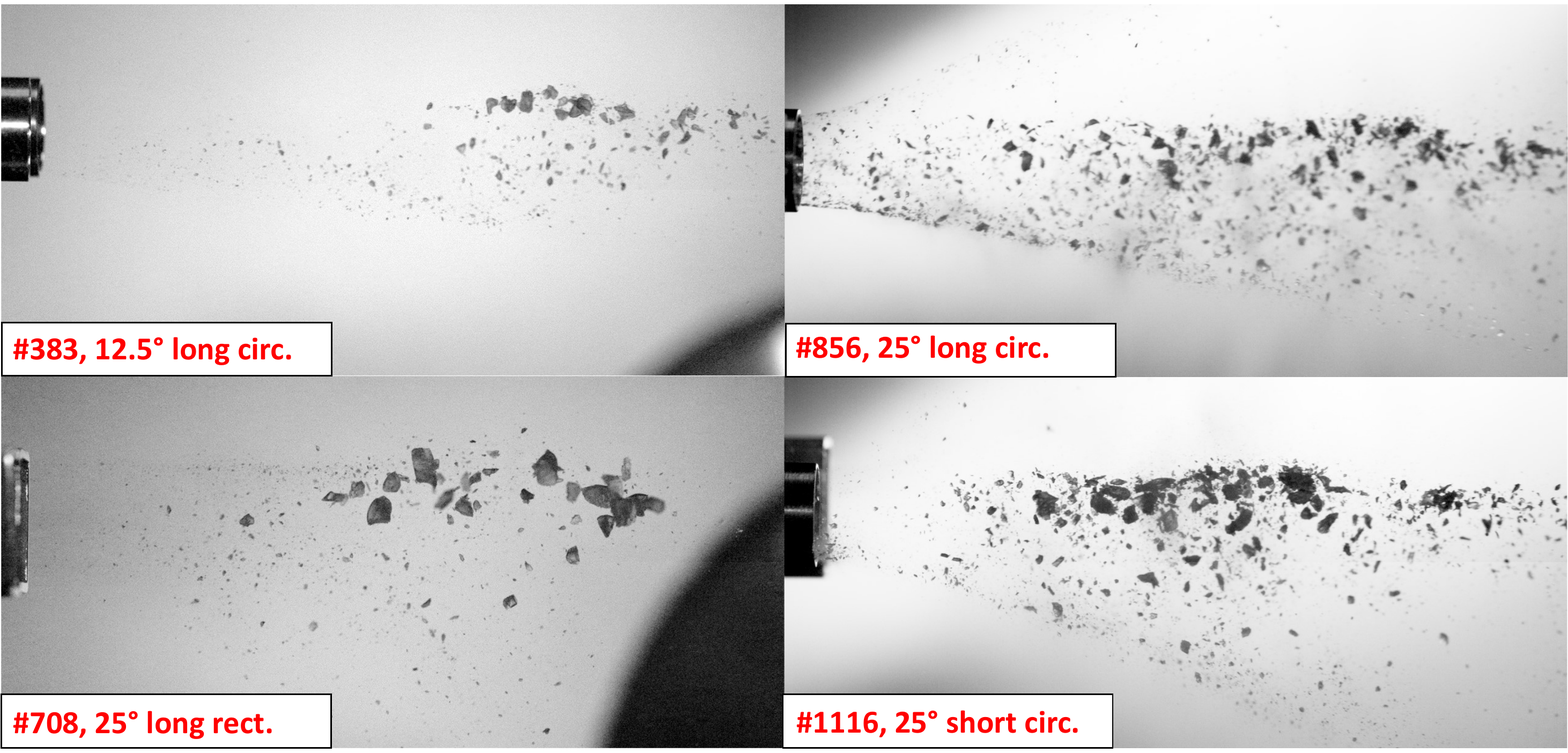}
     \caption{Example fragment sprays with different shatter heads.}
     \label{f:spray_examples}
\end{figure}

\begin{figure}[!h]
     \centering
     \includegraphics[clip, trim = 0 0 0 0,width=0.8\textwidth]{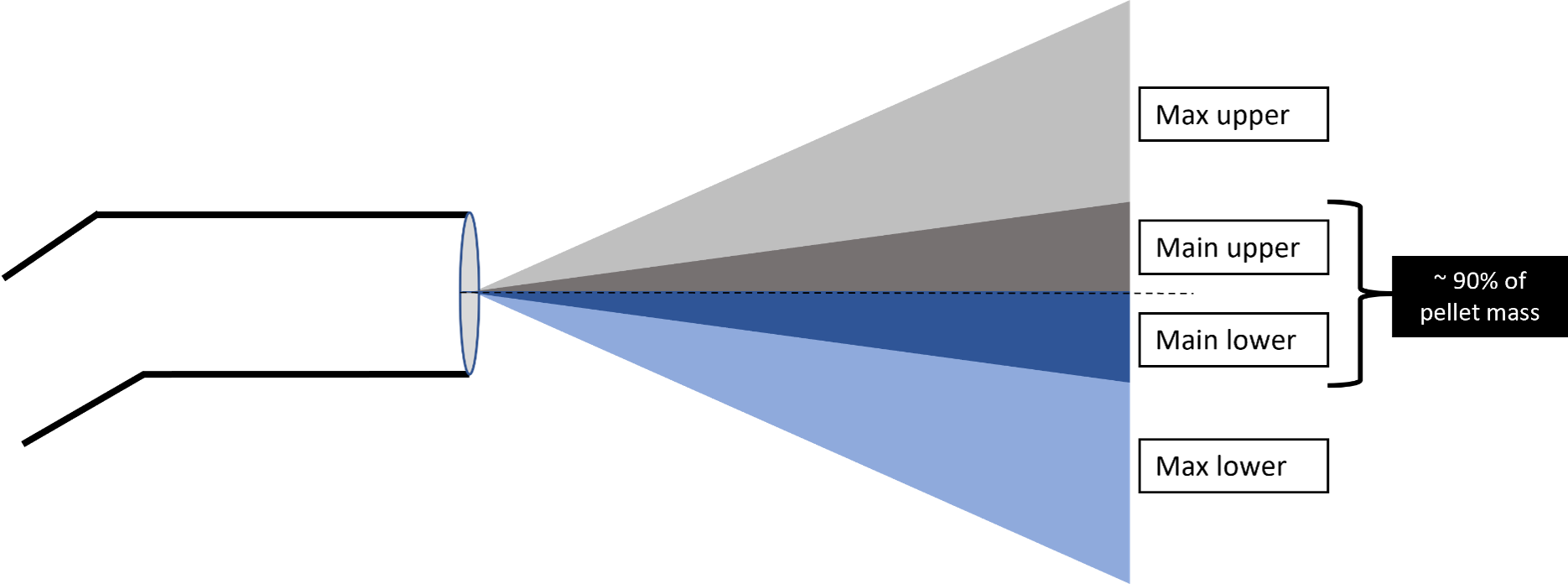}
     \caption{Opening angles for the spatial spray characterization.}
     \label{f:plume_angle}
\end{figure}

\noindent
During the course of the SPI laboratory testing, multiple shatter geometries, angles and post shatter lengths were used. 
We observed that the spatial distribution of the fragment spray differed strongly based on the choice of these parameters.
Some examples for this are presented in fig.~\ref{f:spray_examples}.
Therefore, I characterized the fragment plume for multiple settings by defining four plume regions, shown in fig.~\ref{f:plume_angle}.

\FloatBarrier
\noindent
The maximum upper, main upper, main lower and maximum lower plume regions were measured by the respective opening angle with respect to the center axis of the tube.
The main regions were defined to contain aproximately 90\% of pellet mass.
The maximum upper and lower region both contain outliers.

\begin{table}[!h]
     \centering
     \captionsetup{width=\textwidth}
     \caption{Spatial spray analysis results for different combinations shatter angle, tube geometry, and post shatter length, measured with different pellet diameters.
     Post shatter lengths of 78 mm and 46 mm are defined as long and short.
     Max u. and main u. denote the upper regions indicated in fig.~\ref{f:plume_angle}.
     For some tube designs, a perpendicular view (camera axis on bend plane) is also given.} 
     \begin{tabular}{|c|c|c|c|c|c|}
     \hline
     \textbf{Pellet diam.}&\textbf{Tube design} & \textbf{Max u.} & \textbf{Main u.} & \textbf{Main l.} & \textbf{Max l.}\\
     \hline
     4 mm & 12.5\degree~long circ.&3\degree&3\degree& 5\degree& 15\degree\\
     \hline
     4 mm &15\degree~long circ. &5\degree&5\degree&10\degree&18\degree\\
     \hline
     4 mm &25\degree~long circ. &28\degree&4\degree&17\degree&17\degree\\
     \hline
     4 mm &30\degree~long circ. &28\degree&4\degree&17\degree&17\degree\\
     \hline
     4 mm &25\degree~long rect. &3\degree&3\degree&6\degree&28\\
     \hline
     8 mm &12.5\degree~long rect. &6\degree&6\degree&6\degree&17\degree\\
     \hline
     8 mm &12.5\degree~long circ. &6\degree&6\degree&10\degree&18\degree\\
     \hline
     8 mm &25\degree~long rect. &5\degree&5\degree&6\degree&30\degree\\
      &perp. view&13\degree&13\degree&13\degree&13\degree\\
     \hline
     8 mm &25\degree~long circ. &26\degree&3\degree&13\degree&13\degree\\
      &perp. view&25\degree&7\degree&7\degree&25\degree\\
     \hline
     8 mm &30\degree~long circ. &21\degree&2\degree&6\degree&15\degree\\
     \hline
     8 mm &25\degree~short circ. &56\degree&4\degree&8\degree&32\degree\\
     \hline
     8 mm &25\degree~short rect. &6\degree&6\degree&5\degree&37\degree\\
     \hline

     \end{tabular} 
     \label{t:plume_angle}
\end{table}

\noindent
The results of the spatial spray analysis are summarized in tab.~\ref{t:plume_angle}, where I list the opening angles of these regions with respect to the center axis.
Regions were selected manually for multiple videos of using one tube design, the corresponding opening angles were calculated and the average values were recorded.
Some pellets were fired with shatter tubes that were rotated by 90\degree, so the camera axis lies on the bend plane.
This is specified as ``perp. view'' in tab.~\ref{t:plume_angle}.
Furthermore, some tubes were shortened from 78 mm (long) to a post shatter length of 46 mm (short) to prevent so-called ``double bouncing'', where a portion of the fragments hits the shatter tube a second time after shattering.\\

\noindent
Since measurements with the same tube and different pellet diameters agree to $\pm$ 5\degree, it seems that this parameter does not directly affect the spatial fragment distribution.
As a general trend, circular tubes were found to produce a slightly broader main plume and much larger regions containing outliers.
This was especially true for the shortened 25\degree circular tube.
Based on this analysis, a long 12.5\degree~ rectangular, a long 25\degree~ rectangular and a short 25\degree circular shatter tube were chosen and installed in the ASDEX Upgrade SPI system. 
This way, both wide and narrow spatial distributions can be achieved and similar perpendicular velocities can be reached with different pellet speeds by using the 12.5\degree~and long 25\degree~tubes.
The question wether a wider ore more narrow spray is better for disruption mitigation is one of the things that will be investigated during tokamak experiments with the ASDEX-Upgrade SPI.\\

\newpage

\FloatBarrier
\section{Comparison of rectangular and circular shatter tubes}
\label{s:Comparison of rectangular and circular shatter tubes}

\begin{figure}[!h]
     \centering
     \includegraphics[clip, trim = 0 0 0 0,width=0.9\textwidth]{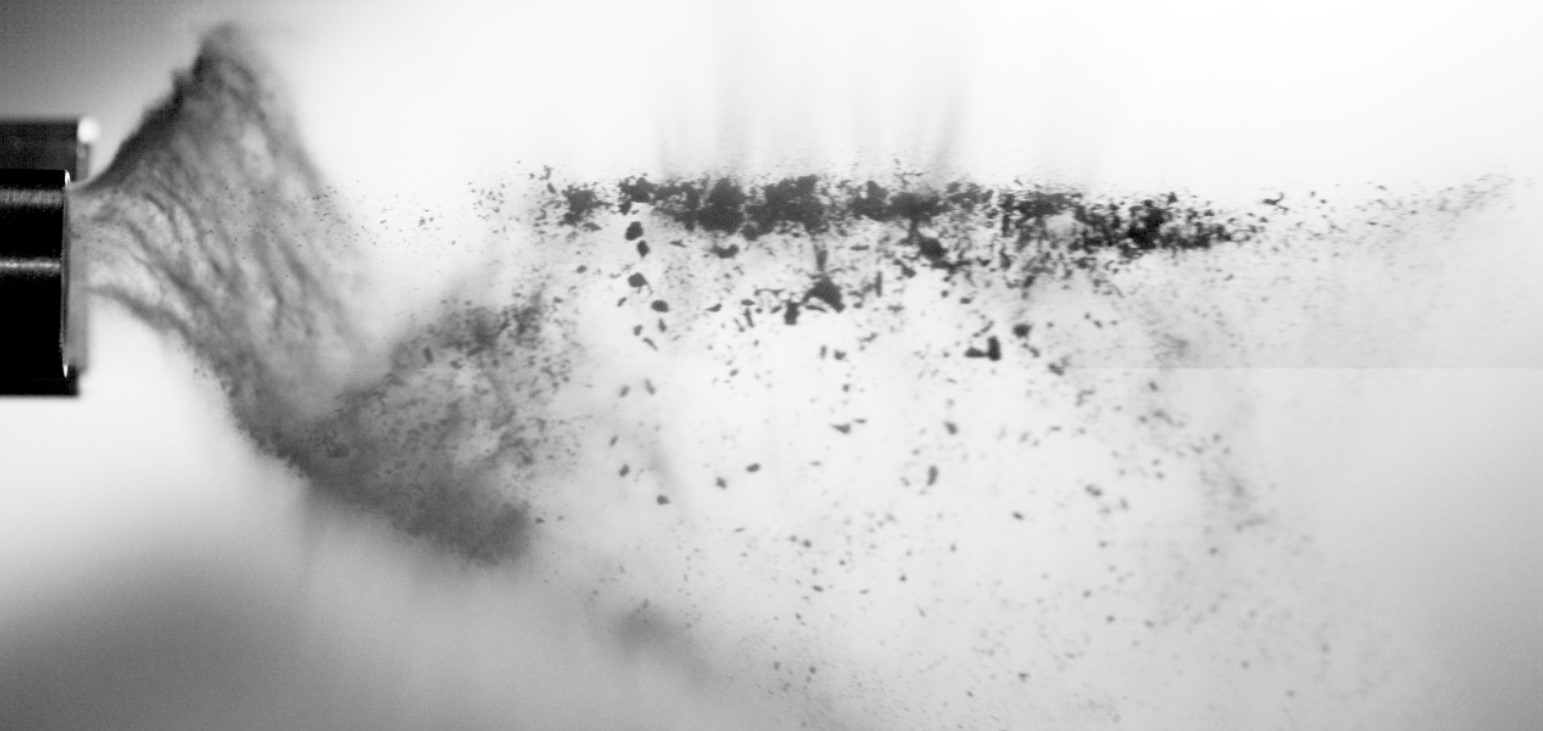}
     \caption{Frame showing a pronounced spiral dust patter produced by pellet \#1098 (8 mm diameter) in a circular 25\degree~shatter tube.}
     \label{f:spiral_dust}
\end{figure}

\noindent
Motivated by unusual observations during the spatial fragment distribution analysis, I decided to also study the impact of shatter head geometry on the fragment size distribution.
In some cases, circular tubes seemed to produce a helical dust cloud following the bulk of the fragment plume, as shown in fig.~\ref{f:spiral_dust}.
Furthermore, in these cases, fragments seemed larger when compared to similar shots using rectangular tubes.
In this section, a comparison of fragment sizes produced by rectangular and circular geometries is shown and some theoretical investigations of the observed differences are discussed.
\\

\begin{figure}[!h]
     \centering
     \includegraphics[clip, trim = 120 30 270 10,width=0.9\textwidth]{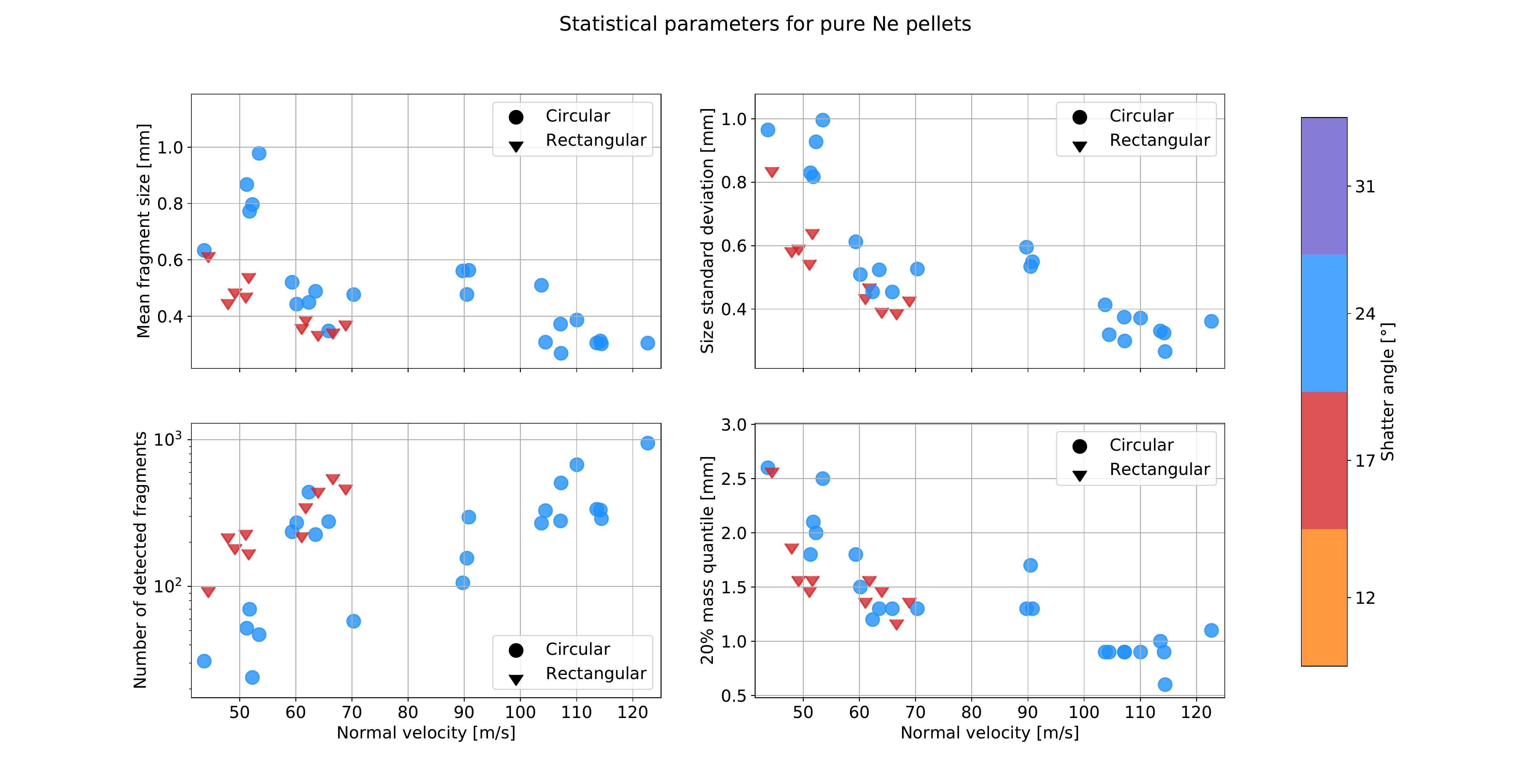}
     \caption{Statistical fragment size parameters for pure Ne pellets. Blue circles indicate circular tube cross-section, red rectangles indicate rectangular.}
     \label{f:ne_circ_rect}
\end{figure}

\noindent
Figure~\ref{f:ne_circ_rect} shows the previously presented statistical parameters for pure Ne pellets with rectangular and circular shatter head geometry.
To rule out any influence of shatter angle, only results for 25\degree~circular and rectangular shatter tubes (blue circles and red rectangles) are shown.

\noindent
Rectangular shatter heads generally seem to produce smaller fragments when compared to circular shatter tubes at similar speeds and form a ``soft lower limit'' in mean size over all observed pellet compositions.
Furthermore, the mean fragment size seems to have a smaller spread with rectangular heads. 
Similar behavior is found in the standard deviation. 
Circular shatter geometries generally produce fragment distributions with higher standard deviation compared to rectangular ones at similar speeds.
The number of fragments was also found to be sensitive to shatter tube geometry. 
As a general trend, rectangular cross-sections produce more fragments, forming an upper limit for the circular case.
Smaller fragment sizes for constant pellet size result in a higher number of fragments.
The 20\% mass quantile also follows a similar trend as the other parameters. 
Since rectangular designs produce smaller fragments, the quantile for those cases again forms a ``soft lower limit'' for the circular one.\\

\begin{figure}[!h]
     \centering
     \includegraphics[clip, trim = 0 150 70 130,width=0.9\textwidth]{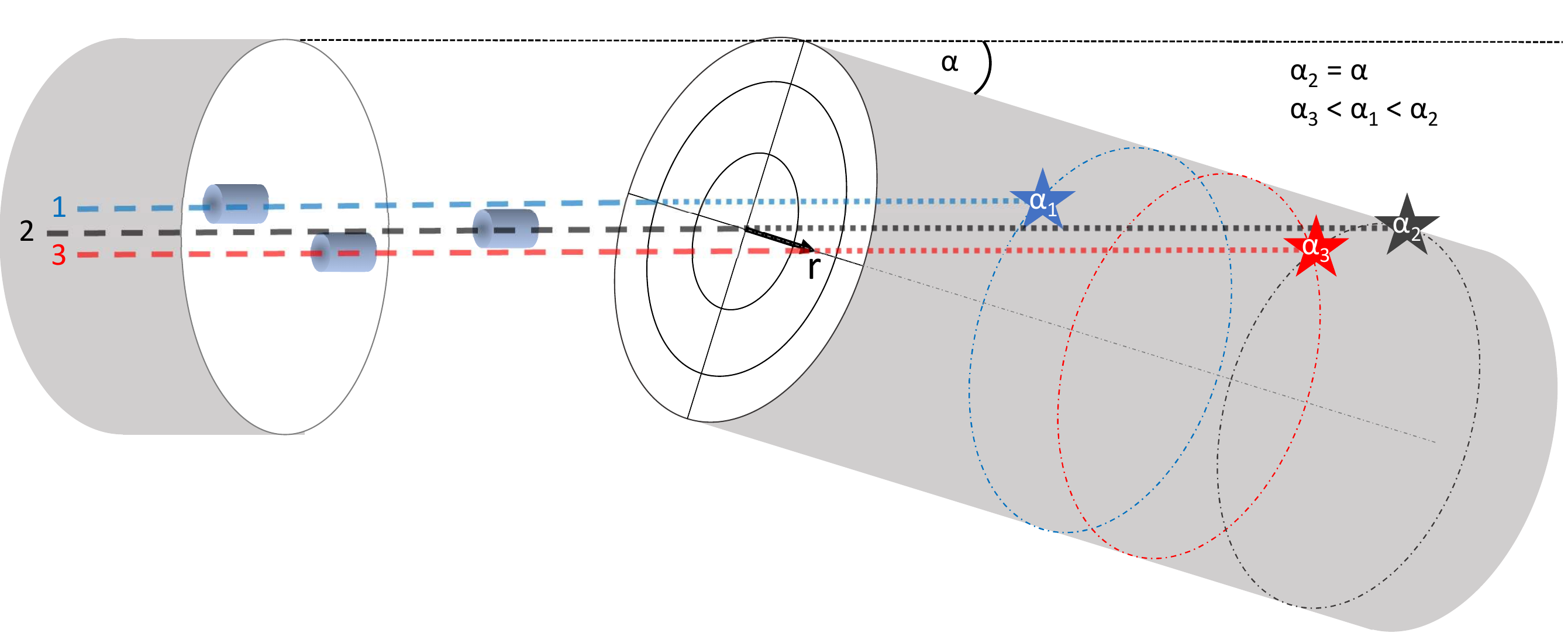}
     \caption{Schematic image of three radially different pellet flight paths in a circular shatter tube with shatter angle $\alpha$. The distance orthogonal to the cylinder axis at the bend is denoted by $r$. Note that only the distance perpendicular to the plane of $\alpha$ affects the effective shatter angle. The higher $r$, the smaller the impact angle.}
     \label{f:circular_impact}
\end{figure}

\noindent
A possible explanation for the different behaviors is that the effective shatter angle in a circular tube is dependent on radial pellet location during impact relative to the bend plane of the angle $\alpha$.
As fig.~\ref{f:circular_impact} illustrates, this can lead to shallow effective shatter angles if the projectile hits off-center.
Therefore, only pellets hitting a circular tube perfectly along the cylindrical axis ($\alpha_2$ in fig.~\ref{f:circular_impact}) experience the full angle of incidence.

\begin{figure}[!h]
     \centering
     \includegraphics[clip, trim = 30 0 50 20, width=0.7\textwidth]{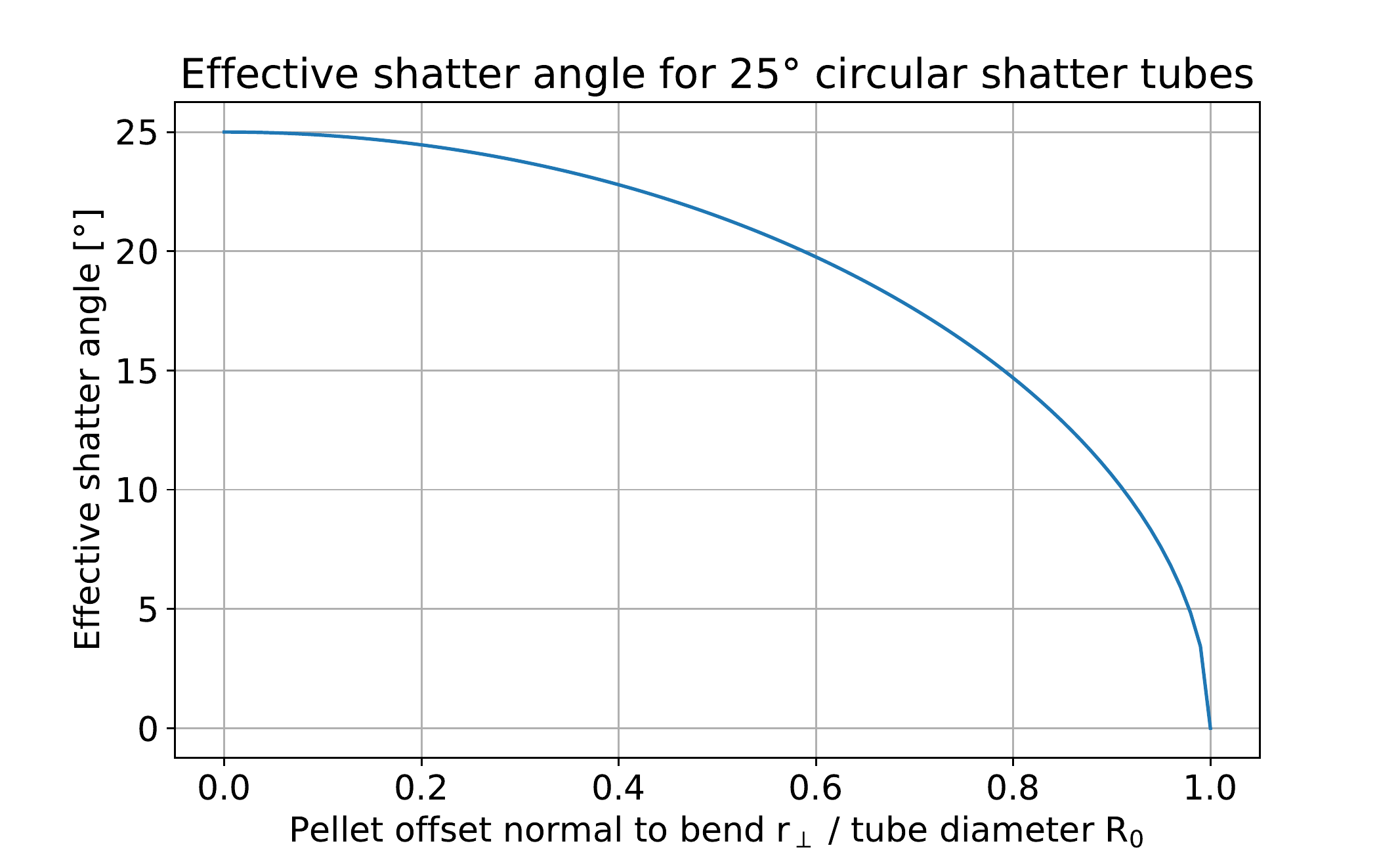}
     \caption{Dependence of the effective shatter angle on the ratio of normal offset $r_\perp$ and $R_0$.}
     \label{f:eff_angle}
\end{figure}

\FloatBarrier
\noindent
For a 25\degree~circular tube, the dependence of the effective shatter on the pellet offset normal to the bend plane $r_\perp$ relative to the tube radius $R_0$ is shown in fig.~\ref{f:eff_angle}.
It can be derived as follows.
Let the tube bend $\alpha$ lie in the $x$-$z$-plane of a Cartesian coordinate system and assume that the part of the shatter tube after the bend has the x-axis as the cylinder axis.
The incoming pellet is described by the offsets $r_\perp$ and $r_\parallel$ from the cylinder axis and a normalized velocity vector $\hat{v}$ with 

\begin{gather}
\hat{v} = 
\begin{pmatrix}   
-\cos(\alpha)\\
0\\
\sin(\alpha)
\end{pmatrix}
\, .
\label{e:pellet_vec}
\end{gather}

\noindent
As the pellet can not be located outside the tube, $-R_0 \leq r_\perp \leq R_0$ and $-R_0 \leq r_\parallel \leq R_0$.
Since the pellet does not move in $y$-direction, the $z$-component of the impact location is given by $\sqrt{R_0^2 - r_\perp^2}$.
An offset in the bend plane $r_\parallel$ changes only the $x$-component of the point of impact.
Therefore, the normal vector $\hat{n}$ to the surface of impact is given by

\begin{gather}
     \hat{n} = \frac{1}{R_0}
     \begin{pmatrix}   
     0\\
     r_\perp\\
     \sqrt{R_0^2 - r_\perp^2}
     \end{pmatrix}
     \, .
     \label{e:norm_vec}
\end{gather}

\noindent
The effective angle of impact $\alpha_{\text{eff}}$ is then given by

\begin{equation}
     \alpha_{\text{eff}} = 90 - \frac{\cos^{-1} \left(\hat{n} \cdot \hat{v} \right)}{\pi} * 180 \, .
\end{equation}

\noindent
In rectangular shatter heads however, the shatter angle is not sensitive to pellet position.
The direction of the helical dust pattern in fig.~\ref{f:spiral_dust} indicates that the pellet hit the tube off-center, similar to path \#3 in fig.~\ref{f:circular_impact}.
This effect might be an explanation for the large spread in mean fragment size that we observe for circular tube shots with similar velocities.
The decreasing effective shatter angle with circular tubes also explains the observation that the rectangular shatter tube results form a ``soft lower limit'' in mean fragment size for circular cases. \par \noindent
Furthermore, a spiral-like dust ejection pattern like shown in fig.~\ref{f:spiral_dust} was observed for some pellets fired at circular tubes.
This could be the outcome of an off-center pellet impact on the tube, which causes the debris to follow a helical path along the shatter tube wall.\\

\FloatBarrier

\noindent
As radial pellet impact location is a parameter that is hard to control, the findings of this thesis suggest that the use of rectangular shatter geometries is advantageous to the circular counterpart.

\section{Fragment volume vs. pellet volume}
\label{s:volume}

\begin{figure}[!h]
     \centering
     \includegraphics[clip, trim = 0 0 0 0,width=0.85\textwidth]{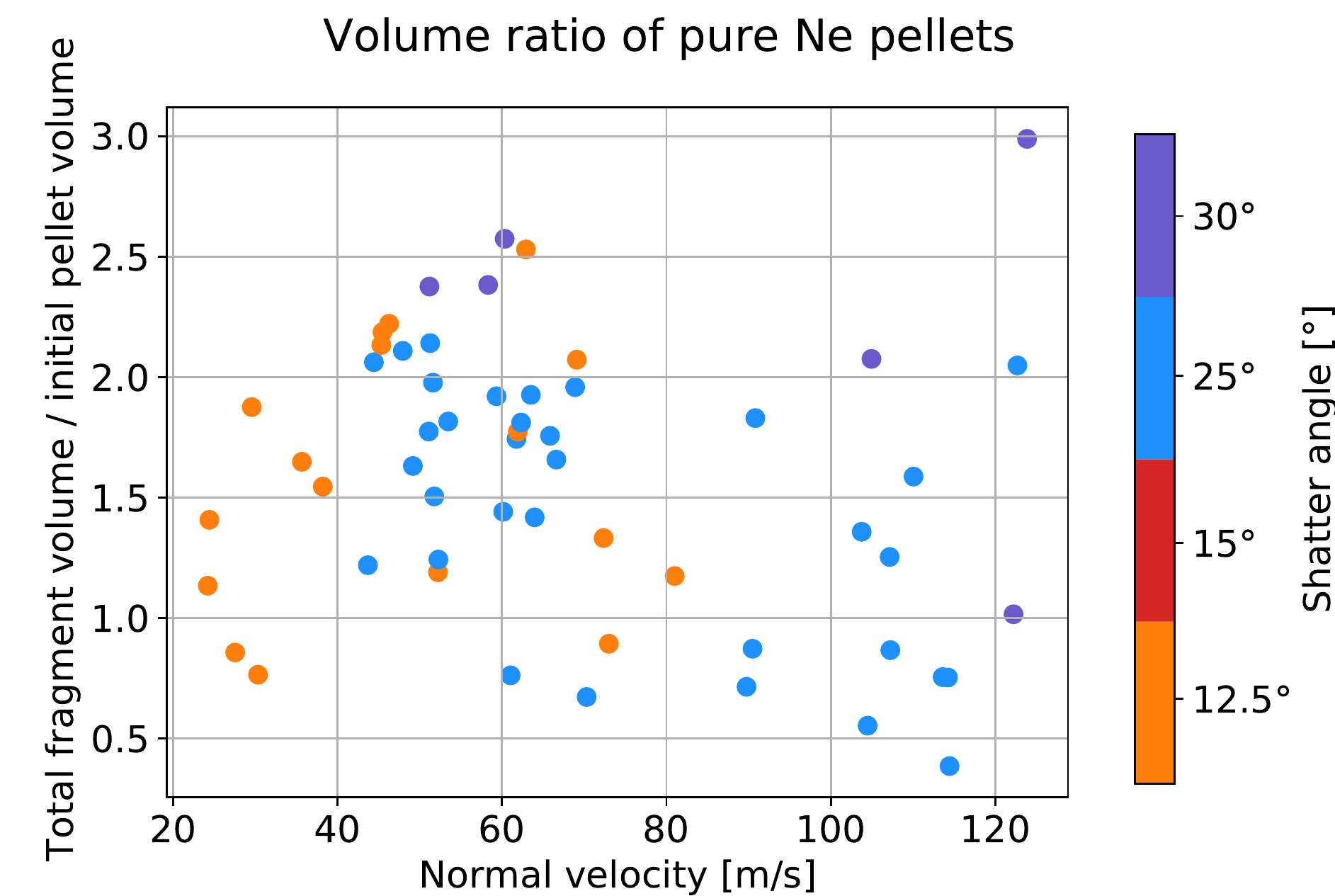}
     \caption{Ratio of cumulative fragment volume and initial pellet volume for pure Ne pellets.}
     \label{f:vol_comp}
\end{figure}

The limitations of one-perspective size measurements of 3-dimensional objects can be showcased by looking at the ratio of total detected fragment volume and initial pellet volume.
Figure~\ref{f:vol_comp} shows a scatter plot of this ratio, plotted over the perpendicular velocity.

\noindent
While it is good that this ratio is of the order of 1, it is still concerning that for most of the pellets, more material was detected after fragmentation than before.
Various possible reasons for these deviations, such as fragment overlap or the need for manual adjustment of preprocessing parameters have been listed in section \ref{s:limitations}.
Concerning the fragment volume, any error/uncertainty $\epsilon$ in the computation of the fragment area will result in an error in the volume that is proportional to $\epsilon^{3/2}$.
For example, an overlap of two fragments with the same visible area, that just barely touch, leads to an increase of the detected volume by 82\% compared to spatially separated shards. 
Since fragment overlap occurs frequently, especially inside the bulk of the fragment plume, this might explain the pronounced deviation from the expected result.

\section{Experimental fragment velocity distributions}
\label{s:speeds}

\begin{figure}[!h]
     \centering
     \begin{subfigure}[b]{0.49\textwidth}
          \centering
          \includegraphics[clip, trim = 100 20 130 10, width=\textwidth]{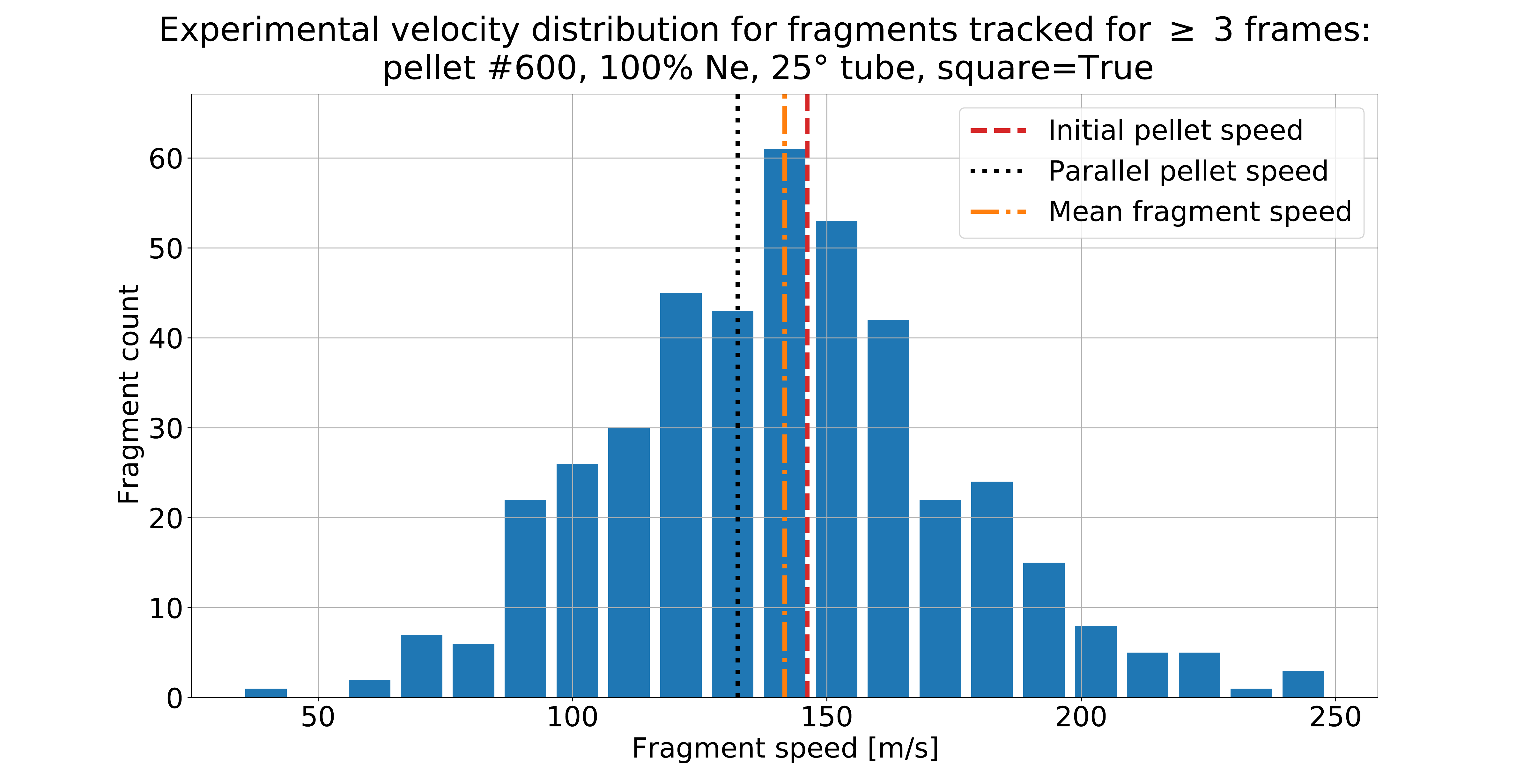}
          \caption{}
          \label{f:speed1}
      \end{subfigure}
      \hfill
     \begin{subfigure}[b]{0.49\textwidth}
         \centering
         \includegraphics[clip, trim = 100 20 130 10, width=\textwidth]{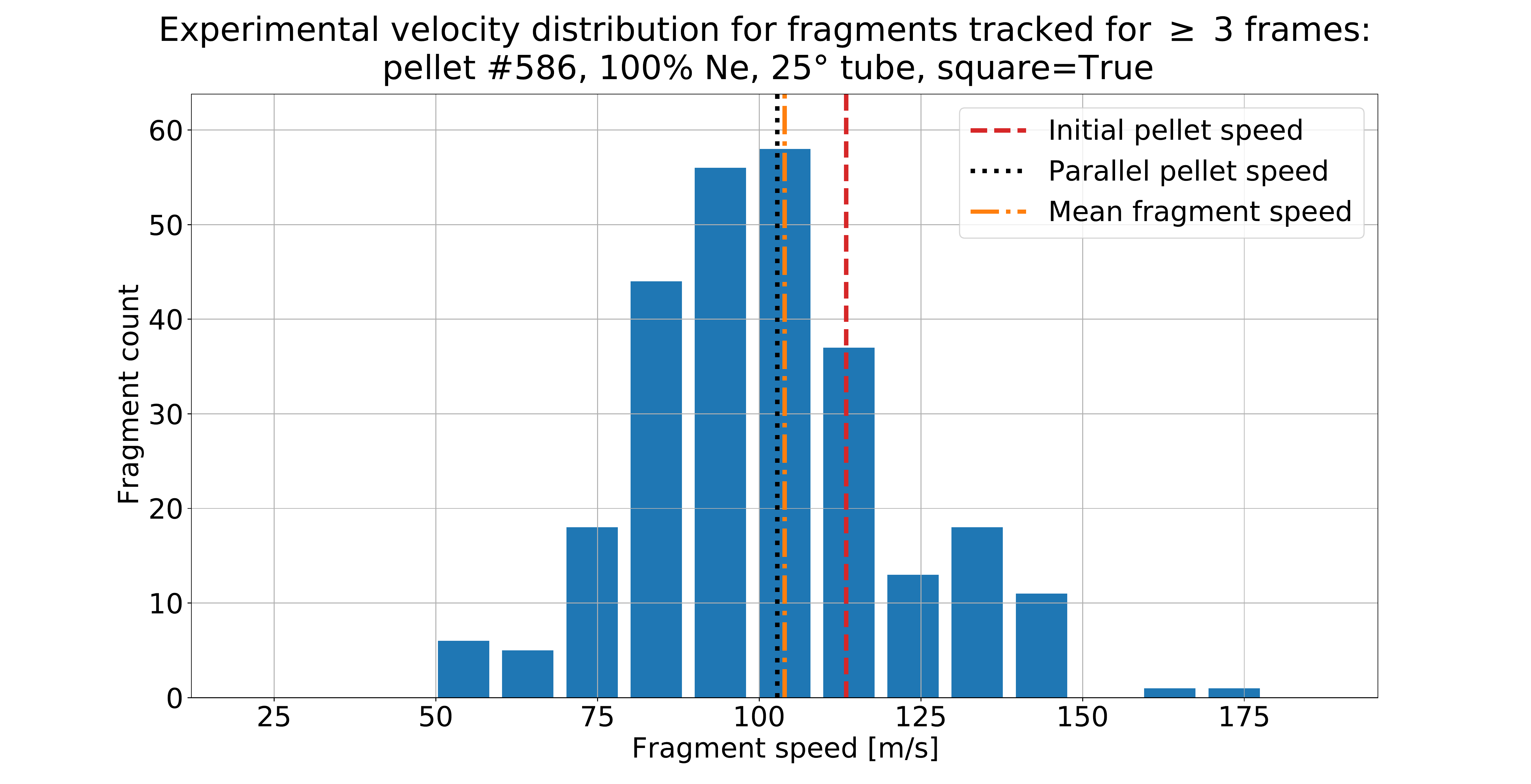}
         \caption{}
         \label{f:speed2}
     \end{subfigure}
     \vskip\baselineskip
     \begin{subfigure}[b]{0.49\textwidth}
         \centering
         \includegraphics[clip, trim = 100 20 130 10, width=\textwidth]{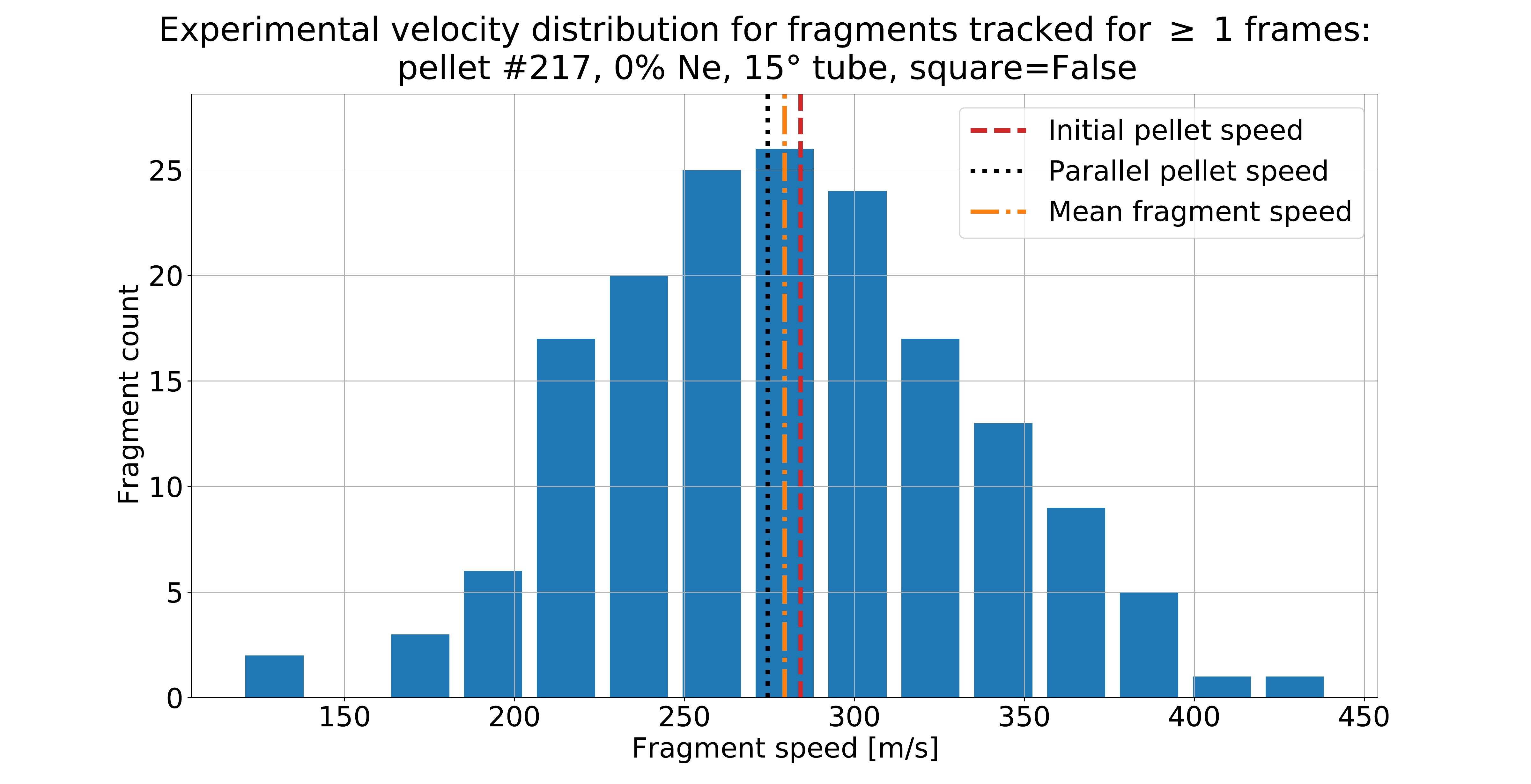}
         \caption{}
         \label{f:speed3}
     \end{subfigure}
     \begin{subfigure}[b]{0.49\textwidth}
          \centering
          \includegraphics[clip, trim = 100 20 130 10, width=\textwidth]{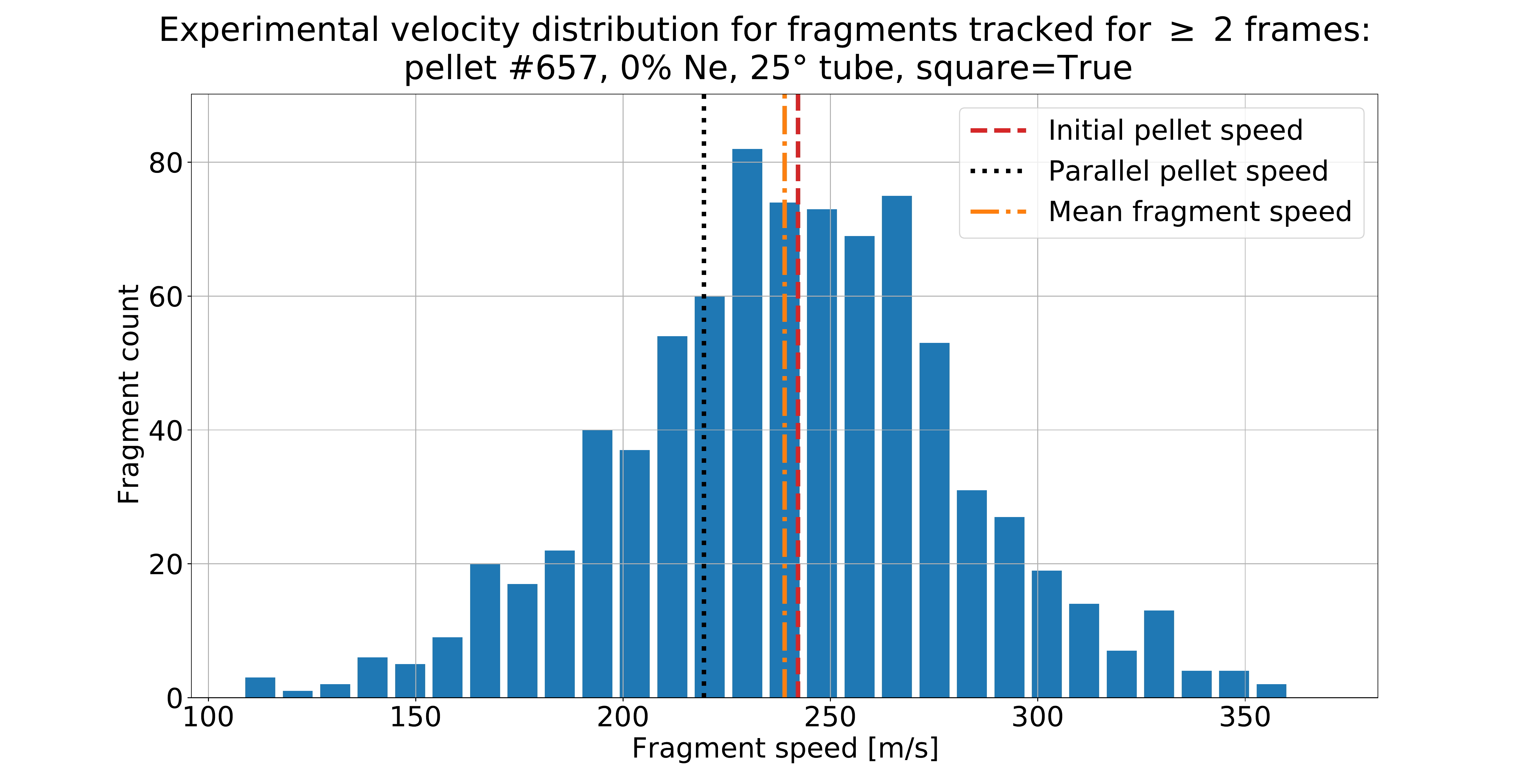}
          \caption{}
          \label{f:speed4}
      \end{subfigure}
     \caption{Example fragment velocity distributions for (a, b) Ne, (c,d) D$_2$ pellets.
     Only fragments that were tracked for more than 1/3 of the plume duration were included.}
        \label{f:speed}
\end{figure}

With the algorithm presented in section~\ref{s:analysis_software}, fragments were tracked across multiple frames and their velocity was computed by multiplying their average displacement with the framerate. 
Although this thesis is not focused on fragment velocity, some example fragment velocity distributions are shown in fig.~\ref{f:speed}.
These histograms include all fragments that were tracked for more than 1/3 of the duration of plume visibility.
This duration was estimated as the time it would take the initial pellet to travel through the FOV.
Since this differs for different shots, the exact number of frames is noted in the title of the respective plots.\\

\noindent
In the figures, the initial pellet speed, the pellet speed parallel to the shatter plane and the mean fragment speed are marked as dashed vertical lines in red, black and orange.
Initial speed was computed from the flight time from the integrity diagnostic to the end of the shatter tubes.
Figures~\ref{f:speed1} and \ref{f:speed2} show fragment velocities produced by pure Ne pellets, fired at 25\degree~rectangular shatter tubes.
The lower images, fig.~\ref{f:speed3} and~\ref{f:speed4}, show results for pure D$_2$ pellets, fired at different shatter tubes.
In each figure, the mean fragment speed (orange) lies between initial and parallel pellet speed, suggesting an inelastic collision of the pellet with the shatter tube.
While these results are not yet a quantitative analysis of fragment velocity over a large parameter space, they motivate further investigation of the topic with the methods presented in this thesis.

%
%
%

\chapter{Conclusions and Outlook}

Over the course of this thesis, I contributed to the optical design and the video diagnostics of the ASDEX-Upgrade SPI test setup. 
I developed, tested and applied software for pellet integrity assessment and fragment tracking and participated in person in the laboratory testing.
Furthermore, I generated fragment size distributions for 170 pellets both from the analysis of recorded videos and from simulations using a theoretical fragmentation model.
I compared theoretical predictions to experimental results and studied the spatial distribution of the fragments for different shatter tube geometries and bending angles.
Furthermore, I discussed the impact of circular and rectangular shatter geometries.
Additionally, I showed example velocity distributions that were generated as a result of the automated tracking of individual fragments.
The pellet tracking code for the integrity diagnostic is being used to measure incident pellet speed and size in SPI experiments at ASDEX-Upgrade.\\

\noindent
Especially for pure Ne pellets, a clear dependence of the mean fragment size, size standard deviation, number of fragments and 20\% mass quantile on the pellet velocity perpendicular to the shatter plane was discovered.
By comparing the pellet fragmentation model proposed by Parks~\cite{Parks} to  data obtained in the experiment, I found that the theoretical model severely underestimates the amount of fragments below 0.9 mm in diameter.
This is a significant result, since fragment size strongly affects the ablation procedure and impurity deposition in the plasma \cite{breizman2019physics}.
Furthermore, circular tube cross-sections have been found to generate less reproducible size distributions than rectangular ones.
For circular shatter tubes, the effective shatter angle was found to be a function of the radial offset ot the pellet from the bend plane.
These findings suggest that rectangular shatter heads will produce more reliable and collimated fragment sprays in SPI disruption mitigation systems.\\

\noindent
Further steps could include the analysis of 8 mm pellet data to see if and how the observed trends scale with pellet diameter.
Also, by finding efficient and generally applicable pre-processing settings, the analysis procedure could be further optimized.
Since theoretical predictions of fragment sizes were less successful than hoped, it could be beneficial to construct an empirical model for the fragment size distribution from the observed data.
Trends in statistical parameters suggest a dependence on the normal velocity.
One could also try to look at fragment size distributions as a function of the ratio of kinetic impact energy to shatter threshold energy, as done by Gebhart \textit{et al.}~\cite{Gebhart} and Baylor \textit{et al.}~\cite{Baylor_2019}, thereby enabling comparison between multiple pellet materials.
Another interesting approach, which is already being tackled by ITER, is multi-camera analysis of the fragment plume.
Multiple observation angles might improve the estimation of fragment volume, at the price of adding another layer of complexity to the analysis.
The effects of incident speed and shatter angle on the velocity distribution width could also be investigated further to gain estimates for the arrival time of fragments.
The combination of fragment size distributions, position measurements and fragment velocity distributions could be used to calculate a fragment mass flow, providing realistic boundary conditions for future disruption mitigation simulations.

\bibliography{Master_thesis.bib}
\addcontentsline{toc}{chapter}{Bibliography}
\bibliographystyle{unsrt}

\newpage
\setcounter{page}{1} \pagenumbering{Roman}

\chapter{Appendix}

\begin{figure}[!h]
     \centering
     \includegraphics[clip, trim = 120 30 100 10,width=\textwidth]{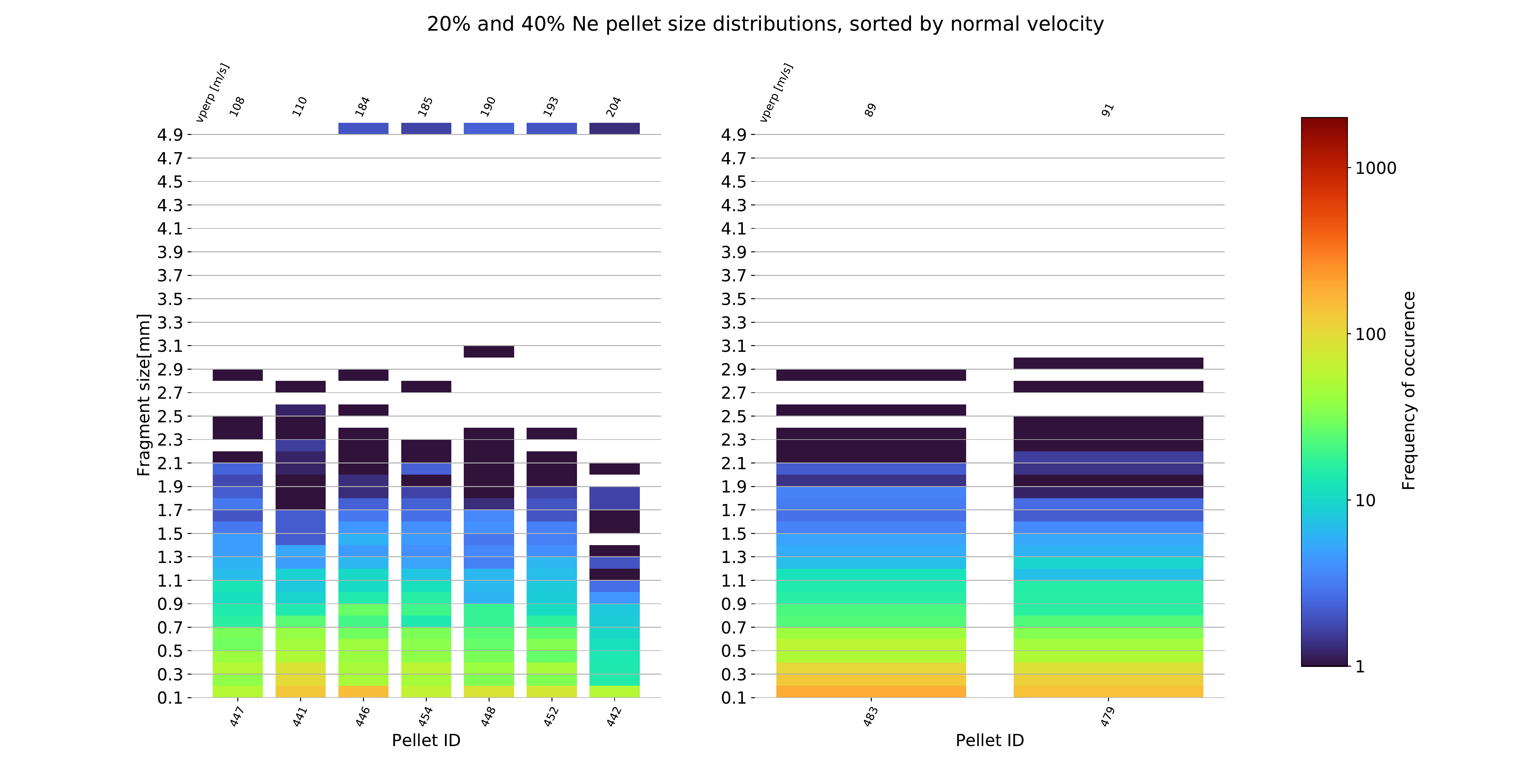}
     \caption{Experimental size distribution for 20\% Ne pellets (left) and 40\% Ne pellets (right), sorted by normal velocity (top). On the bottom, the corresponding pellet ID is given.}
     \label{f:mix2040_exp_vperp}
\end{figure}

\begin{figure}[!h]
     \centering
     \includegraphics[clip, trim = 120 30 100 10,width=\textwidth]{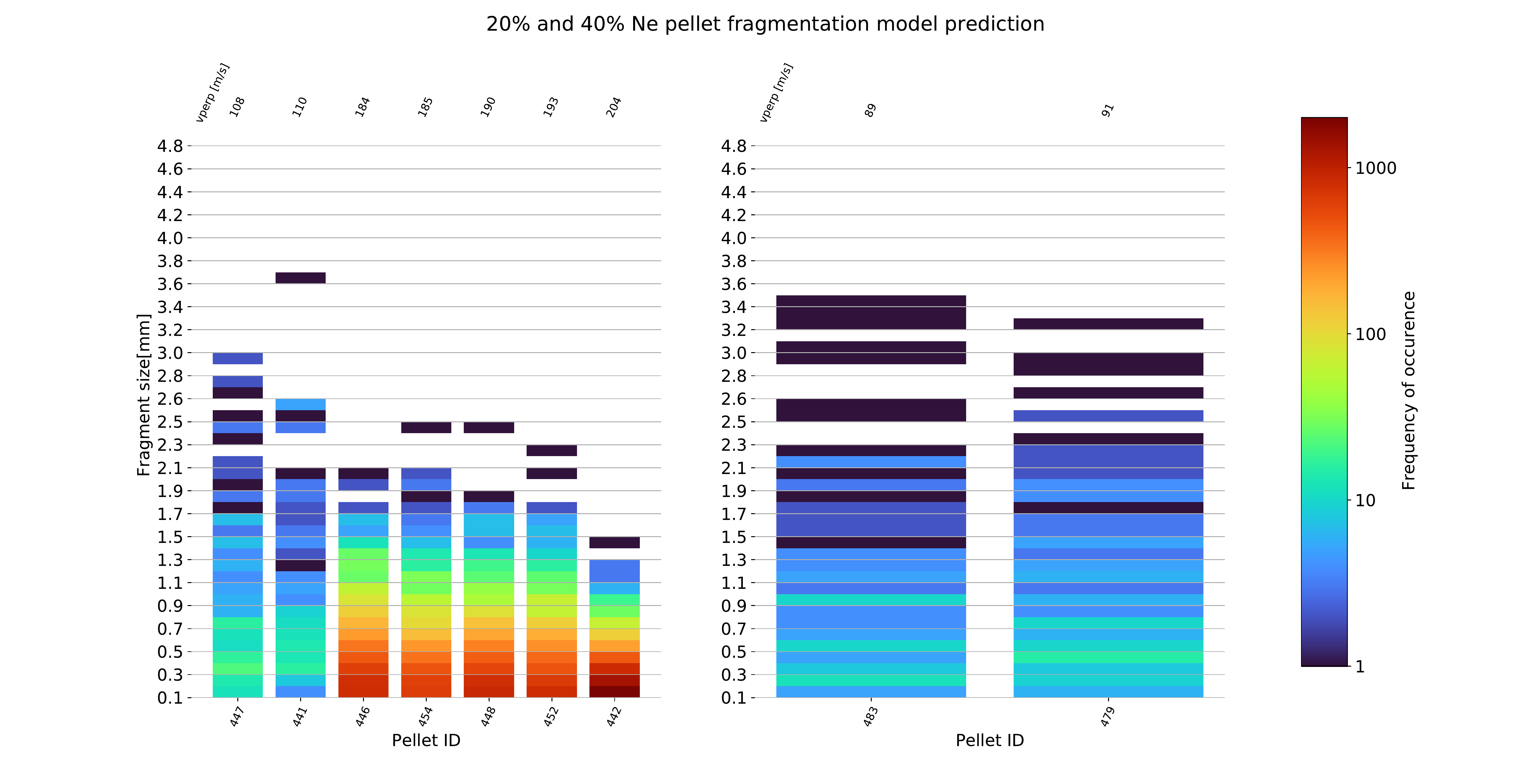}
     \caption{Modeled size distribution for 20\% Ne pellets (left) and 40\% Ne pellets (right) pellets, sorted by normal velocity (top). On the bottom, the corresponding pellet ID is given.}
     \label{f:mix2040_model_vperp}
\end{figure}

\begin{figure}[!h]
     \centering
     \includegraphics[clip, trim = 120 30 100 10,width=\textwidth]{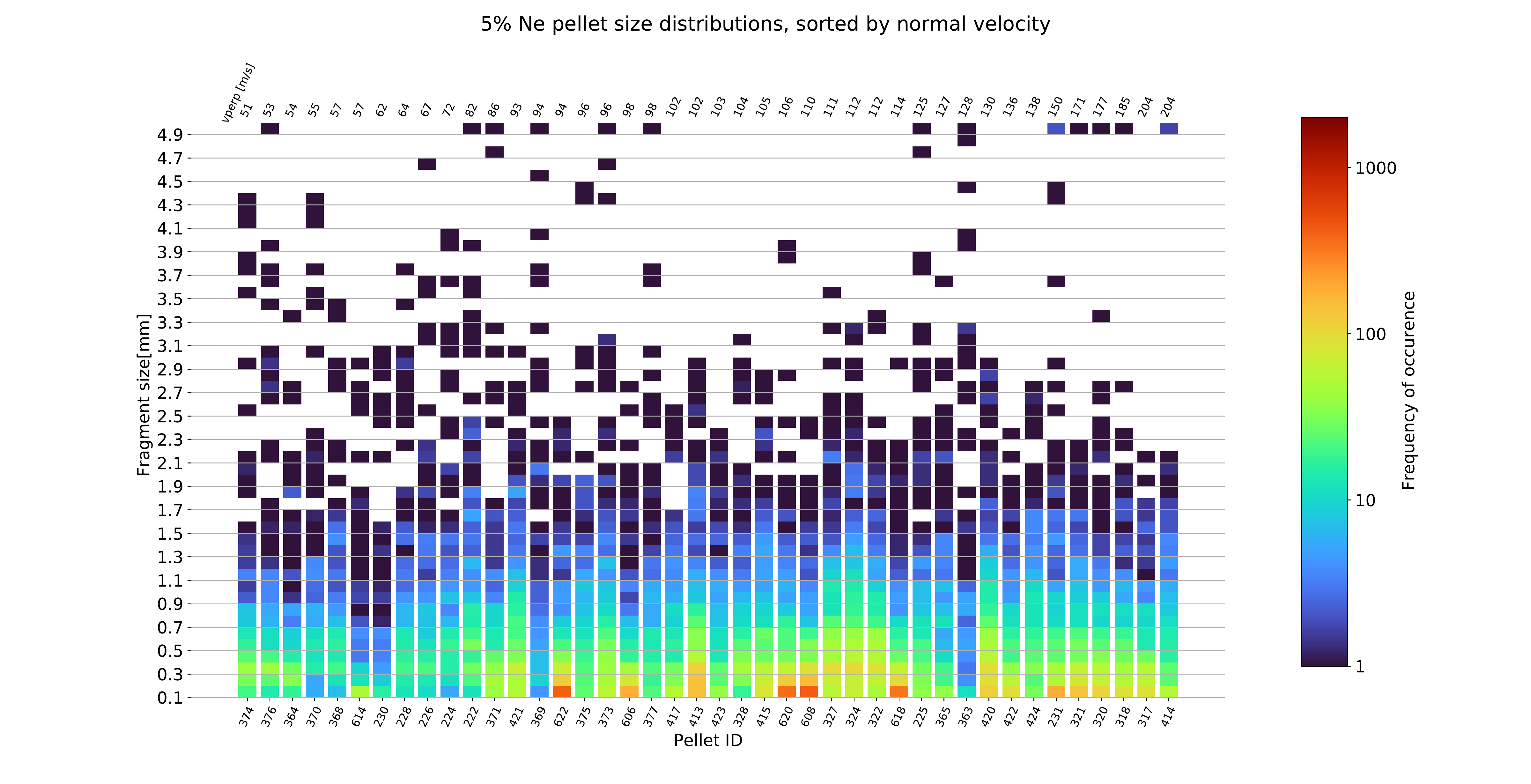}
     \caption{Experimental size distribution for 5\% Ne pellets, sorted by normal velocity (top). On the bottom, the corresponding pellet ID is given.}
     \label{f:mix5_exp_vperp}
\end{figure}

\begin{figure}[!h]
     \centering
     \includegraphics[clip, trim = 120 30 100 10,width=\textwidth]{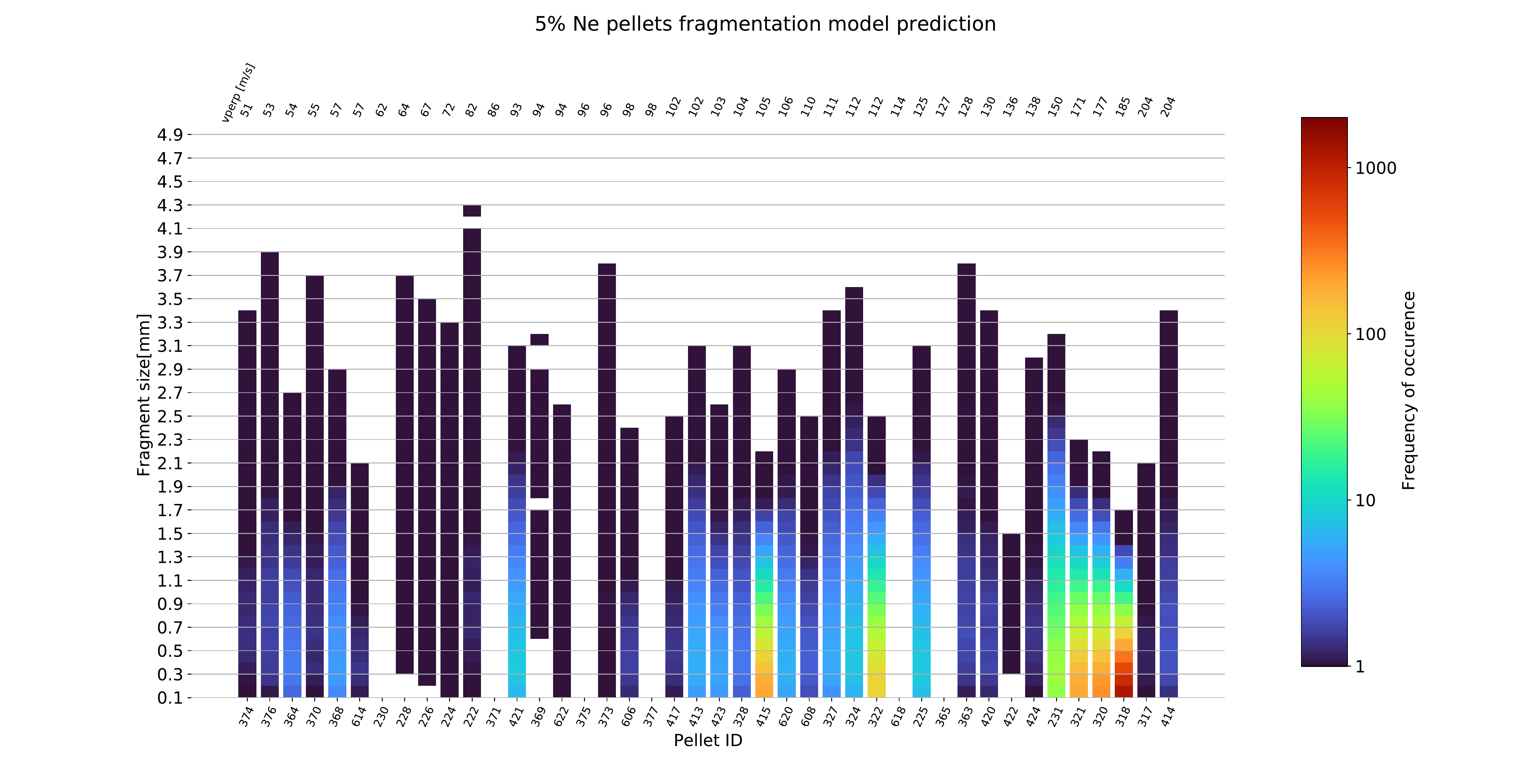}
     \caption{Modeled size distribution for 5\% Ne pellets, sorted by normal velocity (top). On the bottom, the corresponding pellet ID is given.}
     \label{f:mix5_model_vperp}
\end{figure}

\begin{figure}[!h]
     \centering
     \includegraphics[clip, trim = 110 30 90 10,width=\textwidth]{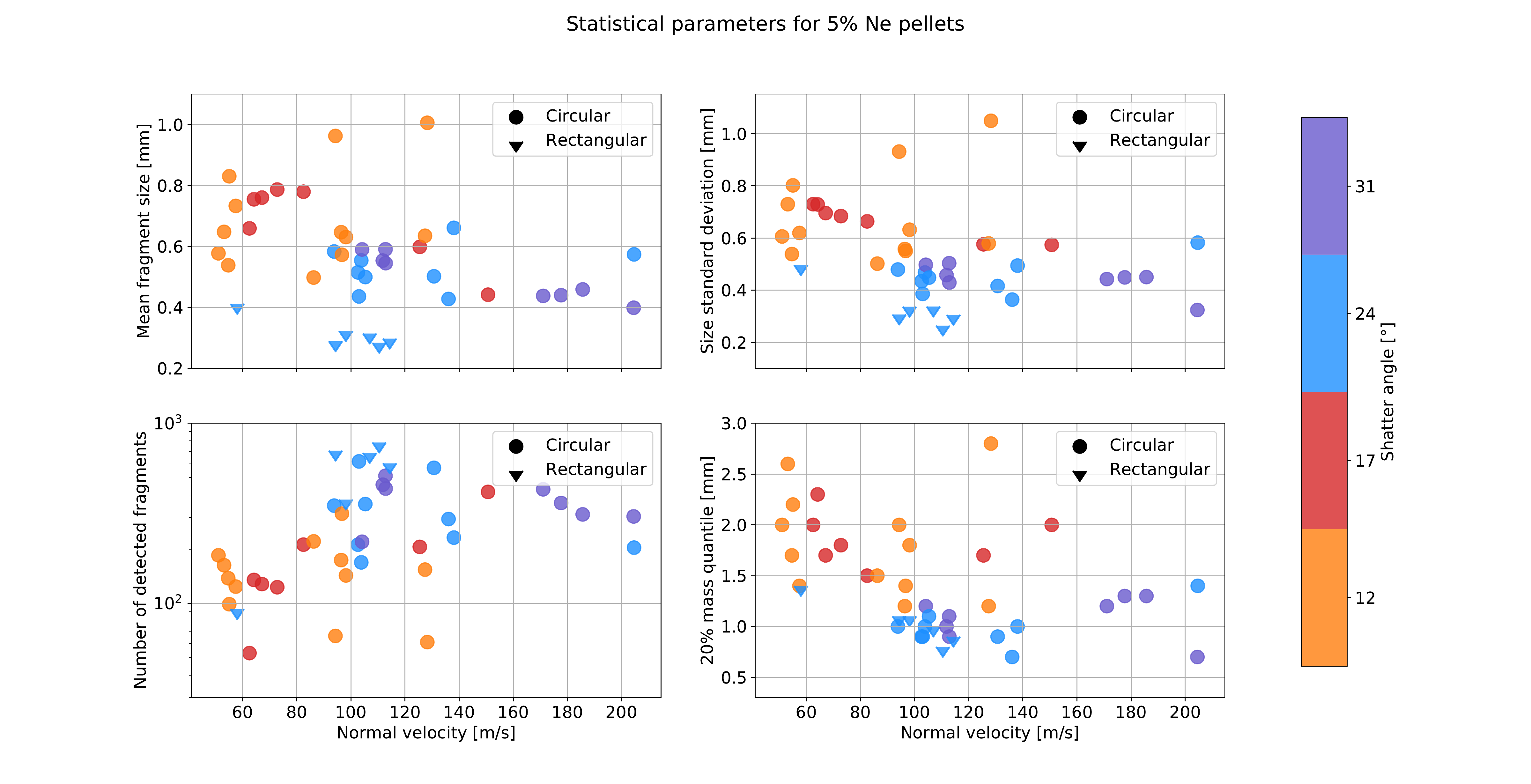}
     \caption{Experiment: Mean fragment size, standard deviation, number of fragments and the 20\% mass quantile plotted against perpendicular velocity for 5\% Ne pellets.}
     \label{f:mix5_exp_mean}
\end{figure}

\begin{figure}[!h]
     \centering
     \includegraphics[clip, trim = 110 30 90 10,width=\textwidth]{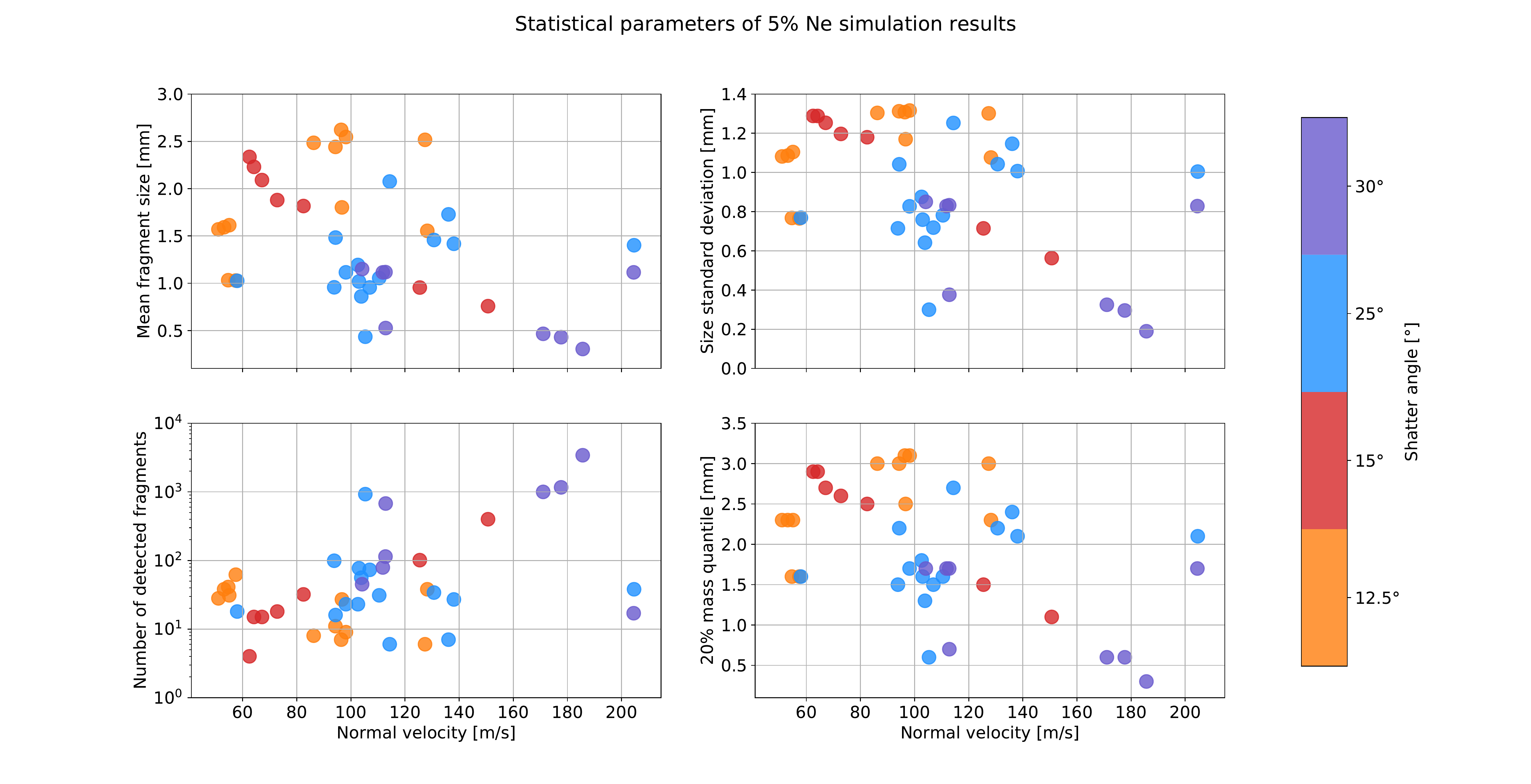}
     \caption{Simulation: Mean fragment size, standard deviation, number of fragments and the 20\% mass quantile plotted against perpendicular velocity for 5\% Ne pellets.}
     \label{f:mix5_model_mean}
\end{figure}

\begin{figure}[!h]
     \centering
     \includegraphics[clip, trim = 120 30 100 10,width=\textwidth]{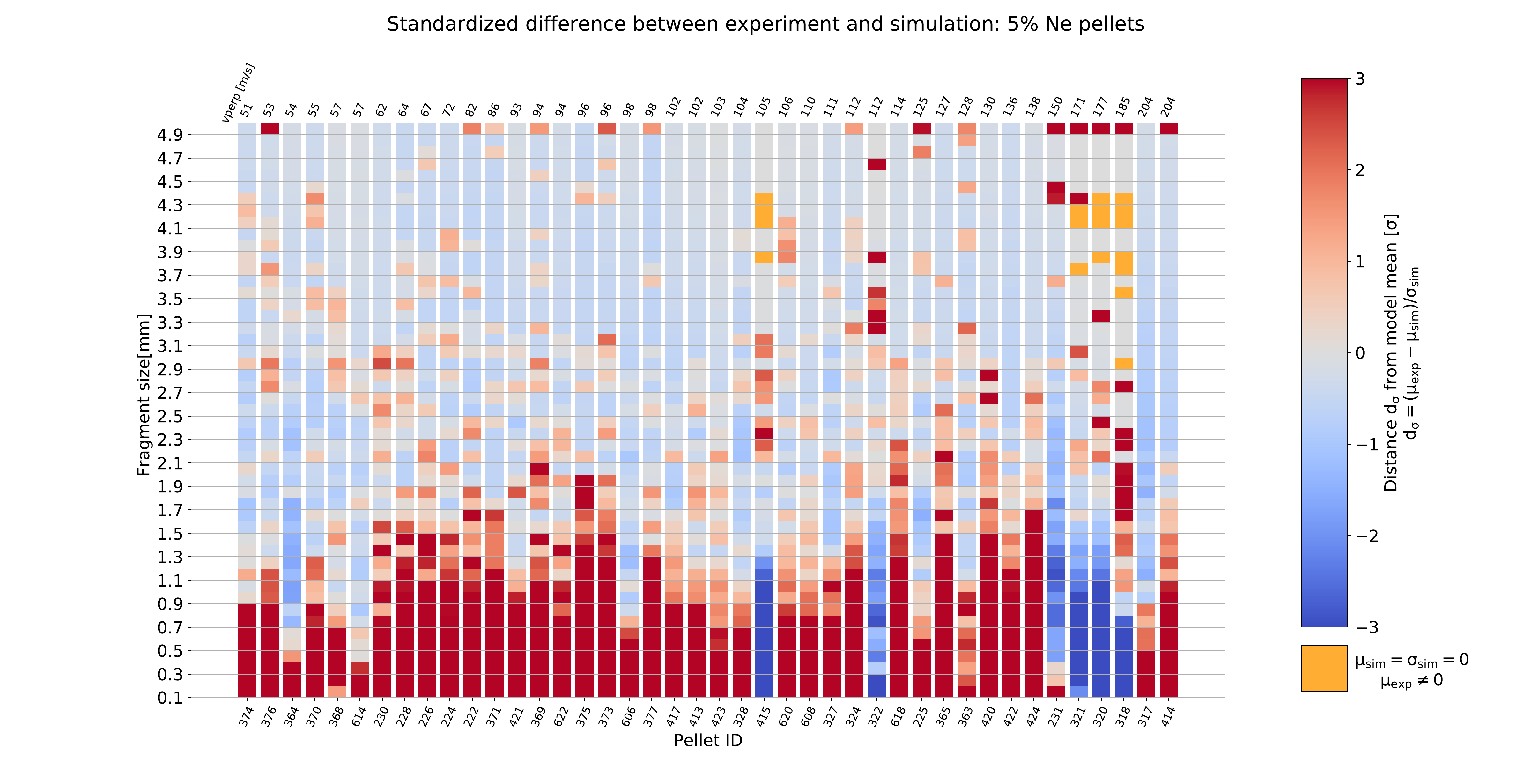}
\caption{Normalized distance of experimental and simulated size distributions for 5\% Ne pellets. Red indicates that the model predicts significantly less fragments for a bin, blue indicates that the model predicts significantly more fragments than observed.}
     \label{f:mix5_comp}
\end{figure}

\begin{figure}[!h]
     \centering
     \includegraphics[clip, trim = 120 30 100 10,width=\textwidth]{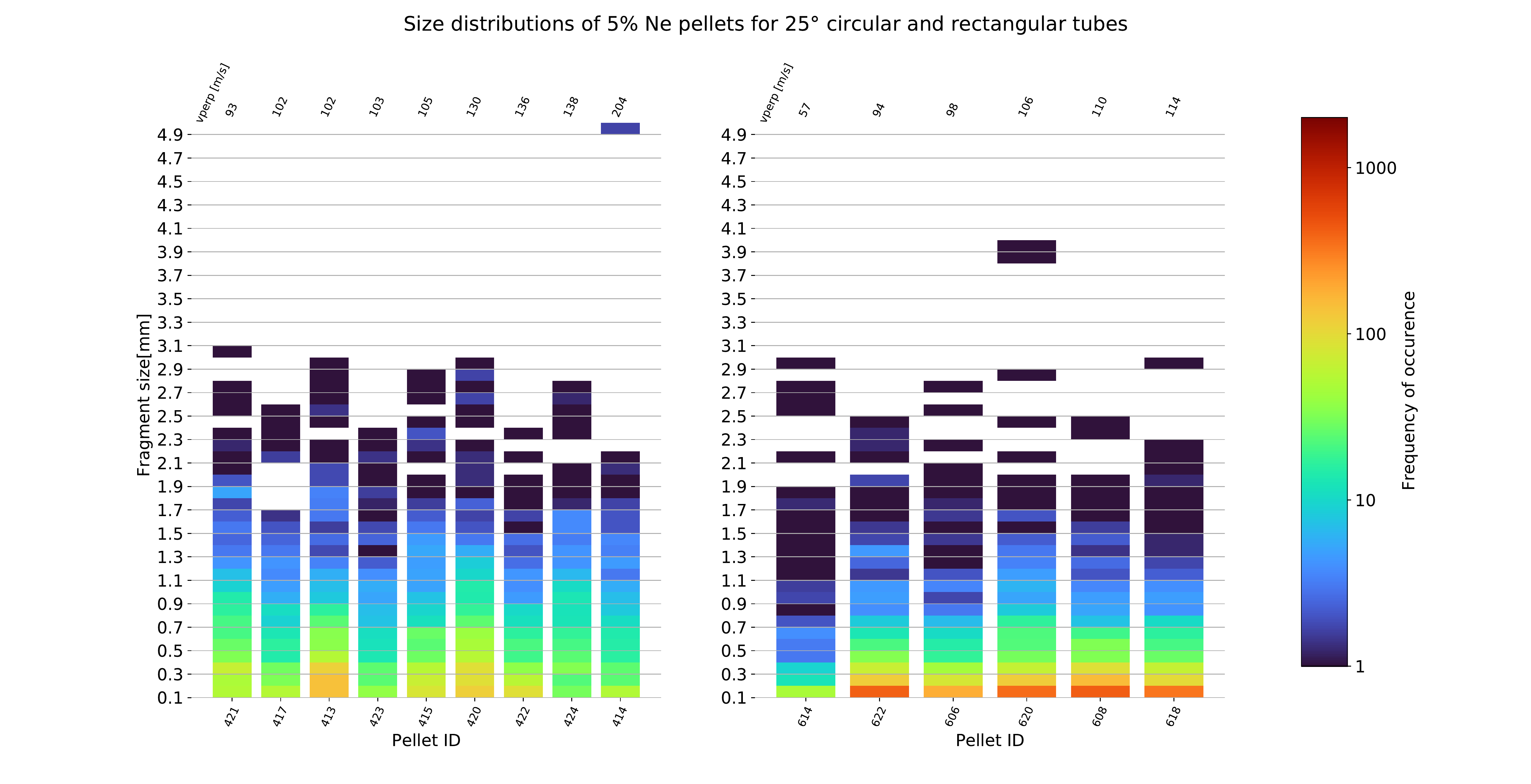}
     \caption{Experimental size distributions produced with 5\% Ne pellets, fired through a 25\degree circular (left) and rectangular (right) shatter tube.}
     \label{f:mix5_circ_rect}
\end{figure}

\begin{figure}[!h]
     \centering
     \includegraphics[clip, trim = 120 30 100 10,width=\textwidth]{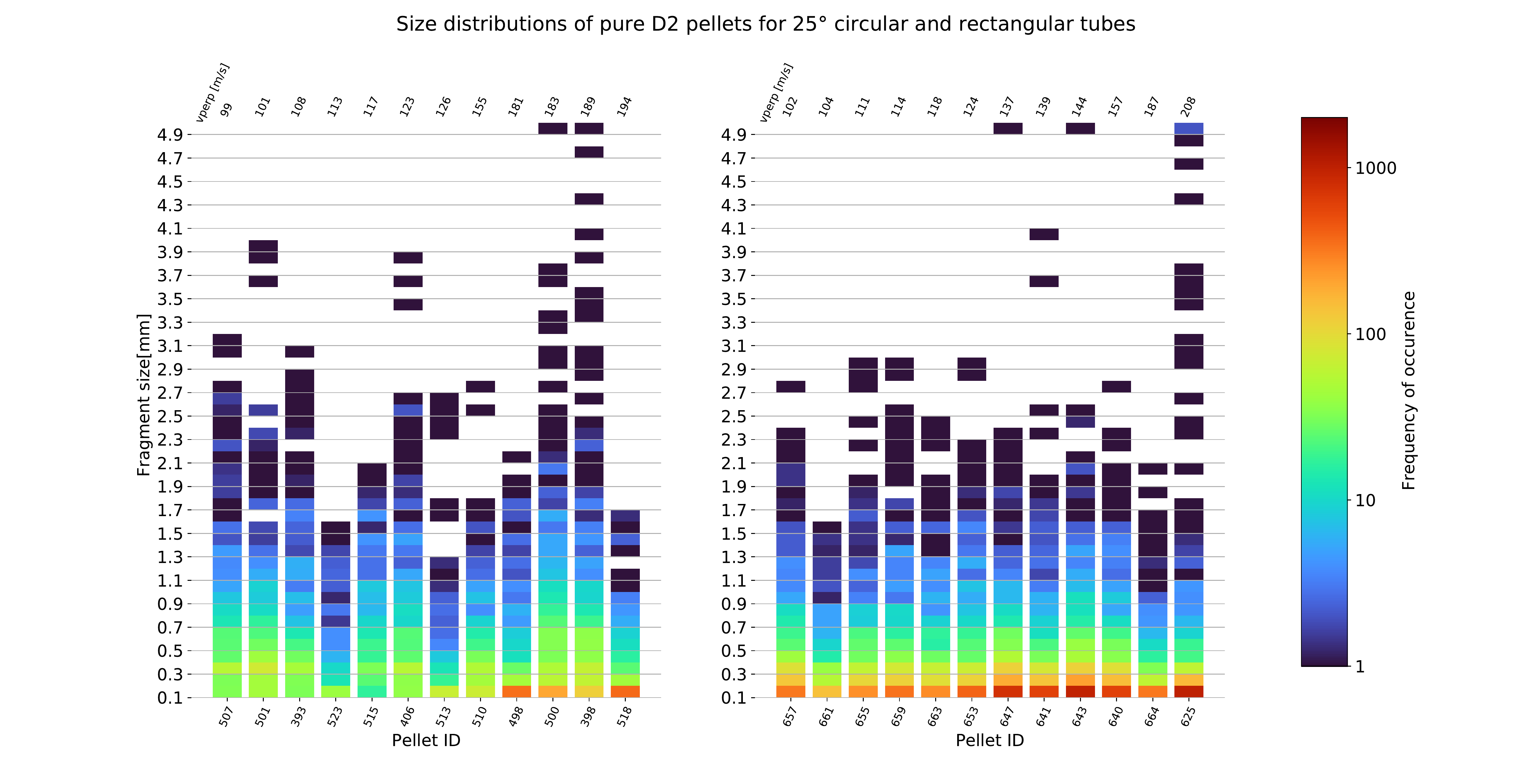}
     \caption{Experimental size distributions produced with pure D$_2$ pellets, fired through a 25\degree circular (left) and rectangular (right) shatter tube.}
     \label{f:de_circ_rect}
\end{figure}

\end{document}